# Polarimetric characterization of light and media
## Physical quantities involved in polarimetric phenomena

José J. Gil

*Department of Applied Physics. University of Zaragoza, Pedro Cerbuna 12, 50009 Zaragoza (Spain)*

ppgil@unizar.es

An analysis of the matrix models representing the polarimetric properties of light and material media is carried out by using the concept of the coherency matrix, which leads to the identification and definition of their corresponding physical quantities. For light, cases of homogeneous and inhomogeneous wavefront are analyzed, and a model for 3D polarimetric purity is formulated. For linear passive material media, a general model is developed on the basis that any physically realizable linear transformation of Stokes vectors is equivalent to an ensemble average of passive, deterministic, nondepolarizing transformations. Through this framework, the relevant physical quantities, including the indices of polarimetric purity, are identified and decoupled. Some decompositions of the whole system into a set of well-defined components are considered, as well as techniques for isolating the unknown components by means of new procedures for subtracting coherency matrices. These results and methods constitute a powerful tool for analyzing and exploiting experimental and industrial polarimetry. Some particular application examples are indicated.

## 1 Introduction

A proper description of the polarization properties of electromagnetic waves relies on the concept of the coherency matrix. This mathematical formulation is applicable regardless the particular band of the electromagnetic spectrum considered. In this paper we will frequently refer to "light" but, except for particular cases that are identified from the context, this term can be substituted by "electromagnetic radiation". In fact, there are many important industrial and research subjects involving polarization of electromagnetic radiation beyond the optical range such as microwaves, X-rays and gamma-rays.

Given a fixed point in space, the electric field of an electromagnetic wave describes an ellipse (the *polarization ellipse*) in a plane perpendicular to the propagation direction. For the ideal case of monochromatic light, both the shape of the polarization ellipse and the plane containing it (polarization plane) are fixed. In general, polychromatic light behaves as monochromatic for time intervals longer than the natural period $T_0$ of the wave and shorter than the coherence time $\tau$, so that the following different situations can be distinguished:

*In general, it is more rigorous to refer to the polarization plane than to the direction of propagation. Since a polarization a state is defined for a fixed point in space, the polarization plane is intrinsic of the state, while the direction of propagation is not intrinsic. In fact, certain types of evanescent waves have a 3D character (i.e. the polarization plane is not well defined) but with a well defined direction of propagation.*

- Both the polarization plane and the shape of the polarization ellipse are stable; in consequence, the light beam is fully polarized. For most purposes, and from a polarimetric point of view, quasimonochromatic totally polarized light can be considered equivalent to monochromatic light.

- The polarization plane is fixed, whereas the shape of the polarization ellipse fluctuates; in consequence, the light beam is partially polarized. It is well-known that these states are polarimetrically equivalent to a statistical mixture of two pure states with orthogonal polarizations. We will show that mixed states are also equivalent to statistical mixtures of non-orthogonal pure states.

- Both the polarization plane and the shape of the polarization ellipse fluctuate. These genuine *three-dimensional polarization states* require a particular treatment, different than what is commonly used for light with fixed polarization plane.

In general, concerning the optical region of the electromagnetic spectrum, the measurement time $T$ (i.e. the response time of the detector) is much longer than the coherence time. Typical values of the indicated time intervals are the following: $T_0 \equiv 1/\nu_0 \cong 10^{-15}\,\text{s}$ ($\nu_0$ being the central frequency of the spectral profile of the wave), $10^{-9}\,\text{s} \leq \tau \leq 10^{-4}\,\text{s}$ and $T \cong 10^{-4}\,\text{s}$ [1,2]. Thus, for most purposes the assumption of quasimonochromaticity is justified.

In the second-order optics approach, the Stokes parameters [3] provide a complete mathematical characterization of the different states of polarization. The physical meaning of these four real parameters has been studied by some authors [4-7].

When a material medium is illuminated by an electromagnetic wave, molecular electric charges are set in oscillatory motion by the electric field of the wave, producing secondary radiation, so that the overall effects of the combined basic interactions (scattering in its essential sense) [8] results in refracted, reflected, diffracted or scattered light. The effects of linear passive media can be represented by linear transformations of the electric field variables. By *passive* we refer to the property of not amplifying the light intensity.

Depending on: (1) the nature and particular conditions of the linear interaction; (2) the spectral profile of the incident light beam; (3) the chromatic properties of the material sample, and (4) the specific polarizing properties of the sample, the waves emerging from different homogeneous parts of the material target can have different degrees of mutual coherence. Coherent interactions can be represented through the Jones formalism [9], whereas for the general case of incoherent interactions of polychromatic light, the Stokes-Mueller model is required.

Concerning the transformations of the state of polarization due to the interaction of the wave with linear material media, Jones introduced a simple method applicable for deterministic interactions [9]. In this model, the polarization states are represented by two-component complex *Jones vectors* and the medium is characterized by a 2x2 complex *Jones matrix*.

The classic contributions of authors such as Soleillet [10], Perrin [11], Mueller [12] and Parke III [13,14], have led to a more general model, in which the action of media is represented through linear transformations of the Stokes parameters. Although the 4x4 real matrices characterizing these transformations are usually called *Mueller matrices* it should be noted that it was Soleillet who first introduced implicitly this concept [10]. The lecture notes where Mueller used these matrices remained unpublished except for a brief note [12]. The connection between the Jones and the Soleillet-Mueller models was studied by Parke III [13,14]. An interesting compilation of seminal works on polarization optics can be found in Ref. [15].

Some works have dealt with the case of deterministic nondepolarizing media (polarimetrically pure media), leading to clear physical interpretations of the seven independent parameters that appear in this case [16,17]. In the general case, a complete polarimetric characterization of the medium requires considering up to sixteen independent parameters.

Given the essential nature of polarization phenomena, polarimetry is a general-purpose technique which can be applied in a great variety of industrial, medical and scientific environments. Thus, different interaction conditions can be arranged in order to obtain information on material samples from polarimetric measurements. These techniques have been widely used for many years. One particular technique, called *ellipsometry* [18,19], is based on determining some properties of material samples from the changes produced on totally polarized light. The interaction transforms the *input polarization ellipse*, characterized by a complex number that contains information on the corresponding azimuth and ellipticity, into a reflected *output polarization ellipse*. The ratio between these numbers results in two ellipsometric real parameters. Ellipsometric techniques provide very fast in-situ and non-destructive control of some industrial processes. Nevertheless, it should be noted that the term *ellipsometry* is also frequently used as a synonym for *polarimetry* and, hence, refers to general techniques for measuring polarimetric properties.

In general, interactions not only produce changes in the ellipsometric parameters, but also cause selective changes in the transmitted intensity as well as depolarizing effects, so that, in these cases, a complete polarimetry is required. For the measurement of all the sixteen elements of a Mueller matrix, the material sample is placed between a generator and an analyzer of polarization states. The generator should be arranged into at least four independent configurations for each one of which the analyzer should in turn be disposed into at least four independent configurations. Usually, the generator includes a total polarizer followed by a retarder placed in front of the light source, whereas the analyzer includes a retarder followed by a total polarizer before the detector. By means of rotating wave-plates, Pockels cells or other kinds of controllable retarders, these polarimeters can be automatically operated [20-28].

The analyzer described, used alone, can be used as a Stokes polarimeter, enabling the Stokes parameters of the incoming light to be measured [29]. Other interesting devices are based on the use of polarization gratings [30,31].

Different kinds of polarimeters are used in many different environments such as remote sensing [32]; scattering applications [33,8,34]; ITER&Tokamak [35]; Astronomy [36], etc. For some applications, spectral polarimetry and interferometry are combined into a whole device [35,37-40].

Polarimetric techniques are powerful tools for the study and analysis of material samples because, given the specific interaction conditions (reflection, refraction, scattering…) and given the spectral and spatial properties of the light probe, up to sixteen independent parameters can be measured. These measurable quantities are the elements of the Mueller matrix, which characterize the polarimetric properties of the sample with respect to the interaction conditions mentioned. Nevertheless, important limitations of these techniques arise from the absence of a complete understanding of the information contained in the Mueller matrix. The physical information is structured in a complicated manner and, thus, the obtainment of a set of sixteen quantities with clear physical meaning is not straightforward. In fact, as we will see, some new parameters should be introduced in order to get the indicated objective.

Consequently, the mathematical characterization of Mueller matrices is a key matter in the exploitation of powerful polarimetric techniques. In general, the values of the elements of Mueller matrices are restricted by certain quadratic and bilinear constraining inequalities, which have been the main subject of a number of papers.

In this review, we consider the mathematical description of the polarimetric properties of light and media through a unified model based on the concept of coherency matrices. The measurable quantities arise as the coefficients of the expansion of the coherency matrix in a set of Hermitian trace-orthogonal matrices constituted by the generators of the SU(*n*) group plus the $n \times n$ identity matrix. The physical parameters with direct physical meaning are defined from the corresponding measurable quantities.

The different sections of this paper include relevant results quoted in the literature, as well as some new results, which are treated on the basis of the unified framework indicated [41]. Thus, for example, in the light of the respective *indices of polarimetric purity*, the polarimetric purity of 3D states of polarized light is analyzed, as well as the polarimetric purity of media. Moreover, the retarding, polarizing and depolarizing properties of material media are characterized through a set of sixteen independent parameters with direct physical meaning. We will find that ten of these parameters are physically-invariant quantities.

The following sections consider successively the polarization states of plane waves, the 3D states of polarization and the polarimetric properties of material media. In general, when possible, the common notations and names have been preserved. In certain cases, particular notations have been introduced for the sake of clarity.

## 2 Polarized light

Given a point in space, the state of polarization of a light beam that propagates in a fixed direction $Z$ is given by the temporal evolution of the electric field of the electromagnetic wave, which lies in a *polarization plane* perpendicular to the propagation direction. Let $\mathbf{E}(z,t)$ be the electric field at a point $z$, at time $t$, of a quasimonochromatic plane wave propagating in an isotropic medium, and let $(\mathbf{e}_1, \mathbf{e}_2, \mathbf{e}_3)$ be a reference basis of orthonormal vectors along the respective axes *XYZ*. The components of the electric field are

$$E_x(z,t) = A_x(t) \cos[kz - \omega t + \beta_x(t)]$$
$$E_y(z,t) = A_y(t) \cos[kz - \omega t + \beta_y(t)]$$
(1)

where $k = 2\pi/\lambda_0$ ($\lambda_0$ is the wavelength for the vacuum); $\omega$ is the angular frequency, $\omega = 2\pi\nu$ ($\nu$ is the natural frequency); $\beta_x, \beta_y$ are the respective phases, and $A_x, A_y$ are the respective amplitudes.

In this case of quasimonochromatic light, i.e. when the bandwidth $\Delta\nu$ is very narrow with respect to the central frequency $\bar{\nu}$ of the spectral profile, the components of the electric field can be expressed as

$$E_x(z,t) = A_x(t)\cos\left[\bar{k}\,z - \bar{\omega}t + \beta_x(t)\right]$$
$$E_y(z,t) = A_y(t)\cos\left[\bar{k}\,z - \bar{\omega}t + \beta_y(t)\right] \tag{2}$$

where $\bar{k}, \bar{\omega}$ are the respective mean values of $k, \omega$.

As usual in polarization optics, under this assumption of quasimonochromaticity, it is very advantageous to use the analytic signal representation, where the components are described through their respective complex variables [6].

Let us consider the analytical signal representations $\eta_x(t), \eta_y(t)$ of the two mutually orthogonal components of the electric field in the polarization plane

$$\eta_x(t) = E_x(t) + i\tilde{E}_x(t) = A_x(t)\,e^{i[u(t)+\beta_x(t)]},$$
$$\eta_y(t) = E_y(t) + i\tilde{E}_y(t) = A_y(t)\,e^{i[u(t)+\beta_y(t)]}, \tag{3}$$

where, $\tilde{E}_x(t), \tilde{E}_y(t)$, are the Hilbert transforms of the real components of the electric field, and $u(t) = \bar{k}z - \bar{\omega}t$.

These two components can be arranged as the components of the following 2x1 complex vector

$$\boldsymbol{\eta}(t) = \begin{pmatrix} \eta_1 \\ \eta_2 \end{pmatrix} = e^{i[u(t)+\beta_x(t)]}\begin{pmatrix} A_x(t) \\ A_y(t)\,e^{i[\beta_y(t)-\beta_x(t)]} \end{pmatrix}, \tag{4}$$

and, by avoiding the global phase factor (without physical meaning for an isolated state of polarization), the *instantaneous Jones vector* is defined as

$$\boldsymbol{\varepsilon}(t) = \begin{pmatrix} A_x(t) \\ A_y(t)\,e^{i\delta_y(t)} \end{pmatrix} \tag{5}$$

where $\delta_y(t) = \beta_y(t) - \beta_x(t)$ is the relative phase.

This vector includes all the information relative to the temporal evolution of the electric field. It is called *instantaneous* because the amplitudes and relative phase are time-dependent variables. In the case of a polychromatic wave, the instantaneous Jones vector has slow time dependence with respect to the coherence time, so that, for time intervals shorter than the coherence time, the shape of the polarization ellipse can be considered fixed. For time intervals longer than the coherence time of the light wave, the instantaneous Jones vector can vary, resulting in partially polarized light.

Thus, when the following quantities have not time dependence

$$\frac{A_y(t)}{A_x(t)} = \text{constant}, \quad \delta_y(t) = \text{constant}, \tag{6}$$

so that the electric field describes a stable well-defined ellipse (Fig. 1), the Jones vector is defined as [9]

$$\boldsymbol{\varepsilon} = \begin{pmatrix} A_x \\ A_y e^{i\delta_y} \end{pmatrix}. \tag{7}$$

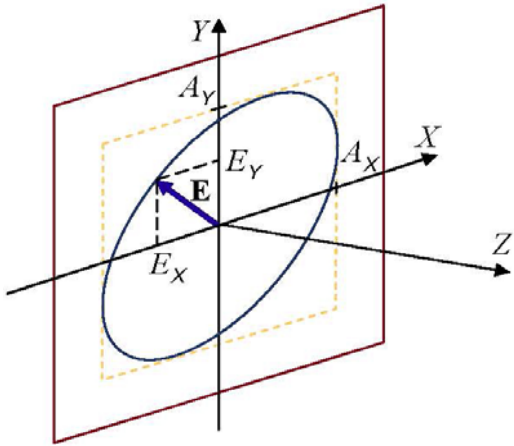   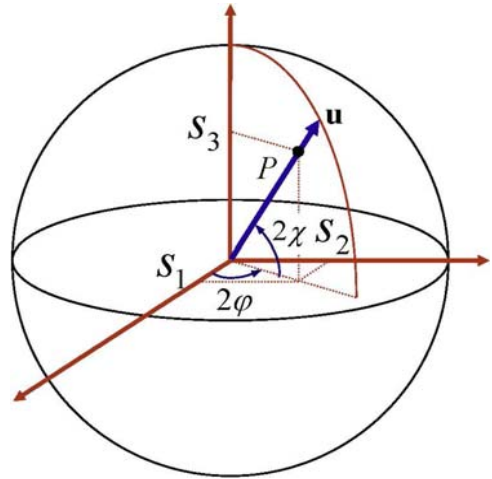

Fig. 1. Polarization ellipse and polarization plane       Fig. 2. Poincaré sphere

It should be noted that the conditions given by Eq. (6) are compatible with intensity fluctuations. In fact, totally polarized light maintains the azimuth and ellipticity of the polarization ellipse fixed whereas the size of the ellipse fluctuates resulting in a mean intensity over the measurement time. Moreover, slow time variations of the Jones vector with respect to the measurement time can be represented by this model [42].

We see that the Jones vector provides a basic model for representing the state of polarization when light is totally polarized. The state of polarization of the coherent superposition of totally polarized light beams is given by the sum of the corresponding Jones vectors.

In the case of partially polarized light, the azimuth and (or) ellipticity of the polarization ellipse varies during the measurement time. In consequence, the treatment of partially polarized light requires a different mathematical model in order to take into account all the parameters that characterize completely the state of polarization.

### 2.1 The coherency matrix

The *coherency matrix* (or *polarization matrix*) $\boldsymbol{\Phi}$ [43-45,4,6,13] of a light beam contains all the measurable information on its state of polarization (including intensity). This Hermitian 2x2 matrix is defined as

$$\boldsymbol{\Phi} = \langle \boldsymbol{\varepsilon}(t) \otimes \boldsymbol{\varepsilon}^\dagger(t) \rangle = \begin{pmatrix} \langle \varepsilon_1(t)\varepsilon_1^*(t) \rangle & \langle \varepsilon_1(t)\varepsilon_2^*(t) \rangle \\ \langle \varepsilon_2(t)\varepsilon_1(t) \rangle & \langle \varepsilon_2(t)\varepsilon_2^*(t) \rangle \end{pmatrix} \tag{8}$$

where $\boldsymbol{\varepsilon}$ is the *instantaneous Jones vector* whose two components are the analytic signals of the electric field of the wave; $\boldsymbol{\varepsilon}^\dagger$ is the conjugate transposed vector of $\boldsymbol{\varepsilon}$; $\varepsilon_i^*$ represents the complex conjugate of $\varepsilon_i$; $\otimes$ stands for the Kronecker product, and the brackets indicate time averaging over the measurement time

$$\langle X(t) \rangle = \lim_{T \to \infty} \frac{1}{T} \int_0^T X(t) \, dt \,. \tag{9}$$

Therefore, $\mathbf{\Phi}$ is a covariance matrix whose elements are the second-order moments of the zero-mean analytic signals $\varepsilon_i(t)$; $i = 1, 2$. Under the assumption that they are stationary and ergodic, the brackets can alternatively be considered as ensemble averaging of $\mathbf{\varepsilon} \otimes \mathbf{\varepsilon}^\dagger$, where $\mathbf{\varepsilon}$ are simple realizations.

Due to its statistical nature, $\mathbf{\Phi}$ is characterized by the fact that its two eigenvalues are nonnegative. These two constraints determine a complete set of necessary and sufficient conditions for a Hermitian matrix $\mathbf{\Phi}$ to be a coherency matrix, i.e. to represent a particular state of polarization of a light beam.

The statistical properties of $\mathbf{\Phi}$ appear explicitly when its elements are written in terms of the corresponding standard deviations $\sigma_1, \sigma_2$ and the complex degree of mutual coherence $\mu$

$$\mathbf{\Phi} = \begin{pmatrix} \sigma_1^2 & \mu \sigma_1 \sigma_2 \\ \mu^* \sigma_1 \sigma_2 & \sigma_2^2 \end{pmatrix} \tag{10}$$

As some authors have pointed out [46, 2] the normalized matrix

$$\hat{\mathbf{\Phi}} \equiv \mathbf{\Phi} / \operatorname{tr} \mathbf{\Phi} \tag{11}$$

is the corresponding *polarization density matrix*, which contains information on the populations and coherences of the polarization states.

## 2.2 The Stokes parameters

The coherency matrix $\mathbf{\Phi}$ can be expressed as a linear expansion, with real coefficients, in the following basis, constituted by the three Pauli matrices plus the identity matrix

$$\mathbf{\sigma}_0 = \begin{pmatrix} 1 & 0 \\ 0 & 1 \end{pmatrix}, \quad \mathbf{\sigma}_1 = \begin{pmatrix} 1 & 0 \\ 0 & -1 \end{pmatrix}, \quad \mathbf{\sigma}_2 = \begin{pmatrix} 0 & 1 \\ 1 & 0 \end{pmatrix}, \quad \mathbf{\sigma}_3 = \begin{pmatrix} 0 & -i \\ i & 0 \end{pmatrix} \tag{12}$$

It should be noted that, although the habitual notations for the variances ($\sigma_1, \sigma_2$) and for the Pauli matrices ($\mathbf{\sigma}_i$) have been preserved, this should not lead to confusion because $\mathbf{\sigma}_i$ are matrices (bold), whereas the variances are scalar quantities. Observe also that in quantum physics and quantum optics the Pauli matrices $\mathbf{\sigma}_2$ and $\mathbf{\sigma}_3$ are frequently labeled as $\mathbf{\sigma}_3$ and $\mathbf{\sigma}_2$ respectively.

This set of linearly independent matrices $\mathbf{\sigma}_i$ constitutes an appropriate basis for the space of 2×2 Hermitian matrices. They are Hermitian $\mathbf{\sigma}_i = \mathbf{\sigma}_i^\dagger$; trace-orthogonal $\operatorname{tr}(\mathbf{\sigma}_i \mathbf{\sigma}_j) = 2\delta_{ij}$, and satisfy $\mathbf{\sigma}_i^2 = \mathbf{D}(1,1)$, [$\mathbf{D}(1,1) \equiv \operatorname{diag}(1,1)$ being the 2×2 identity matrix]. Thus, these matrices are unitary and, except for $\mathbf{\sigma}_0$, are traceless. This *Pauli basis* allows $\mathbf{\Phi}$ to be expressed as the following linear expansion [4,47].

$$\mathbf{\Phi} = \frac{1}{2} \sum_{i=0}^{3} s_i \mathbf{\sigma}_i \tag{13}$$

where the real coefficients $s_i$ are given by

$$s_i = \mathrm{tr}(\mathbf{\Phi}\, \boldsymbol{\sigma}_i), \quad i = 0, 1, 2, 3. \tag{14}$$

The *Stokes parameters* $s_0, s_1, s_2, s_3$ are measurable quantities which satisfy the restrictions given by the inequalities

$$s_0 \geq 0, \quad s_0^2 \geq s_1^2 + s_2^2 + s_3^2, \tag{15}$$

which are equivalent to the nonnegativity of the coherency matrix $\mathbf{\Phi}$.

These parameters, arranged as a 4x1 vector, constitute the *Stokes vector* **s**. Although in this section we will use the Stokes vector, we also introduce the following alternative notation in order to compare some expressions with others that will appear in further sections

$$\mathbf{S} = \begin{pmatrix} s_{00} & s_{01} \\ s_{10} & s_{11} \end{pmatrix} \equiv \begin{pmatrix} s_0 & s_1 \\ s_2 & s_3 \end{pmatrix}. \tag{16}$$

## 2.3 The purity criterion

Let us consider now the Euclidean norms of $\mathbf{\Phi}$ and $\mathbf{S}$

$$\|\mathbf{\Phi}\|_2 \equiv \sqrt{\sum_{i,j=1}^{2} |\phi_{ij}|^2} = \sqrt{\mathrm{tr}(\mathbf{\Phi}^\dagger \mathbf{\Phi})} = \sqrt{\mathrm{tr}\,\mathbf{\Phi}^2}, \qquad \|\mathbf{S}\|_2 \equiv \sqrt{\sum_{i=0}^{3} s_i^2} = \sqrt{\mathrm{tr}(\mathbf{S}^T \mathbf{S})}. \tag{17}$$

Moreover, taking into account that $\mathbf{\Phi}$ is a positive semidefinite Hermitian matrix, we can define the following norm

$$\|\mathbf{\Phi}\|_0 \equiv \mathrm{tr}\,\mathbf{\Phi} = \left\|\sqrt{\mathbf{\Phi}}\right\|_2^2. \tag{18}$$

It is easy to show that the above norms satisfy the following relations

$$\|\mathbf{\Phi}\|_2^2 = \frac{1}{2}\|\mathbf{S}\|_2^2, \tag{19}$$

$$\|\mathbf{\Phi}\|_0 = s_0, \tag{20}$$

$$\frac{1}{2}\|\mathbf{\Phi}\|_0^2 \leq \|\mathbf{\Phi}\|_2^2 \leq \|\mathbf{\Phi}\|_0^2. \tag{21}$$

For pure states, $\|\mathbf{\Phi}\|_2^2 = \|\mathbf{\Phi}\|_0^2$. This equality constitutes an objective purity criterion, while the other limit $\|\mathbf{\Phi}\|_2^2 = \|\mathbf{\Phi}\|_0^2/2$ occurs in the case of unpolarized light (equiprobable mixture of states).

## 2.4 Physical quantities characterizing the polarized light

The Stokes vector can be written as

$$\mathbf{s} = I \begin{pmatrix} 1 \\ P\mathbf{u} \end{pmatrix}, \tag{22}$$

where the characteristic physical quantities, namely (1) the intensity $I$; (2) the degree of polarization $P$; and (3) the unit vector **u** (*Pauli vector*) that determines univocally the azimuth $\varphi\,(0 \leq \varphi < \pi)$ and the ellipticity angle $\chi\,(-\pi/4 \leq \chi \leq \pi/4)$ of the average polarization ellipse, appear explicitly.

$$I = s_0, \quad P \equiv \frac{\sqrt{s_1^2 + s_2^2 + s_3^2}}{s_0}, \quad \mathbf{u} \equiv \frac{1}{IP}\begin{pmatrix} s_1 \\ s_2 \\ s_3 \end{pmatrix} = \frac{1}{IP}\begin{pmatrix} \cos 2\chi \cos 2\varphi \\ \cos 2\chi \sin 2\varphi \\ \sin 2\chi \end{pmatrix}. \tag{23}$$

In the second-order approach, the Poincaré sphere (Fig. 2) provides a meaningful geometric representation of all possible states of polarization. Taking into account the characteristic properties of the Stokes parameters, they define a 4D cone, whose generatrix is the $s_0$ axis. The cut of this cone with the plane $s_0 = 1$ leads to the 3D representation given by the Poincaré sphere, which is a solid sphere of radius 1 where all states have intensity equal to 1. Points on the surface represent totally polarized states (pure states) and points inside the sphere represent partially polarized states (mixed states). The origin represents second-order unpolarized light.

### 2.5 Decomposition of mixed states

From Eq. (22), the following *characteristic decomposition* (or *trivial decomposition*) of **s** as a convex sum of a pure state and a unpolarized state can be immediately obtained [6,44]

$$\mathbf{s} = I\begin{pmatrix} 1 \\ P\mathbf{u} \end{pmatrix} = I\left[P\begin{pmatrix} 1 \\ \mathbf{u} \end{pmatrix} + (1-P)\begin{pmatrix} 1 \\ \mathbf{0} \end{pmatrix}\right], \tag{24}$$

Another alternative decomposition with different physical meaning is [44]

$$\mathbf{s} = I\begin{pmatrix} 1 \\ P\mathbf{u} \end{pmatrix} = I\left[\frac{1+P}{2}\begin{pmatrix} 1 \\ \mathbf{u} \end{pmatrix} + \frac{1-P}{2}\begin{pmatrix} 1 \\ -\mathbf{u} \end{pmatrix}\right]. \tag{25}$$

This *spectral decomposition*, represented graphically in Fig.3, expresses a mixed state as a convex sum of two orthogonal pure states (their respective Jones vectors are orthogonal), which have the same direction in the Poincaré sphere and are represented by antipodal points (these states have the equal azimuths and their ellipticities differ by $\pi/2$). The origin of the name *spectral decomposition* is derived from the fact that the above decomposition of a Stokes vector is equivalent to the spectral decomposition of its corresponding coherency matrix.

Nevertheless, given a mixed state, there exist infinite possibilities for decomposing it as a convex sum of two pure states

$$\mathbf{s} = I\begin{pmatrix} 1 \\ P\mathbf{u} \end{pmatrix} = I[c\mathbf{x} + (1-c)\mathbf{y}]; \quad \mathbf{x} \equiv \begin{pmatrix} 1 \\ \mathbf{v} \end{pmatrix}, \quad \mathbf{y} \equiv \begin{pmatrix} 1 \\ \mathbf{w} \end{pmatrix}, \tag{26}$$

where $\mathbf{v}, \mathbf{w}$ are linearly independent unit vectors, and $0 < c < 1$.

Given a mixed state **s**, any pure state **x** can be considered an incoherent component of **s**. In fact, once **x** is chosen, the coefficient $c$ and the remaining pure component **y** are univocally determined

$$c = \frac{1-P^2}{2(1-P\mathbf{u}^T\mathbf{v})}, \quad 0 < c < 1; \quad \mathbf{w} = \frac{P\mathbf{u} - c\mathbf{v}}{1-c}. \tag{27}$$

Provided **s** is a mixed state, it is straightforward to show that condition $0 < c < 1$ is always satisfied.

To illustrate this *arbitrary decomposition* by means of an explicit example, let us consider the mixed state $\mathbf{s} = (1,1/2,0,0)^T$ and let us chose the pure state $\mathbf{x} = (1,0,1,0)^T$ as a component. We find that $\alpha = 3/8$ and $\mathbf{y} = (1,4/5,-3/5,0)^T$.

Although in a further section a demonstration of the arbitrary decomposition of $3\times 3$ coherency matrices is included, which can also be applied to $n\times n$ coherency matrices, we have considered it useful to include this demonstration in terms of Stokes vectors, because it provides a more intuitive view of this kind of decomposition.

The existence of different possible decompositions of a mixed state into pure states has particular importance due to its implications in the possible *target decompositions* used in radar polarimetry, where the spectral decomposition is frequently considered to be the only way to achieve this [48].

Fig.4 illustrates the arbitrary decomposition by means of the corresponding representation in the Poincaré sphere.

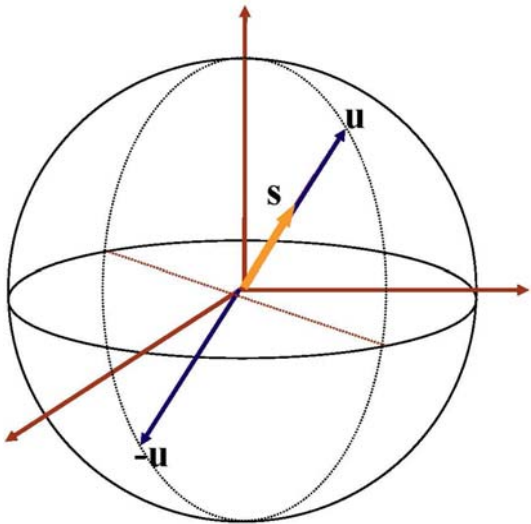
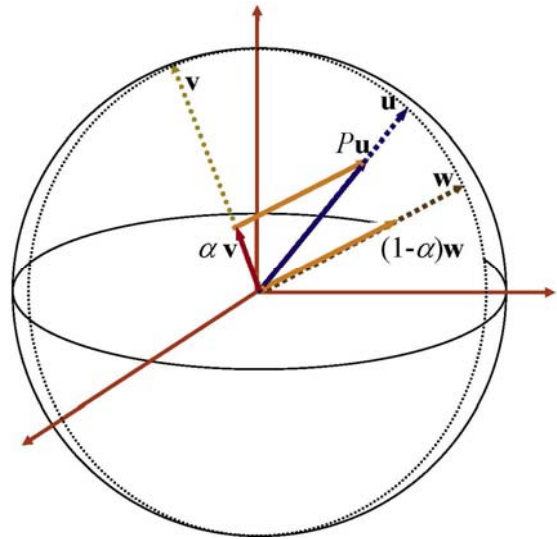

Fig. 3. Spectral decomposition of a mixed state.

Fig. 4. Arbitrary decomposition of a mixed state into a convex combination of two pure states.

## 2.6 Physical invariants and degree of polarization

Although the Stokes vectors and the Poincaré sphere models are particularly appropriate for representing the above decompositions, it is useful to consider again the coherency matrix model because, as it will be shown in later sections, the coherency matrix formalism can be generalized to represent the 3D states of polarization as well as the polarimetric effects of linear media (4D).

The 2x2 coherency matrix $\boldsymbol{\Phi}$ can be written as follows in terms of the physical parameters $I, P, \chi, \varphi$

$$\boldsymbol{\Phi} = \frac{I}{2}\begin{pmatrix} 1 + P\cos 2\varphi \cos 2\chi & P(\sin 2\varphi \cos 2\chi - i\sin 2\chi) \\ P(\sin 2\varphi \cos 2\chi + i\sin 2\chi) & 1 - P\cos 2\varphi \cos 2\chi \end{pmatrix}, \quad (28)$$

and, from this expression, it is straightforward to reproduce the characteristic decomposition.

The diagonalization of $\boldsymbol{\Phi}$ leads to the particular case of the spectral decomposition, expressed in terms of coherency matrices [44,49].

The quantities $I$ and $P$ are invariant in the sense that they remain unchanged under unitary transformations of the coherency matrix [50] and, thus, they are invariant with respect to changes of the reference system *XY*. In fact, $I$ and $P$ are directly related with the eigenvalues of $\mathbf{\Phi}$ [44]

$$\lambda_1 = \frac{1}{2}I(1+P), \ \lambda_2 = \frac{1}{2}I(1-P); \quad I = \lambda_1 + \lambda_2 = \operatorname{tr}\mathbf{\Phi}, \ P = \frac{\lambda_1 - \lambda_2}{\operatorname{tr}\mathbf{\Phi}}. \tag{29}$$

In accordance to the statistical nature of $\mathbf{\Phi}$, which is a covariance matrix (hence, it is a positive semidefinite Hermitian matrix), $\lambda_1, \lambda_2$ are nonnegative. Pure states are characterized by rank-one polarization matrices (only one nonzero eigenvalue, i.e., $P=1$), whereas rank-two polarization matrices correspond to mixed states ($P<1$).

Appropriate expressions for $P$ that are useful in certain analyses are

$$P_{2D} = \sqrt{\frac{2\operatorname{tr}\mathbf{\Phi}^2}{\operatorname{tr}^2\mathbf{\Phi}} - 1} = \sqrt{\frac{2\|\mathbf{\Phi}\|_2^2}{\|\mathbf{\Phi}\|_0^2} - 1}, \tag{30}$$

$$P_{2D} = \sqrt{\frac{\sum_{i,j=0}^{1} s_{ij}^2}{s_{00}^2} - 1} = \sqrt{\frac{\operatorname{tr}(\mathbf{S}^T\mathbf{S})}{s_{00}^2} - 1}, \tag{31}$$

where the subscript $2D$ has been added in order to compare these expressions with other that will appear concerning higher order coherency matrices.

As another interesting way of studying $P$, Wolf, in a classic paper [6], showed that there always exist two orthogonal reference directions such that the degree of mutual coherence $\mu$ reaches its maximum value, which coincides with $P$.

There are various random distributions which correspond to unpolarized light. As Ellis and Dogariu have shown [51], the measurement of the correlations of the Stokes parameters allows us to distinguish between these different types of unpolarized light. Moreover, for the case of non-Gaussian light, a quantitative determination of the degree of polarization of type-II unpolarized light by means of the distance between the polarization distribution and the uniform distribution associated with type-I unpolarized light, has been introduced by Luis [52].

**2.7 Polarization Entropy**

Let us consider the concept of the von Neumann entropy $S$ applied to electromagnetic waves [46,49,53-55,2], which is defined as follows in terms of the polarization density matrix $\hat{\mathbf{\Phi}}$

$$S = -\operatorname{tr}\left(\hat{\mathbf{\Phi}}\log_2\hat{\mathbf{\Phi}}\right) \tag{32}$$

This quantity is a measure of the difference in the amount of information between a pure state and a mixed state (both with the same intensity) and it has a close relation to $P$ by means of

$$S = -\operatorname{tr}\left(\hat{\mathbf{\Phi}}\log_2\hat{\mathbf{\Phi}}\right) = -\sum_{i=1}^{2}\left(\hat{\lambda}_i\log_2\hat{\lambda}_i\right) \tag{33}$$

where $\hat{\lambda}_i = \lambda_i/\operatorname{tr}\mathbf{\Phi}$ are the eigenvalues of $\hat{\mathbf{\Phi}}$. Thus, in the case of $2\times 2$ coherency matrices, $S$ can be written as follows as a function of $P$ [54]

$$S(P) \equiv -\left\{ \frac{1}{2}(1+P)\log_2\left[\frac{1}{2}(1+P)\right] + \frac{1}{2}(1-P)\log_2\left[\frac{1}{2}(1-P)\right] \right\}. \tag{34}$$

Therefore, $S$ is characterized univocally by $P$ and decreases monotonically as $P$ increases. The maximum $S = \log_2 2 = 1$ corresponds to $P = 0$, whereas the minimum $S = 0$ is reached for $P = 1$, (i.e. when light is totally polarized, regardless of its spectral profile) [2].

We observe that $S$ is nondimensional. Its relation with the concept of *specific radiation entropy* introduced by Plank, formulated explicitly by von Laue and extended by Barakat for *n* pencils of radiation [53], has been studied by Barakat and Brosseau [54], who have expressed the normalized specific radiation entropy as the difference of two von Neumann entropies.

Moreover, from the von Neumann entropy, Brosseau and Bicout [56] have defined a *polarization temperature* which can be expressed as a monotonic decreasing function of the degree of polarization.

Although a detailed study of the different measures related with entropy and the relations among them falls outside the scope of this review, other relevant approaches to polarization entropy are briefly discussed in Sec. 3.6. and Sec. 3.7.

As we will see in further sections, the concept of *polarization entropy* [57], when related with the depolarizing effects of a material medium, is commonly used to analyze measurements and images [58] by remote sensing, by lidar detection and by synthetic aperture radar polarimetry (SAR Polarimetry), in order to detect spatial heterogeneity in a great variety of terrestrial and oceanic surface targets [59]. It also will be shown that a meaningful parameter defined as the degree of polarimetric purity can be used as a quantity alternative to entropy.

## 2.8 Scope of the coherency matrix model and some extensions

The Stokes parameters are defined through second-order moments of the field variables. Therefore, the statistical nature of light underlies the mathematical description of the states of polarization, so that the study of the fluctuations of the field variables in different physical situations is an important subject. For the cases of non-Gaussian probability distributions of these variables, higher order moments should be considered and, hence, the Stokes parameters do not provide a complete description of the state of polarization. A number of authors have dealt with this matter, so that a significant amount of knowledge has been achieved [5,7,60-65]. A good compilation, including certain contributions of the author, can be found in a book by C. Brosseau [2].

The measurement of the polarization properties of light beams with non-Gaussian spectral profiles should requires particular measurement arrangements involving interferometers and fast detectors.

Coherence properties have a close relation with polarization description, and some authors have dealt with this important subject [45,66-73].

A particular definition of Stokes parameters at the wavelet scale has been introduced by Castañeda [74]. These parameters take into account the correlation properties of the random electromagnetic field on the aperture plane.

A generalization of the Stokes parameters of a random electromagnetic beam has been introduced by Ellis and Dogariu [71] in the space-time domain and has also been studied by Korotkova and Wolf [75] in the space-frequency domain on the basis of the Wolf's unified theory of coherence and polarization [76]. Whereas the usual Stokes parameters depend on one spatial variable, these *two-point Stokes parameters*, which can be defined in the space–frequency domain from the cross-spectral density matrix that characterizes the correlations at two points, depend on two spatial variables and contain additional information on the coherence properties.

Another relevant approach to polarization description is the model of *spatial-angular Stokes parameters* developed by Luis [77,78]. Using Wigner functions [79], these parameters are introduced for generalized rays including spatial and angular dependence and allow considering their evolution upon propagation.

Both above-mentioned generalizations have their corresponding counterpart in Mueller representations, i.e. Mueller matrices transforming two-point Stokes parameters [80,81] and spatial-angular Mueller matrices transforming spatial-angular Stokes parameters [78].

The above paragraphs deal with a classical description of polarization. The Stokes parameters have been used traditionally to describe certain quantum systems [46,82]. Nevertheless, it should be noted that a quantum description of polarization, performed by replacing classical amplitudes by bosonic operators, results in some properties that cannot be fully described by the classical Stokes parameters [83]. Thus, for instance, the electric field vector never describes a definite ellipse (due to uncertainty relationships), while there are states with $P=0$ that cannot be regarded as unpolarized.

The quantum description of polarization states is necessary for the study and characterization of coherent, squeezed, number, and phase states, and for the design of experiments in order to show fundamental properties and applications of quantum theory such as entanglement, complementarity, information processing, cryptography, teleportation and Bell inequalities.

The characterization of states of polarization through probability distributions on the Poincaré sphere allows to get the definition of the quantum degree of polarization as the distance between the corresponding polarization distribution and the uniform distribution representing unpolarized light. This definition can be particularized as a trace distance between the coherency matrix and the identity matrix (unpolarized light), and it is an important and open subject that provides results applicable to the description of quantum field states [83-86].

## 3. 3D Polarized light

The two-dimensional formalism considered in the previous section is valid when the polarization plane of light is constant, which is the commonest physical situation of interest in polarimetry. In the most general case, the three components of the electric field vector of the light wave should be considered in order to describe the evolution of the end point of the electric field vector (which determines the polarization state).

This section deals with a model for a general description of the polarization properties of light, applicable to waves whose polarization plane is not stable in time. Thus, from the extension of the concept of the Jones vector, the 3D polarization matrix as well as the 3D Stokes parameters, are defined. The main physical quantities characterizing the 3D states of polarization are obtained and interpreted, with special attention being given to the description of polarimetric purity by means of two invariant parameters.

### 3.1. 3D Jones vector

Let us consider a quasimonochromatic wave of arbitrary form, propagating in an isotropic medium, and let $\left(\mathbf{e}_1, \mathbf{e}_2, \mathbf{e}_3\right)$ be a reference basis of orthonormal vectors along the respective axes *XYZ*. Given a point $\mathbf{r}$ in space, the analytic signals of the three components of the electric field of the electromagnetic wave can be arranged as the 3x1 complex vector

$$\boldsymbol{\eta}(t) \equiv \begin{pmatrix} \eta_1 \\ \eta_2 \\ \eta_3 \end{pmatrix} = \begin{pmatrix} \eta_x(t) \\ \eta_y(t) \\ \eta_z(t) \end{pmatrix} = e^{i[u(t)+\beta_x(t)]} \begin{pmatrix} A_x(t) \\ A_y(t) e^{i[\beta_y(t)-\beta_x(t)]} \\ A_z(t) e^{i[\beta_z(t)-\beta_x(t)]} \end{pmatrix} \quad (35)$$

where $u(t) = \overline{\mathbf{k}} \cdot (\mathbf{r}/|\mathbf{r}|) - \overline{\omega} t$, $\overline{\mathbf{k}}$ being the mean wave-vector at point **r**.

The corresponding *3D instantaneous Jones vector* is defined as

$$\boldsymbol{\varepsilon}(t) = \begin{pmatrix} \varepsilon_1(t) \\ \varepsilon_2(t) \\ \varepsilon_3(t) \end{pmatrix} = \begin{pmatrix} A_x(t) \\ A_y(t) e^{i\delta_y(t)} \\ A_z(t) e^{i\delta_z(t)} \end{pmatrix}. \quad (36)$$

In this expression, the possible time dependence of amplitude and phases of the three components of the wave are indicated. Since only the phase differences have physical meaning, $\delta_x(t)$ has been taken as reference, so that a global phase factor has been avoided.

For totally-polarized sates, the following conditions are satisfied

$$\frac{A_y(t)}{A_x(t)} = \text{constant}, \quad \frac{A_z(t)}{A_x(t)} = \text{constant},$$

$$\delta_y(t) = \text{constant}, \quad \delta_z(t) = \text{constant}, \quad (37)$$

and the generalized Jones vector is expressed as

$$\boldsymbol{\varepsilon} = \begin{pmatrix} A_x \\ A_y e^{i\delta_y} \\ A_z e^{i\delta_z} \end{pmatrix}. \quad (38)$$

In this case, the polarization plane remains constant and the end of the electric field vector describes a well-defined ellipse in the polarization plane. The 2D model is then easily reproduced by taking the polarization plane normal to the *Z* reference axis, so that the third component of the 3D Jones vector vanishes.

### 3.2. 3D coherency matrix

*For updated details, see*

J. J. Gil, "Interpretation of the coherency matrix for three-dimensional polarization states," Phys. Rev. A **90**, 043858 (2014).

J. J. Gil, T. Setälä, I San Jose, A. Norrman, A. T. Friberg, "Sets of orthogonal three-dimensional polarization states and their physical interpretation," Phys. Rev. A **100**, 033824 (2019).

J. J. Gil, "Intrinsic Stokes parameters of 3D and 2D polarization states," J. Eur. Opt. Soc. RP **10**, 15054 (2015).

A number of authors have dealt with the 3D description of the polarization state. From a chronological point of view, it is worth emphasizing the work of Roman [87], who defined a set of generalized Stokes parameters from a particular basis of 3x3 matrices. These matrices are not trace-orthogonal and, consequently, do not constitute the best choice for building a basis. As Fano pointed out [46], an appropriate basis for $n \times n$ coherency matrices is a set of Hermitian trace-orthogonal operators. Thus, for our purposes, the best choice is the basis constituted by the 3x3 identity matrix together with the eight Gell-Mann matrices. We will see that this choice, based on the generators of the SU(3) group, is consistent with a generalization of the polarization algebra to $n \times n$ coherency matrices.

The so-called *coherency matrix* [42,6,45] or *polarization matrix*, which contains all measurable information on the state of polarization (including intensity) of an electromagnetic wave, is defined as the 3x3 Hermitian matrix

$$\mathbf{R} = \langle \boldsymbol{\varepsilon}(t) \otimes \boldsymbol{\varepsilon}^\dagger(t) \rangle = \begin{pmatrix} \langle \varepsilon_1(t)\,\varepsilon_1^*(t)\rangle & \langle \varepsilon_1(t)\,\varepsilon_2^*(t)\rangle & \langle \varepsilon_1(t)\,\varepsilon_3^*(t)\rangle \\ \langle \varepsilon_2(t)\,\varepsilon_1^*(t)\rangle & \langle \varepsilon_2(t)\,\varepsilon_2^*(t)\rangle & \langle \varepsilon_2(t)\,\varepsilon_3^*(t)\rangle \\ \langle \varepsilon_3(t)\,\varepsilon_1^*(t)\rangle & \langle \varepsilon_3(t)\,\varepsilon_2^*(t)\rangle & \langle \varepsilon_3(t)\,\varepsilon_3^*(t)\rangle \end{pmatrix}, \tag{39}$$

whose elements $r_{ij}$ are the second-order moments $r_{ij} = \langle \varepsilon_i(t)\,\varepsilon_j^*(t)\rangle$ $(i,j=1,2,3)$ of the zero-mean analytic signals $\varepsilon_i(t)$, $i=1,2,3$ ($\varepsilon_i^*$ represents the complex conjugate of $\varepsilon_i$ and the brackets indicate time averaging over the measurement time), so that $\mathbf{R}$ is a covariance matrix and, therefore, $\mathbf{R}$ is fully characterized by the fact that it is a positive semidefinite Hermitian matrix [88].

The expression of $\mathbf{R}$ in terms of the amplitudes and relative phases of the field components is

$$\mathbf{R} = \begin{pmatrix} \langle A_x^2(t)\rangle & \langle A_x(t)A_y(t)e^{-i\delta_y(t)}\rangle & \langle A_x(t)A_z(t)e^{-i\delta_z(t)}\rangle \\ \langle A_x(t)A_y(t)e^{i\delta_y(t)}\rangle & \langle A_y^2(t)\rangle & \langle A_y(t)A_z(t)e^{i(\delta_y(t)-\delta_z(t))}\rangle \\ \langle A_x(t)A_z(t)e^{i\delta_z(t)}\rangle & \langle A_y(t)A_z(t)e^{-i(\delta_y(t)-\delta_z(t))}\rangle & \langle A_z^2(t)\rangle \end{pmatrix}, \tag{40}$$

As we have observed, $\mathbf{R}$ is characterized by the fact that its three eigenvalues are nonnegative. These three constraints constitute a complete set of necessary and sufficient conditions for a Hermitian matrix $\mathbf{R}$ to be a 3D coherency matrix, i.e. to represent a particular state of polarization of a light beam. An equivalent set of conditions is derived from the fact that the Hermitian matrix $\mathbf{R}$ has three nested nonnegative principal minors [88].

The statistical properties of $\mathbf{R}$ arise clearly when its elements $r_{ij}$ $(i,j=1,2,3)$ are written in terms of the corresponding standard deviations $\sigma_i$ and degrees of mutual coherence $\mu_{ij}$,

$$r_{ij} \equiv \mu_{ij}\sigma_i\sigma_j \;, \tag{41}$$

where $\sigma_i^2 = \langle \varepsilon_i \varepsilon_i^* \rangle = r_{ii}$, $\mu_{ij} = r_{ij}/\sigma_i\sigma_j$, i.e.

$$\mathbf{R} = \begin{pmatrix} \sigma_1^2 & \mu_{12}\sigma_1\sigma_2 & \mu_{13}\sigma_1\sigma_3 \\ \mu_{12}^*\sigma_1\sigma_2 & \sigma_2^2 & \mu_{23}\sigma_2\sigma_3 \\ \mu_{13}^*\sigma_1\sigma_3 & \mu_{23}^*\sigma_2\sigma_3 & \sigma_3^2 \end{pmatrix} \tag{42}$$

[*For updated details on the mutual degrees of coherence, see* J. J. Gil, "Degrees of mutual coherence of a 3D polarization state," J. Mod. Opt. 63, 1055-1058 (2018)].

From the nonnegativity of three nested principal minors, we can write a set of necessary and sufficient explicit conditions for $\mathbf{R}$ to be a covariance matrix and, hence, to be a coherency matrix representing a 3D state of polarization

$$r_{11} \geq 0 \,; \tag{43.a}$$

$$1 \geq \rho_{12}\,; \tag{43.b}$$

$$\det \mathbf{R} = 1 + 2\rho_{12}\rho_{23}\rho_{13}\cos\left(\beta_{12} + \beta_{23} - \beta_{13}\right) - \rho_{12}^2 - \rho_{23}^2 - \rho_{13}^2 \geq 0. \tag{43.c}$$

where $\rho_{ij} \equiv |\mu_{ij}|$, $\mu_{ij} \equiv \rho_{ij} e^{i\beta_{ij}}$.

An interesting geometric interpretation of the 3D coherency matrix has been presented by Dennis [89] through the decomposition of $\mathbf{R}$ into a real symmetric positive definite matrix (interpreted as the moment of inertia of the ensemble, and represented by means of the corresponding ellipsoid) and a real axial vector (corresponding to the mean angular momentum of the ensemble). Another kind of representation of 3D states of polarization (but only valid for pure states), based on the Majorana sphere representation has been presented by Hannay [90] who studied its relation to the Berry phase. Moreover, Saastamoinen and Tervo [91] and Ellis and Dogariu [92] have introduced geometric representations of the invariants of $\mathbf{R}$ in the eigenvalue space.

As a basis for the expansion of $\mathbf{R}$, let us now consider the following set of Hermitian, trace-orthogonal matrices constituted by the Gell-Mann matrices plus the identity matrix

$$\boldsymbol{\omega}_{11} \equiv \begin{pmatrix} 1 & 0 & 0 \\ 0 & 1 & 0 \\ 0 & 0 & 1 \end{pmatrix} \quad \boldsymbol{\omega}_{12} \equiv \sqrt{\frac{3}{2}}\begin{pmatrix} 0 & 1 & 0 \\ 1 & 0 & 0 \\ 0 & 0 & 0 \end{pmatrix} \quad \boldsymbol{\omega}_{13} \equiv \sqrt{\frac{3}{2}}\begin{pmatrix} 0 & 0 & 1 \\ 0 & 0 & 0 \\ 1 & 0 & 0 \end{pmatrix}$$

$$\boldsymbol{\omega}_{21} \equiv \sqrt{\frac{3}{2}}\begin{pmatrix} 0 & -i & 0 \\ i & 0 & 0 \\ 0 & 0 & 0 \end{pmatrix} \quad \boldsymbol{\omega}_{22} \equiv \sqrt{\frac{3}{2}}\begin{pmatrix} 1 & 0 & 0 \\ 0 & -1 & 0 \\ 0 & 0 & 0 \end{pmatrix} \quad \boldsymbol{\omega}_{23} \equiv \sqrt{\frac{3}{2}}\begin{pmatrix} 0 & 0 & 0 \\ 0 & 0 & 1 \\ 0 & 1 & 0 \end{pmatrix} \tag{44}$$

$$\boldsymbol{\omega}_{31} \equiv \sqrt{\frac{3}{2}}\begin{pmatrix} 0 & 0 & -i \\ 0 & 0 & 0 \\ i & 0 & 0 \end{pmatrix} \quad \boldsymbol{\omega}_{32} \equiv \sqrt{\frac{3}{2}}\begin{pmatrix} 0 & 0 & 0 \\ 0 & 0 & -i \\ 0 & i & 0 \end{pmatrix} \quad \boldsymbol{\omega}_{33} \equiv \frac{1}{\sqrt{2}}\begin{pmatrix} 1 & 0 & 0 \\ 0 & 1 & 0 \\ 0 & 0 & -2 \end{pmatrix}$$

The notation used for these matrices is justified for the sake of simplicity as well as to emphasize the symmetry in some future mathematical expressions.

Matrices $\boldsymbol{\omega}_{ij}$ are Hermitian $\boldsymbol{\omega}_{ij} = \boldsymbol{\omega}_{ij}^\dagger$ and trace-orthogonal $\text{tr}\left(\boldsymbol{\omega}_{ij}\boldsymbol{\omega}_{kl}\right) = 3\delta_{ik}\delta_{jl}$ and constitute a basis that allows for $\mathbf{R}$ to be expressed as the following linear combination [2]

$$\mathbf{R} = \frac{1}{3}\sum_{i,j=1}^{3} q_{ij}\boldsymbol{\omega}_{ij}, \quad q_{ij} = \text{tr}\left(\mathbf{R}\boldsymbol{\omega}_{ij}\right), \tag{45}$$

where the nine real coefficients $q_{ij}$ can be properly called the *3D Stokes parameters.*

In order to compare this expansion with that presented by other authors, the following observations should be taken into account:

o For the sake of symmetry and simplicity in the mathematical relations, the order chosen for the Gell-Mann matrices differs from that used by Gell-Mann and by some authors [93-95]. Except for the change $\boldsymbol{\omega}_{22} \leftrightarrow \boldsymbol{\omega}_{33}$, the order coincides with the one used by Brosseau [2].

o The Gell-Mannn matrices have been normalized in order to ensure that their weight (Euclidean norm) coincide with that of the 3x3 identity matrix, which is also included in the basis. This choice differs from those used in the cited references, as well as in Ref. [96].

Obviously, these options have no physical consequences and do not affect the results expressed in terms of invariant quantities (i.e. derived from the eigenvalues of $\mathbf{R}$), but lead to different expressions for the 3D Stokes parameters and for their relations with other parameters.

As we will see, the normalization performed is consistent with a generalized model for the polarimetric properties of light and media based on $n$-dimensional coherency matrices.

### 3.3. 3D Stokes parameters

*For updated details on the mutual degrees of coherence, see*
J. J. Gil, "Intrinsic Stokes parameters of 3D and 2D polarization states," J. Eur. Opt. Soc. RP **10**, 15054 (2015).

According to the definitions introduced in the previous section, the 3D Stokes parameters take the form

$$\begin{aligned}
q_{11} &\equiv r_{11} + r_{22} + r_{33} & q_{12} &\equiv \sqrt{\frac{3}{2}}(r_{12} + r_{21}) & q_{13} &\equiv \sqrt{\frac{3}{2}}(r_{13} + r_{31}) \\
q_{21} &\equiv i\sqrt{\frac{3}{2}}(r_{12} - r_{21}) & q_{22} &\equiv \sqrt{\frac{3}{2}}(r_{11} - r_{22}) & q_{23} &\equiv \sqrt{\frac{3}{2}}(r_{23} + r_{32}) \\
q_{31} &\equiv i\sqrt{\frac{3}{2}}(r_{13} - r_{31}) & q_{32} &\equiv i\sqrt{\frac{3}{2}}(r_{23} - r_{32}) & q_{33} &\equiv \frac{1}{\sqrt{2}}(r_{11} + r_{22} - 2r_{33})
\end{aligned} \qquad (46)$$

and the coherency matrix can be expressed in terms of $q_{ij}$ as follows

$$\mathbf{R} = \frac{1}{6} \begin{pmatrix} (2q_{11} + \sqrt{6}q_{22} + \sqrt{2}q_{33}) & (\sqrt{6}q_{12} - i\sqrt{6}q_{21}) & (\sqrt{6}q_{13} - i\sqrt{6}q_{31}) \\ (\sqrt{6}q_{12} + i\sqrt{6}q_{21}) & (2q_{11} - \sqrt{6}q_{22} + \sqrt{2}q_{33}) & (\sqrt{6}q_{23} - i\sqrt{6}q_{32}) \\ (\sqrt{6}q_{13} + i\sqrt{6}q_{31}) & (\sqrt{6}q_{23} + i\sqrt{6}q_{32}) & (2q_{11} - 2\sqrt{2}q_{33}) \end{pmatrix}. \qquad (47)$$

It is obvious that the Poincaré sphere representation is not applicable to 3D states of polarization. The constraining inequalities among $q_{ij}$ (derived from the nonnegativity of the eigenvalues of $\mathbf{R}$) are of greater complexity than that of the 2D Stokes parameters. In fact, these restrictions involves, not only $\operatorname{tr}\mathbf{R}$ and $\operatorname{tr}\mathbf{R}^2$, but also $\operatorname{tr}\mathbf{R}^3$.

Now we examine the physical meaning of the 3D Stokes parameters. To do so, we first consider the reduction to the case of plane waves, which can be performed through a transformation of the reference axes $XYZ$ to $X'Y'Z'$, $Z'$ being normal to the polarization plane. The resulting coherency matrix $\mathbf{R}'$ is

$$\mathbf{R}' = \frac{1}{6} \begin{pmatrix} (2q'_{11} + \sqrt{6}q'_{22} + \sqrt{2}q'_{33}) & (\sqrt{6}q'_{12} - i\sqrt{6}q'_{21}) & 0 \\ (\sqrt{6}q'_{12} + i\sqrt{6}q'_{21}) & (2q'_{11} - \sqrt{6}q'_{22} + \sqrt{2}q'_{33}) & 0 \\ 0 & 0 & 0 \end{pmatrix}, \qquad (48)$$

where the Stokes parameters can be interpreted by comparing them with the corresponding 2D Stokes parameters

$$s_0 = \frac{2}{3}\left(q'_{11} + \frac{1}{\sqrt{2}}q'_{33}\right), \quad s_1 = \sqrt{\frac{2}{3}}q'_{22}, \quad s_2 = \sqrt{\frac{2}{3}}q'_{12}, \quad s_3 = \sqrt{\frac{2}{3}}q'_{21}. \qquad (49)$$

We see that, due to the particular normalization $\text{tr}(\boldsymbol{\omega}_{ij}\boldsymbol{\omega}_{kl}) = 3\delta_{ik}\delta_{jl}$ chosen for $\boldsymbol{\omega}_{ij}$, and due to the intrinsic asymmetry of the Gell-Mann matrices, which have only two nonzero elements except for the diagonal matrix $\boldsymbol{\omega}_{33}$ (and also $\boldsymbol{\omega}_{11}$ in the extended set we are using), the intensity $s_0$ includes a contribution from $q'_{33}$. Thus, as Carozzi, Karlsson, and Bergman [93], and Setälä, Shevchenko, Kaivola and Friberg [94] have pointed out, in general, the parameter $q_{22}$ represents the predominance of the $X$ component ($q_{22} > 0$) or of the $Y$ component ($q_{22} < 0$). Attending to the projections of the field components on the $XY$ plane, the value $q_{22} = 0$ occurs, for example, in the cases of circularly polarized light and unpolarized light. Analogously, the parameters $q_{12}, q_{21}$ refer respectively to the predominance of projections on the positive ($q_{12} > 0$), or negative ($q_{12} < 0$), parts of the bisector axis of $XY$ and of right handed ($q_{21} > 0$), or left handed ($q_{21} < 0$), circularly polarized light.

Analogous meanings correspond to the parameters $q_{13}, q_{31}$ (for projections on the $XZ$ plane), and $q_{23}, q_{32}$ (for projections on the $YZ$ plane). Finally, parameter $q_{33}$ represents the intensity in the $XY$ plane additional to that in the $Z$ direction [94].

A particular analysis devoted to plane waves in the 3D spectral density tensor formalism is presented in Ref. [93], where the angles that determine the plane containing the polarization ellipse are obtained from **R**. This analysis can also be applied to characterize the *spectral polarization ellipse* corresponding to each frequency contained in the light wave. In the time domain, and under the classic assumptions, the electric field describes the polarization ellipse, but, in general, both its shape and its plane vary with time. For the spectral density tensor, the dual pseudovector associated with its antisymmetric part is given by $i(-q_{32}, q_{23}, -q_{31})$, so that the real vector $\mathbf{v} \equiv (q_{32}, -q_{23}, q_{31})$ determines the plane of the polarization ellipse [93].

In the quantum model, these interpretations do not longer hold because the polarization ellipse is not well defined [97].

As some authors have pointed out [98-100], the measurement of 3D Stokes parameters of near fields is an important subject for the study of coherent light diffracted by microstructures (producing subwavelength features) [99], fluorescence, multiphoton microscopy, imaging objects embedded in dense scattering media, nanophotonics [100], etc. Dändliker et al. [99] have dealt with the measurement of the transverse component of the electric field by means of a heterodyne scanning probe optical microscope (heterodyne SPOM) and, more recently, Ellis and Dogariu [100] have suggested an original experimental setup for full 3D polarimetric measurements.

### 3.4. Purity criterion for 3x3 coherency matrices

*For updated details, see*

J. J. Gil, "Interpretation of the coherency matrix for three-dimensional polarization states," Phys. Rev. A **90**, 043858 (2014).

J. J. Gil, A. Norrman, A. T. Friberg, T. Setälä, "Polarimetric purity and the concept of degree of polarization", Phys. Rev. A **97**, 023838 (2018).

J. J. Gil, A. Norrman, T. Setälä, A. T. Friberg, "Intensity and spin anisotropy of a three-dimensional polarization state," Opt. Lett. **44**, 3578-3581 (2019).

The expression of the 3x3 coherency matrix **R** in terms of the statistical parameters, given by Eq. (42), shows that, as Ellis, Dogariu, Ponomarenko and Wolf have pointed out [101], total polarimetric purity corresponds to complete correlation between the field variables, so that the values of the absolute values of the mutual degrees of coherence are equal to 1, $|\mu_{ij}| = 1$.

In order to express some relevant relations in terms of invariant metric parameters, let us now consider the Euclidean norms (also called Frobenius norms) of **R** and $\mathbf{Q} \equiv (q_{ij})$

$$\|\mathbf{R}\|_2 \equiv \sqrt{\sum_{i,j=1}^{3} |r_{ij}|^2} = \sqrt{\operatorname{tr}(\mathbf{R}^\dagger \mathbf{R})} = \sqrt{\operatorname{tr}\mathbf{R}^2}\,, \tag{50}$$

$$\|\mathbf{Q}\|_2 \equiv \sqrt{\sum_{i,j=1}^{3} q_{ij}^2} = \sqrt{\operatorname{tr}(\mathbf{Q}^T \mathbf{Q})}\,, \tag{51}$$

and, as in the 2D case, we define the following norm for $\mathbf{R}$

$$\|\mathbf{R}\|_0 \equiv \operatorname{tr}\mathbf{R} = \left\|\sqrt{\mathbf{R}}\right\|_2^2. \tag{52}$$

Since $\mathbf{R}$ is a positive semidefinite Hermitian matrix, it is easy to show that $\|\mathbf{R}\|_0$ satisfies all the conditions to be a norm. Furthermore, these norms satisfy the following relations

$$\|\mathbf{R}\|_0 = q_{00}\,, \tag{53}$$

$$\frac{1}{3}\|\mathbf{R}\|_0^2 \leq \|\mathbf{R}\|_2^2 \leq \|\mathbf{R}\|_0^2. \tag{54}$$

For pure states, the equality $\|\mathbf{R}\|_2 = \|\mathbf{R}\|_0$ holds, which constitutes an objective purity criterion formulated in a symmetric manner with respect to that presented for the 2D case. The other limit $\|\mathbf{R}\|_2 = \|\mathbf{R}\|_0/\sqrt{3}$ is reached in the case of unpolarized light (equiprobable mixture of states).

### 3.5. Decompositions of the coherency matrix

In the subsections below, we analyze different procedures for decomposing a coherency matrix $\mathbf{R}$ into a convex linear combination of particular coherency matrices. The *spectral decomposition* is based on the eigenvalue structure of $\mathbf{R}$ and consists of a set of pure components. The *characteristic decomposition* is also based on the eigenvalue structure of $\mathbf{R}$ and consists of a pure component and a scaled set of unpolarized components. Finally, the *arbitrary decomposition* provides a general method for decomposing a mixed state into pure states and, hence, includes the spectral decomposition as a particular case.

### 3.5.1. The spectral decomposition of the coherency matrix

Since the coherency matrices are Hermitian positive semidefinite, they can be diagonalized through a unitary transformation. Therefore,

$$\mathbf{R} = \mathbf{U}\,\mathbf{D}(\lambda_1, \lambda_2, \lambda_3)\,\mathbf{U}^+\,, \tag{55}$$

where $\mathbf{D}(\lambda_1, \lambda_2, \lambda_3)$ represents the diagonal matrix composed of the nonnegative eigenvalues $\lambda_3 \leq \lambda_2 \leq \lambda_1$. The columns $\mathbf{u}_i$ $(i=1,2,3)$ of the unitary matrix $\mathbf{U}$ are the respective eigenvectors. Consequently, $\mathbf{R}$ can be expressed as the following convex linear combination of three rank-one coherency matrices that represent respective pure states with equal intensities $I = \operatorname{tr}\mathbf{R}$

$$\mathbf{R} = \hat{\lambda}_1\left[I\mathbf{U}\mathbf{D}(1,0,0)\mathbf{U}^\dagger\right] + \hat{\lambda}_2\left[I\mathbf{U}\mathbf{D}(0,1,0)\mathbf{U}^\dagger\right] + \hat{\lambda}_3\left[I\mathbf{U}\mathbf{D}(0,0,1)\mathbf{U}^\dagger\right],\ \hat{\lambda}_i \equiv \lambda_i/\operatorname{tr}\mathbf{R}\,, \tag{56}$$

where each term in the sum is only affected by the corresponding eigenvector

$$\mathbf{R} = \sum_{i=1}^{3} \hat{\lambda}_i \left[I\left(\mathbf{u}_i \otimes \mathbf{u}_i^\dagger\right)\right],\quad \sum_{i=1}^{3} \hat{\lambda}_i = 1. \tag{57}$$

As in the 2D case, this *spectral decomposition* shows that any 3D polarization state can be considered an incoherent superposition of three pure states with weights equal to the eigenvalues of *polarization density matrix* $\hat{\mathbf{R}} \equiv \mathbf{R}/\operatorname{tr}\mathbf{R}$. The coherency matrix is a convex linear combination of the coherency matrices generated by the respective eigenvectors. It should be noted that, when one of the eigenvalues has a multiplicity higher than one, then the eigenvectors of the corresponding invariant subspace are not unique.

### 3.5.2. The characteristic decomposition of the coherency matrix

The *characteristic decomposition* (also called *trivial decomposition*) of $\mathbf{R}$ cannot be performed in the form of a sum of a pure state and a 3D unpolarized state but, as we will see at the end of the Sec. 3.6, a generalized characteristic decomposition can be properly realized through a sum of a pure state, a *discriminating state* (i.e. an equiprobable mixture of the eigenstates generated by $\mathbf{u}_1$ and $\mathbf{u}_2$) and a 3D unpolarized state. This agrees with the fact that the nine real independent parameters of $\mathbf{R}$ can be generated by the sum of a pure 3x3 coherency matrix (five real independent parameters), a discriminating 3x3 coherency matrix (three real independent parameters) and a 3D-unpolarized 3x3 coherency matrix (one real independent parameter). In general, pure $n \times n$ coherency matrices contain $2n-1$ independent parameters whereas *m*D-unpolarized $n \times n$ coherency matrices contain $2(n-m)+1$ independent parameters. For more details of the mathematical formulation and physical interpretation of the characteristic decomposition of $\mathbf{R}$, see Section 3.6.

### 3.5.3. The arbitrary decomposition of the coherency matrix

*Along this Section, some details of the calculations have been clarified with respect to those of the original version*

As with the spectral decomposition, an *arbitrary decomposition* can also be applied to 3x3 coherency matrices.

$$\mathbf{R} = \sum_{i=1}^{3} p_i \left[ (\operatorname{tr}\mathbf{R}) \left( \hat{\mathbf{w}}_i \otimes \hat{\mathbf{w}}_i^{\dagger} \right) \right], \quad \sum_{i=1}^{3} p_i = 1, \ 0 \le p_i, \tag{58}$$

where $\hat{\mathbf{w}}_i$ are linearly independent unit vectors. The number $r$ of pure components is equal to $\operatorname{rank}\mathbf{R} = r$. The demonstration of the existence of this kind of decomposition for $n \times n$ coherency matrices, as well as the restrictions on $\hat{\mathbf{w}}_i$, is dealt with in Ref. [102]. In our opinion this is a relevant result because it provides all the ways for representing a mixed state as a superposition of pure states. Nevertheless, as we will see in Sec. 5.6.3, the application of this decomposition to coherency matrices representing material media (4D) requires that the mathematical expression of the superposition of pure states have the form of a convex sum. Thus, given the importance of this kind of decomposition and its potential application for scientific and industrial purposes we include here a new demonstration where $\mathbf{R}$ is expressed as a convex sum of pure components. It is straightforward to extend this demonstration to $n \times n$ coherency matrices.

Given a nonsingular 3×3 coherency matrix $\mathbf{R}$ with $\operatorname{rank}\mathbf{R} = 3$, we consider Eq. (55) and write $\mathbf{R}$ as

$$\mathbf{R} = \mathbf{CC}^{\dagger}, \ \mathbf{C} \equiv \mathbf{U}\sqrt{\mathbf{D}}, \ \mathbf{C}^{\dagger} = \sqrt{\mathbf{D}}\,\mathbf{U}^{\dagger},$$

$$\sqrt{\mathbf{D}} \equiv \mathbf{D}\left(\sqrt{\lambda_1}, \sqrt{\lambda_2}, \sqrt{\lambda_3}\right) \equiv \begin{pmatrix} \sqrt{\lambda_1} & 0 & 0 \\ 0 & \sqrt{\lambda_2} & 0 \\ 0 & 0 & \sqrt{\lambda_3} \end{pmatrix}. \tag{59}$$

From successive steps we are going to decompose $\mathbf{R}$ into a linear convex combination of pure coherency matrices $\mathbf{R}_{pi}$ satisfying $\text{tr}\,\mathbf{R}_{pi} = \text{tr}\,\mathbf{R}$ (equal intensities before being affected by their respective weights in the convex sum). Since $\mathbf{R}_{pi}$ is a pure coherency matrix, then $\text{rank}\,\mathbf{R}_{pi} = 1$.

First we observe that, given any rank-one positive semidefinite Hermitian matrix $\mathbf{R}_{p3}$, the matrix $\mathbf{C}^{-1}\mathbf{R}_{p3}(\mathbf{C}^\dagger)^{-1}$ is also a rank-one Hermitian matrix, and there exist a unitary matrix $\mathbf{V}$ that diagonalizes it [103]

$$\mathbf{V}^\dagger \left[ \mathbf{C}^{-1}\mathbf{R}_{p3}(\mathbf{C}^\dagger)^{-1} \right] \mathbf{V} = \mathbf{D}(0,0,\alpha_3), \ 0 \leq \alpha_3, \tag{60}$$

and, consequently,

$$\mathbf{R}_{p3} = \mathbf{C}\mathbf{V}\mathbf{D}(0,0,\alpha_3)\mathbf{V}^\dagger\mathbf{C}^\dagger, \quad \alpha_3 > 0. \tag{61}$$

We introduce the following notations for $\mathbf{V}$ in order to use them in subsequent calculations

$$\mathbf{V} \equiv \begin{pmatrix} v_{11} & v_{12} & v_{13} \\ v_{21} & v_{22} & v_{23} \\ v_{31} & v_{32} & v_{33} \end{pmatrix} \equiv (\hat{\mathbf{v}}_1, \hat{\mathbf{v}}_2, \hat{\mathbf{v}}_3), \tag{62}$$

where the column vectors $\hat{\mathbf{v}}_i \equiv (v_{1i}, v_{2i}, v_{3i})^T$ are the orthogonal unit eigenvectors of the matrix $\mathbf{C}^{-1}\mathbf{R}_{p3}(\mathbf{C}^\dagger)^{-1}$. Therefore, $\mathbf{R}_{p3}$ can be written as

$$\begin{aligned}\mathbf{R}_{p3} &= \mathbf{C}\left[\alpha_3(\hat{\mathbf{v}}_3 \otimes \hat{\mathbf{v}}_3^\dagger)\right]\mathbf{C}^\dagger = \alpha_3 (\mathbf{U}\sqrt{\mathbf{D}}\,\hat{\mathbf{v}}_3) \otimes (\hat{\mathbf{v}}_3^\dagger \sqrt{\mathbf{D}}\,\mathbf{U}^\dagger) \\ &= \alpha_3 (\mathbf{U}\sqrt{\mathbf{D}}\,\hat{\mathbf{v}}_3) \otimes (\mathbf{U}\sqrt{\mathbf{D}}\,\hat{\mathbf{v}}_3)^\dagger.\end{aligned} \tag{63}$$

It is important to note that, whereas the eigenvector $\hat{\mathbf{v}}_3$, corresponding to the nonzero eigenvalue $a_3$, is fixed by the particular election of $\mathbf{R}_{p3}$, the other two orthonormal eigenvectors $\hat{\mathbf{v}}_1, \hat{\mathbf{v}}_2$, corresponding to the null subspace of $\mathbf{R}_{p3}$, can be chosen freely (provided they are orthogonal to $\hat{\mathbf{v}}_3$).

Let us now consider the trace of $\mathbf{R}_{p3}$, in order to determine the parameter $\alpha_3$ from the condition $\text{tr}\,\mathbf{R}_{p3} = \text{tr}\,\mathbf{R}$. From Eq. (63),

$$\text{tr}\,\mathbf{R}_{p3} = \alpha_3 \left|\mathbf{U}\sqrt{\mathbf{D}}\,\hat{\mathbf{v}}_3\right|^2 \Rightarrow \alpha_3 = \frac{\text{tr}\,\mathbf{R}_{p3}}{\left|\mathbf{U}\sqrt{\mathbf{D}}\,\hat{\mathbf{v}}_3\right|^2} = \frac{\text{tr}\,\mathbf{R}}{\left|\sqrt{\mathbf{D}}\,\hat{\mathbf{v}}_3\right|^2}. \tag{64}$$

Moreover, note that $\mathbf{R}$ can be expressed as

$$\mathbf{R} = \mathbf{C}\mathbf{C}^\dagger = \mathbf{C}(\mathbf{V}\mathbf{V}^\dagger)\mathbf{C}^\dagger = (\mathbf{C}\mathbf{V})\mathbf{D}(1,1,1)(\mathbf{V}^\dagger\mathbf{C}^\dagger). \tag{65}$$

Thus, the simultaneous diagonalization of $\mathbf{R}_{p3}$ and $\mathbf{R}$, allows for performing the subtraction

$$\begin{aligned}\mathbf{R} - p_3\mathbf{R}_{p3} &= \mathbf{C}\mathbf{V}\left[\mathbf{D}(1,1,1) - p_3\mathbf{D}(0,0,\alpha_3)\right]\mathbf{V}^\dagger\mathbf{C}^\dagger = \\ &= \mathbf{C}\mathbf{D}(1,1,1-p_3\alpha_3)\mathbf{V}^\dagger\mathbf{C}^\dagger.\end{aligned} \tag{66}$$

We observe that, by taking $p_3 = 1/\alpha_3$, the subtraction results in

$$\mathbf{R} - p_3\mathbf{R}_{p3} = \mathbf{R}', \ \mathbf{R}' \equiv \mathbf{C}\mathbf{V}\mathbf{D}(1,1,0)\mathbf{V}^\dagger\mathbf{C}^\dagger = \mathbf{C}\left[(\hat{\mathbf{v}}_1 \otimes \hat{\mathbf{v}}_1^\dagger) + (\hat{\mathbf{v}}_2 \otimes \hat{\mathbf{v}}_2^\dagger)\right]\mathbf{C}^\dagger, \tag{67}$$

where $\mathbf{R}'$ is a rank-two, positive semidefinite Hermitian matrix.

Let us now define the unit vector

$$\hat{\mathbf{w}}_3 \equiv \frac{1}{p_3 \operatorname{tr} \mathbf{R}} \mathbf{U} \sqrt{\mathbf{D}}\, \hat{\mathbf{v}}_3, \quad \text{with} \quad p_3 = \frac{1}{\alpha_3} = \frac{\left|\sqrt{\mathbf{D}}\, \hat{\mathbf{v}}_3\right|^2}{\operatorname{tr} \mathbf{R}},$$

so that,

$$\hat{\mathbf{v}}_3 = p_3 \operatorname{tr} \mathbf{R} \left(\sqrt{\mathbf{D}}\right)^{-1} \mathbf{U}^\dagger \hat{\mathbf{w}}_3, \quad \left(\sqrt{\mathbf{D}}\right)^{-1} \equiv \begin{pmatrix} 1/\sqrt{\lambda_1} & 0 & 0 \\ 0 & 1/\sqrt{\lambda_2} & 0 \\ 0 & 0 & 1/\sqrt{\lambda_3} \end{pmatrix}. \tag{68}$$

From the above equations, and taking into account the fact that $\hat{\mathbf{v}}_3$ is a unit vector, the following relations hold

$$1 = \left|\hat{\mathbf{v}}_3\right|^2 = \operatorname{tr}\left(\hat{\mathbf{v}}_3 \otimes \hat{\mathbf{v}}_3^\dagger\right) =$$

$$= (\operatorname{tr} \mathbf{R})\, p_3 \operatorname{tr}\left\{\left(\sqrt{\mathbf{D}}\right)^{-1} \mathbf{U}^\dagger \hat{\mathbf{w}}_3 \otimes \left[\left(\sqrt{\mathbf{D}}\right)^{-1} \mathbf{U}^\dagger \hat{\mathbf{w}}_3\right]^\dagger\right\}$$

$$= (\operatorname{tr} \mathbf{R})\, p_3 \sum_{j=1}^{3} \frac{1}{\lambda_j} \left|\left(\mathbf{U}^\dagger \hat{\mathbf{w}}_3\right)_j\right|^2,$$

and therefore, we get that the coefficient $p_3$ is expressed as follows as a function of vector $\hat{\mathbf{w}}_3$ (generator of the pure component $\mathbf{R}_{p3}$ of $\mathbf{R}$).

$$p_3 = \frac{1}{(\operatorname{tr} \mathbf{R}) \sum_{j=1}^{3} \frac{1}{\lambda_j} \left|\left(\mathbf{U}^\dagger \hat{\mathbf{w}}_3\right)_j\right|^2}. \tag{69}$$

Let us now consider an arbitrary unitary matrix $\mathbf{V}'$ of the form

$$\mathbf{V}' \equiv \begin{pmatrix} v'_{11} & v'_{12} & 0 \\ v'_{21} & v'_{22} & 0 \\ 0 & 0 & 1 \end{pmatrix}. \tag{70}$$

This matrix $\mathbf{V}'$ always satisfies the following matrix equation

$$\mathbf{V}' \mathbf{D}(1,1,0) \mathbf{V}'^\dagger = \mathbf{D}(1,1,0), \tag{71}$$

and, therefore

$$\mathbf{R}' \equiv \mathbf{C} \mathbf{Z} \mathbf{D}(1,1,0) \mathbf{Z}^\dagger \mathbf{C}^\dagger, \quad \mathbf{Z} \equiv \mathbf{V} \mathbf{V}', \tag{72}$$

Where $\mathbf{Z}$ is a unitary matrix with the form

$$\mathbf{Z} \equiv \begin{pmatrix} z_{11} & z_{12} & v_{13} \\ z_{21} & z_{22} & v_{23} \\ v_{31} & v_{32} & v_{33} \end{pmatrix} \equiv \left(\hat{\mathbf{z}}_1, \hat{\mathbf{z}}_2, \hat{\mathbf{z}}_3\right), \tag{73}$$

In order to find the class of rank-one matrices susceptible to be diagonalized simultaneously with $\mathbf{R}'$, we construct the matrix

$$\mathbf{R}_{p2} = \mathbf{C}\mathbf{Z}\mathbf{D}(0,\alpha_2,0)\mathbf{Z}^\dagger\mathbf{C}^\dagger, \tag{74}$$

so that

$$\mathbf{R}_{p2} = \mathbf{C}\left[\alpha_2\left(\hat{\mathbf{z}}_2 \otimes \hat{\mathbf{z}}_2^\dagger\right)\right]\mathbf{C}^\dagger = \alpha_2\left(\mathbf{U}\sqrt{\mathbf{D}}\,\hat{\mathbf{z}}_2\right)\otimes\left(\hat{\mathbf{z}}_2^\dagger\sqrt{\mathbf{D}}\,\mathbf{U}^\dagger\right)$$
$$= \alpha_2\left(\mathbf{U}\sqrt{\mathbf{D}}\,\hat{\mathbf{z}}_2\right)\otimes\left(\mathbf{U}\sqrt{\mathbf{D}}\,\hat{\mathbf{z}}_2\right)^\dagger. \tag{75}$$

where, as we have observed, the vector $\hat{\mathbf{z}}_2$ can be chosen arbitrarily in the subspace orthogonal to $\hat{\mathbf{v}}_3$.

The trace of $\mathbf{R}_{p2}$ is given by

$$\operatorname{tr}\mathbf{R}_{p2} = \alpha_2\left|\mathbf{U}\sqrt{\mathbf{D}}\,\hat{\mathbf{z}}_2\right|^2 \Rightarrow \alpha_2 = \frac{\operatorname{tr}\mathbf{R}_{p2}}{\left|\mathbf{U}\sqrt{\mathbf{D}}\,\hat{\mathbf{z}}_2\right|^2} = \frac{\operatorname{tr}\mathbf{R}}{\left|\sqrt{\mathbf{D}}\,\hat{\mathbf{z}}_2\right|^2}. \tag{76}$$

Due to the non-uniqueness of the eigenvectors corresponding to the subspace given by the degenerate eigenvalue 1, the coherency matrix $\mathbf{R}'$ can be written as

$$\mathbf{R}' = \mathbf{C}\mathbf{Z}\mathbf{D}(1,1,0)\mathbf{Z}^\dagger\mathbf{C}^\dagger = \mathbf{C}\left[\left(\hat{\mathbf{z}}_1\otimes\hat{\mathbf{z}}_1^\dagger\right) + \left(\hat{\mathbf{z}}_2\otimes\hat{\mathbf{z}}_2^\dagger\right)\right]\mathbf{C}^\dagger. \tag{77}$$

Thus, the simultaneous diagonalization of $\mathbf{R}_{p2}$ and $\mathbf{R}'$ allows us to perform the following subtraction

$$\mathbf{R}' - p_2\mathbf{R}_{p2} = \mathbf{C}\mathbf{Z}\left[\mathbf{D}(1,1-p_2\alpha_2,0)\right]\mathbf{Z}^\dagger\mathbf{C}^\dagger, \tag{78}$$

and, by choosing $p_2 = 1/\alpha_2$ we obtain

$$\mathbf{R}'' = \mathbf{R}' - p_2\mathbf{R}_{p2} = \mathbf{C}\mathbf{Z}\mathbf{D}(1,0,0)\mathbf{Z}^\dagger\mathbf{C}^\dagger = \mathbf{C}\left(\hat{\mathbf{z}}_1\otimes\hat{\mathbf{z}}_1^\dagger\right)\mathbf{C}^\dagger. \tag{79}$$

The last pure component is fully determined by the previous choices of $\mathbf{R}_{p3}$ and $\mathbf{R}_{p2}$ and can be written as

$$\mathbf{R}_{p1} = \alpha_1\mathbf{R}'' = \alpha_1\mathbf{C}\left(\hat{\mathbf{z}}_1\otimes\hat{\mathbf{z}}_1^\dagger\right)\mathbf{C}^\dagger, \tag{80}$$

so that by defining the unit vectors

$$\hat{\mathbf{w}}_2 \equiv \frac{1}{p_2\operatorname{tr}\mathbf{R}}\mathbf{U}\sqrt{\mathbf{D}}\,\hat{\mathbf{z}}_2, \tag{81}$$

$$\hat{\mathbf{w}}_1 \equiv \frac{1}{p_1\operatorname{tr}\mathbf{R}}\mathbf{U}\sqrt{\mathbf{D}}\,\hat{\mathbf{z}}_1, \tag{82}$$

and considering the obtained coefficients

$$p_2 = \frac{1}{\alpha_2} = \frac{\left|\sqrt{\mathbf{D}}\hat{\mathbf{z}}_2\right|^2}{\operatorname{tr}\mathbf{R}} = \frac{1}{(\operatorname{tr}\mathbf{R})\sum_{j=1}^3\frac{1}{\lambda_j}\left|\left(\mathbf{U}^\dagger\hat{\mathbf{w}}_2\right)_j\right|^2}, \tag{83}$$

$$p_1 = \frac{1}{\alpha_1} = \frac{\left|\sqrt{\mathbf{D}}\hat{\mathbf{z}}_1\right|^2}{\mathrm{tr}\,\mathbf{R}} = \frac{1}{\left(\mathrm{tr}\,\mathbf{R}\right)\sum_{j=1}^{3}\frac{1}{\lambda_j}\left|\left(\mathbf{U}^\dagger\hat{\mathbf{w}}_1\right)_j\right|^2}, \tag{84}$$

the arbitrary decomposition of $\mathbf{R}$ is obtained as follows

$$\mathbf{R} = \sum_{i=1}^{3} p_i \mathbf{R}_{pi}, \quad p_i = \frac{1}{\sum_{j=1}^{3}\frac{1}{\hat{\lambda}_j}\left|\left(\mathbf{U}^\dagger\hat{\mathbf{w}}_i\right)_j\right|^2},$$

$$\mathbf{R}_{pi} = \left(\mathrm{tr}\,\mathbf{R}\right)\left(\hat{\mathbf{w}}_i \otimes \hat{\mathbf{w}}_i^\dagger\right), \quad \sum_{i=1}^{3} p_i = 1, \quad \hat{\lambda}_j \equiv \frac{\lambda_j}{\mathrm{tr}\,\mathbf{R}}. \tag{85}$$

For the case $\mathrm{rank}\,\mathbf{R}=2$ the same procedure can be applied by starting from the expression of the coherency matrix $\mathbf{R}'$.

This result can be interpreted as follows: any mixed state of 3D polarized light is physically equivalent to an arbitrary incoherent superposition of up to three independent pure sates. When $\mathrm{rank}\,\mathbf{R}=3$, any pure state can be considered as a component and, once chosen, the election of the second component is not completely arbitrary, whereas the third component is fully determined by the previous choices. When $\mathrm{rank}\,\mathbf{R}=2$, the polarization plane is fixed, and the mixed state corresponds to a 2D state of partially polarized light. This agrees with the arbitrary decomposition seen in the section devoted to 2D polarized light.

Obviously, the spectral decomposition is a particular case of the arbitrary decomposition. Moreover, we see that any mixed state with $\mathrm{rank}\,\mathbf{R}>2$ can be expressed as a convex sum of an arbitrary pure state and a corresponding mixed state.

As we will show, this new result is also relevant when applied to coherency matrices representing the polarimetric properties of material media.

### 3.6. Polarimetric purity of 3x3 coherency matrices

*For updated details, see*

I. San José, J. J. Gil, "Invariant indices of polarimetric purity. Generalized indices of purity for n×n covariance matrices," Opt. Commun. **284**, 38-47 (2011).

J. J. Gil, "Interpretation of the coherency matrix for three-dimensional polarization states," Phys. Rev. A **90**, 043858 (2014).

J. J. Gil, "Components of purity of a three-dimensional polarization state," J. Opt. Soc. Am. A **33**, 40-43 (2016).

J. J. Gil, A. T. Friberg, T. Setälä, I. San José, "Structure of polarimetric purity of three-dimensional polarization states," Phys. Rev. A **95**, 053856 (2017).

J. J. Gil, A. Norrman, A. T. Friberg, T. Setälä, "Nonregularity of three-dimensional polarization states", Opt. Lett. **43**, 4611-4614 (2018).

J. J. Gil, A. Norrman, A. T. Friberg, T. Setälä, "Polarimetric purity and the concept of degree of polarization", Phys. Rev. A **97**, 023838 (2018).

J. J. Gil, A. Norrman, T. Setälä, A. T. Friberg, "Intensity and spin anisotropy of a three-dimensional polarization state," Opt. Lett. **44**, 3578-3581 (2019).

A first proper definition of the 3D degree of polarization was presented by Samson in 1973 [104] within the scope of geophysical studies of ultra-low frequency magnetic fields. This result was also obtained by Barakat by formulating the degree of polarization in terms of scalar invariants of the coherency matrix [53]. Nevertheless, the study of the 3D degree of polarization has recently attracted attention due to the advances in optical nanotechnologies and from the necessity of understanding polarization phenomena in fluctuating near fields and evanescent waves.

A complete formulation of the purity of the 3D states of polarization based on two relative differences of the eigenvalues of the coherency matrix (indices of purity) has been introduced by Gil, Correas, Melero and Ferreira [41]. More recently, Ellis, Dogariu, Ponomarenko and Wolf [105] have also presented two parameters based on two relative differences of the eigenvalues of the coherency matrix and, from this result, Hioe [106] introduced a parameter called the *degree of isotropy*. We refer also to the relevant approaches obtained by Réfrégier, Roche and Goudail [107] who have emphasized the necessity of three quantities to characterize the invariant properties of 3x3 coherency matrices and have shown that, for light with Gaussian fluctuations, a set of three invariant parameters is enough to characterize the *polarimetric contrast*, so that different sets of three invariant parameters defined from the 3D coherency matrix have been considered, as well as their relations with the different *degrees of polarization* defined by Barakat [53] and by Samson [104]. Réfrégier and Goudail have also and proposed the Kullback relative entropy as an invariant parameter that, together with the intensity and the 3D degree of polarization, completes a set of invariant parameters [108].

Another purity parameter has recently been introduced by Dennis [109] by means of averaging the 3D state of polarization due to a dipole over all scattering directions, which leads to a purity measure which is not invariant under unitary transformations of the coherency matrix.

Particularly relevant contributions concerning the concept of the 3D degree of polarization have been presented by Setälä, Shevchenko, Kaivola and Friberg [98,94] and Ellis and Dogariu [92].

Quantum and classical approaches to the 3D degree of polarization concept, defining it as a distance between distributions [83] and between correlation matrices have been presented by Luis [95,110] The connection between the degree of polarization and the effective degree of coherence has been studied by Vahimaa and Tervo [111].

The *3D degree of polarization* can be defined as [104,53,98]

$$P_{3D} = \sqrt{\frac{1}{2}\left(\frac{3\mathrm{tr}\mathbf{R}^2}{\mathrm{tr}^2\mathbf{R}} - 1\right)}. \tag{86}$$

This invariant parameter is limited to the interval $0 \leq P_{3D} \leq 1$, in such a manner that $P_{3D} = 1$ corresponds to the case that $\mathbf{R}$ has only one nonzero eigenvalue (total polarimetric purity: the polarization plane is constant and the electric field describes a well-defined polarization ellipse), whereas $P_{3D} = 0$ is reached when the three eigenvalues of $\mathbf{R}$ are equal (equiprobable mixture of states and zero correlation between the electric field components).

Note that $P_{3D}$ is not other than the 3D version of the depolarization index, or 4D degree of polarimetric purity $P_{4D} \equiv P_\Delta$, defined in [J. J. Gil, E. Bernabéu, "Depolarization and polarization indices of an optical system," Optica Acta **33**, 185-189 (1986)] for 4D coherency matrices associated with Mueller matrices.

For reasons that will be made clear below, we prefer using the term *degree of polarimetric purity* (which obviously refers to polarimetric purity), rather than *degree of polarization*, for $P_{3D}$.

The expression of $P_{3D}$ as a function of the 3D Stokes parameters is

$$P_{3D} = \frac{1}{\sqrt{2}} \sqrt{\frac{\sum_{i,j=0}^{2} q_i^2}{q_{00}^2} - 1} = \sqrt{\frac{1}{2}\left(\frac{\mathrm{tr}(\mathbf{Q}^T\mathbf{Q})}{q_{00}^2} - 1\right)}, \tag{87}$$

and, in terms of both norms of $\mathbf{R}$

$$P_{3D} = \sqrt{\frac{1}{2}\left(\frac{3\|\mathbf{R}\|_2^2}{\|\mathbf{R}\|_0^2} - 1\right)}. \tag{88}$$

As Setälä et al. have pointed out [98,94], $P_{3D}$ takes into account not only the purity of the mean polarization ellipse, but also the stability of the plane that contains the instantaneous components of the electric field of the wave. Thus, for unpolarized light whose propagation direction remains fixed, $P_{2D} = 0$ whereas $P_{3D} = 1/2$. It is clear that in the 3D description of polarization, new relevant quantities and peculiar properties arise that do not appear in the 2D model. Therefore, the existence of three eigenvalues leads to the fact that, unlike the 2D model, the overall degree of purity $P_{3D}$ does not provide complete information of the polarimetric purity properties. In order to find appropriate invariant nondimensional quantities that provide a complete description of the purity of 3D polarization states, we return to $P_{2D}$ and observe that it can be defined as a relative difference between the two eigenvalues, so that $0 \leq P_{2D} \leq 1$. Thus, in the light of the structure of the algebraic expressions of the eigenvalues of $\mathbf{R}$, and by inspecting the various relative differences between them, we see that a convenient pair of *indices of polarimetric purity* (IPP) is defined as [41]

$$P_1 = \frac{\lambda_1 - \lambda_2}{\mathrm{tr}\,\mathbf{R}}, \quad P_2 = \frac{\lambda_1 + \lambda_2 - 2\lambda_3}{\mathrm{tr}\,\mathbf{R}}. \tag{89}$$

These quantities are restricted by the following limits

$$0 \leq P_1 \leq P_2 \leq 1. \tag{90}$$

From the above equations, the following quadratic relation between $P_{3D}$ and the two indices of purity $P_1, P_2$ is derived

$$P_{3D}^2 = \frac{1}{4}\left(3P_1^2 + P_2^2\right). \tag{91}$$

Another interesting expression of $P_{3D}$, as a homogeneous quadratic measure of all the relative differences between the eigenvalues, is the following

$$P_{3D}^2 = \frac{1}{2} \sum_{\substack{i,j=1 \\ i<j}}^{3} p_{ij}^2, \quad p_{ij} \equiv \frac{\lambda_i - \lambda_j}{\mathrm{tr}\,\mathbf{R}}. \tag{92}$$

Thus, the two indices of purity provide complete information on the polarimetric purity of the corresponding polarization state. This enhanced description based on invariant parameters with values restricted to the range between 0 and 1, has special significance from a physical point of view.

Through a series works (see the references below), it has been demonstrated that the diagonalization of the real part of $\mathbf{R}$ leads to a set of physical parameters that are invariant under rotation transformations of the Cartesian reference frame. The intrinsic 3D polarization matrix is defined as $\mathbf{R}_O \equiv \mathbf{Q}^T\mathbf{R}\mathbf{Q}$, where $\mathbf{Q}$ is the proper orthogonal matrix that satisfies $\mathbf{Q}(\mathrm{Re}\,\mathbf{R})\mathbf{Q}^T =$ $I \operatorname{diag}(\hat{a}_1, \hat{a}_2, \hat{a}_3)$ (with $\hat{a}_1 \geq \hat{a}_2 \geq \hat{a}_3 \geq 0$ and $\hat{a}_1 + \hat{a}_2 + \hat{a}_3 = 1$), so that $\mathbf{R}_O$ is the representation of $\mathbf{R}$ with respect to its *intrinsic reference frame* $X_O Y_O Z_O$ and has the form

$$\mathbf{R}_O = I \begin{pmatrix} \hat{a}_1 & -i\hat{n}_{O3}/2 & i\hat{n}_{O2}/2 \\ i\hat{n}_{O3}/2 & \hat{a}_2 & -i\hat{n}_{O1}/2 \\ -i\hat{n}_{O2}/2 & i\hat{n}_{O1}/2 & \hat{a}_3 \end{pmatrix}.$$

In accordance to Eq. (46) corresponding intrinsic Stokes parameters are defined as

$$q_{11} = I, \quad q_{22} = \sqrt{3/2}\, I P_l, \quad q_{33} = \sqrt{1/2}\, I P_d, \triangleright$$

$$q_{23} = -\sqrt{3/2}\, I \hat{n}_{O1}, \quad q_{13} = \sqrt{3/2}\, I \hat{n}_{O2}, \quad q_{12} = -\sqrt{3/2}\, I \hat{n}_{O1},$$

$$\left[ P_l \equiv \hat{a}_1 - \hat{a}_2, \; P_d \equiv \hat{a}_1 + \hat{a}_2 - 2\hat{a}_3, \; \hat{\mathbf{n}}_O \equiv (\hat{n}_{O1}, \hat{n}_{O2}, \hat{n}_{O3})^T \right],$$

where $P_l$ is the *degree of linear polarization*, $P_d$ is the *degree of directionality* (i.e., the degree of stability of the polarization plane of $\mathbf{R}$) and $\hat{\mathbf{n}}_O$ is the *intrinsic spin density vector* (i.e., the intensity-normalized intrinsic angular momentum vector of $\mathbf{R}$), whose components determine the orientation of $\hat{\mathbf{n}}_O$ with respect to the intrinsic reference frame $X_O Y_O Z_O$ as well as its the absolute value $\hat{n}_O = \sqrt{\hat{n}_{O1}^2 + \hat{n}_{O2}^2 + \hat{n}_{O3}^2}$ (note that $\hat{n}_O$ is not other than the degree of circular polarization $P_c$ of the state $\mathbf{R}$). While the indices of polarimetric purity (IPP) provide quantitative information of the relative weights of the incoherent components of $\mathbf{R}$, the above parameters provide additional qualitative information on the physical properties of the state $\mathbf{R}$. Parameters $P_l, P_c, P_d$ are the so-called *components of purity* (CP) of $\mathbf{R}$.

It has also been demonstrated that 3D states can be regular (when either $P_1 = P_2$ or the polarization matrix $\mathbf{R}_m$ of the discriminating component of $\mathbf{R}$ is a real matrix [see Eq. (93) and subsequent paragraphs for details about the characteristic decomposition of $\mathbf{R}$]), or nonregular (the general case where $P_1 \neq P_2$ and $\mathbf{R}_m$ is a complex valued matrix). In the case of regular states, $P_1 = \sqrt{P_l^2 + P_c^2}$ and $P_2 = P_d$, while these equalities are no longer satisfied for nonregular states. Thus, contrary to what was stated in the original version of this Review, except for the particular case of regular states $(P_2 = P_d)$, $P_2$ does not give a measure of the degree of stability of the polarization plane (which is always determined by $P_d$). IPP and CP constitute two different ways (quantitative and qualitative, respectively) to interpret the contributions to the overall polarimetric purity of $\mathbf{R}$, and satisfy

$$P_{3D} = \sqrt{\frac{3}{4}P_1^2 + \frac{1}{4}P_2^2} = \sqrt{\frac{3}{4}\left(P_l^2 + P_c^2\right) + \frac{1}{4}P_d^2}\,.$$

*For additional details, see*

J. J. Gil, "Interpretation of the coherency matrix for three-dimensional polarization states," Phys. Rev. A **90**, 043858 (2014).

J. J. Gil, "Intrinsic Stokes parameters of 3D and 2D polarization states," J. Eur. Opt. Soc. RP **10**, 15054 (2015).

J. J. Gil, "Components of purity of a three-dimensional polarization state," J. Opt. Soc. Am. A **33**, 40-43 (2016).

J. J. Gil, A. T. Friberg, T. Setälä, I. San José, "Structure of polarimetric purity of three-dimensional polarization states," Phys. Rev. A **95**, 053856 (2017).

J. J. Gil, A. Norrman, A. T. Friberg, T. Setälä, "Nonregularity of three-dimensional polarization states", Opt. Lett. **43**, 4611-4614 (2018).

J. J. Gil, A. Norrman, A. T. Friberg, T. Setälä, "Polarimetric purity and the concept of degree of polarization", Phys. Rev. A **97**, 023838 (2018).

J. J. Gil, A. Norrman, T. Setälä, A. T. Friberg, "Intensity and spin anisotropy of a three-dimensional polarization state," Opt. Lett. **44**, 3578-3581 (2019).

The physically feasible region in the *purity space* $P_1, P_2$ is shown in Fig.5.

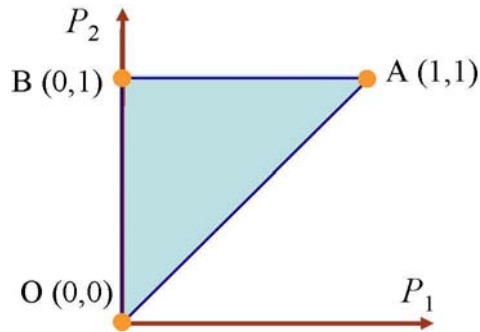

Fig. 5. Feasible region for $P_1, P_2$ in the purity space.

As we will see in the case analysis presented below, $P_1$ and $P_2$ have the following properties:

- If $P_1 = 0$ (fully random polarization ellipse), then $0 \leq P_2 \leq 1$. We see that the only possible contribution to purity is provided by $P_2$.
- If $P_1 = 1$, then $P_2 = 1$ and $P_{3D} = 1$ (pure state).
- If $P_2 = 0$, then $P_1 = 0$ and $P_{3D} = 0$ (3D unpolarized state).

To provide a deeper understanding of the physical meaning of the indices of purity, we analyze different states represented in the corresponding feasible region.

*a)* $P_1 = 0, \ 0 < P_2 < 1$.

$0 < P_{3D} < 1/2; \ \lambda_1 = \lambda_2 > \lambda_3 > 0$. The system is equivalent to an incoherent combination of three pure contributions. The more significant two have equal intensities $\lambda_1 = \lambda_2 > \lambda_3$. These states are represented in the edge OB (vertices O and B excluded) and correspond to light whose polarization plane fluctuates ($P_2 < 1$) with a predominance of an *average polarization plane*, whereas the shape of the polarization ellipse is fully random ($P_1 = 0$).

*b)* $0 \leq P_1 \leq 1, \ P_2 = 1$.

$1/2 \leq P_{3D} \leq 1; \ \lambda_1 \geq \lambda_2 \geq 0, \ \lambda_1 > 0, \ \lambda_3 = 0$. The system is composed of two incoherent beams with orthogonal states of polarization. These states are represented in the edge BA.

   *b.1)* $P_1 = P_2 = 1$.

   $P_{3D} = 1; \ \lambda_1 > 0, \ \lambda_2 = \lambda_3 = 0$. The vertex A represents this state of totally polarized light, i.e. light with fixed polarization plane and polarization ellipse with stable shape (in short times only the intensity fluctuates; whereas the azimuth $\chi$ and the ellipticity $\varphi$ are fixed).

   *b.2)* $P_1 = 0, \ P_2 = 1$.

   $P_{3D} = 1/2; \ \lambda_1 = \lambda_2 > 0, \ \lambda_3 = 0$. The point B represents this *discriminating state* given by equiprobable statistical contributions of the two orthogonally polarized incoherent components.

*c)* $0 < P_2 = P_1 < 1$

$P_{3D} = P_2 = P_1; \ \lambda_1 > \lambda_2 = \lambda_3 > 0$. These states are represented in the segment OA (vertices O, A excluded) and lack discriminating component, so that they are composed of a pure state and a 3D unpolarized state. $P_1$ exhibits its maximum value compatible with $P_2$.

*d)* $P_1 = P_2 = 0$.

$P_{3D} = 0$; $\lambda_1 = \lambda_2 = \lambda_3 > 0$. This state is represented by the vertex O and corresponds to a fully random 3D state of polarization. No predominance exists in the polarization plane and in the polarization ellipse. The only nonzero 3D Stokes parameter is $q_{00}$. It should be stressed that the inequality $P_1 \leq P_2$ indicates that the value of $P_1$ is limited by the value of $P_2$. This result agrees with the expected physical behavior.

*e)* $0 < P_1 < P_2 < 1$.

$0 < P_{3D} < 1$; $\lambda_1 > \lambda_2 > \lambda_3 > 0$. These states are represented by points inside the triangle OBA and exhibit partially random fluctuations of the polarization plane and of the shape of the instantaneous polarization ellipse (with different randomness $P_1 \neq P_2$). It should be noted again that the degree of purity of the fluctuations of the polarization ellipse ($P_1$) can never exceed the purity of the fluctuations of the direction orthogonal to the plane containing the polarization ellipse ($P_2$).

The above description can be compared with the conclusions of the study performed by Ellis and Dogariu concerning the concept of degree of polarization and the other invariant quantities describing the purity of a polarization state [92]. In the light of the previous analysis, we can consider the decomposition proposed by Holm and Barnes [112], which has also been considered by Cloude and Pottier [113] and by Ellis and Dogariu [92]

$$\begin{aligned}\mathbf{R} &= \mathbf{U}\mathbf{D}(\lambda_1, \lambda_2, \lambda_3)\mathbf{U}^\dagger \\ &= \mathbf{U}\mathbf{D}(\lambda_1 - \lambda_2, 0, 0)\mathbf{U}^\dagger + \mathbf{U}\mathbf{D}(\lambda_2 - \lambda_3, \lambda_2 - \lambda_3, 0)\mathbf{U}^\dagger + \mathbf{U}\mathbf{D}(\lambda_3, \lambda_3, \lambda_3)\mathbf{U}^\dagger.\end{aligned} \quad (93)$$

Therefore, the following *characteristic decomposition* can be always performed as

$$\mathbf{R} = P_1 \mathbf{R}_p + (P_2 - P_1)\mathbf{R}_m + (1 - P_2)\mathbf{R}_{u-3D},$$

$$\mathbf{R}_p \equiv I\,\mathbf{U}\mathbf{D}(1,0,0)\mathbf{U}^\dagger, \quad \mathbf{R}_m \equiv I\frac{1}{2}\mathbf{U}\mathbf{D}(1,1,0)\mathbf{U}^\dagger, \quad \mathbf{R}_{u-3D} \equiv I\frac{1}{3}\mathbf{I}, \quad (94)$$

$$[\mathbf{I} \equiv \mathbf{D}(1,1,1)],$$

where the coherency matrix $\mathbf{R}$ is expressed as the convex sum of the following coherency matrices: $\mathbf{R}_p$, which represents a 2D pure state (vertex A, case *b.1*); $\mathbf{R}_m$, which represents a *discriminating state* (vertex B, case *b.2*), and $\mathbf{R}_{u-3D}$, which represents a 3D unpolarized state (vertex O, case *d*).

Contrary to what was stated in the original version of this review, it has been demonstrated in [J. J. Gil, A. T. Friberg, T. Setälä, I. San José, "Structure of polarimetric purity of three-dimensional polarization states," Phys. Rev. A **95**, 053856 (2017).] that, in general, $P_2 = 1$ (i.e. $\lambda_3 = 0$) doest not imply that this decomposition becomes the 2D characteristic decomposition (this only happens for regular states of polarization).

[see J. J. Gil, A. Norrman, A. T. Friberg, T. Setälä, "Nonregularity of three-dimensional polarization states", Opt. Lett. **43**, 4611-4614 (2018)].

This result can be expressed as follows: "*any 3D polarization state can be considered an incoherent superposition of three polarization states: a pure state, a discriminating state and a 3D unpolarized state*".

In further sections devoted to coherency matrices representing the polarimetric properties of material media, we will see that any physically realizable decomposition of such coherency matrices has the form of a linear convex combination of coherency matrices.

The characteristic decomposition constitutes an appropriate framework for distinguishing the pure component from the random or noise component of the radiation field. In the same sense it allows a proper treatment of the measurement errors when it is known that the polarization state being measured is pure.

### 3.7. Entropy associated with R

From the description of the von Neumann entropy for *n*-dimensional density matrices introduced by Fano [46] and studied by Brosseau [2], the 3D polarization entropy can be defined as

$$S_{3D} = -\mathrm{tr}\left(\hat{\mathbf{R}} \log_3 \hat{\mathbf{R}}\right) = -\sum_{i=1}^{3}\left(\hat{\lambda}_i \log_3 \hat{\lambda}_i\right), \quad (95)$$

where $\hat{\mathbf{R}} \equiv \mathbf{R}/\mathrm{tr}\mathbf{R}$ is the polarization density matrix and $\hat{\lambda}_1, \hat{\lambda}_2, \hat{\lambda}_3$ are the eigenvalues of $\hat{\mathbf{R}}$.

Now, in order to complete the study of the physical parameters involved in 3D polarization states, and in the light of the above results, we consider the entropy of a state in terms of the indices of polarimetric purity of $\mathbf{R}$

$$S_{3D}(P_1, P_2) = -\frac{\left(1+\frac{1}{2}P_2+\frac{3}{2}P_1\right)}{3}\log_3\left[\frac{1}{3}\left(1+\frac{1}{2}P_2+\frac{3}{2}P_1\right)\right]$$
$$-\frac{\left(1+\frac{1}{2}P_2-\frac{3}{2}P_1\right)}{3}\log_3\left[\frac{1}{3}\left(1+\frac{1}{2}P_2-\frac{3}{2}P_1\right)\right] - \frac{(1-P_2)}{3}\log_3\left[\frac{1}{3}(1-P_2)\right]. \quad (96)$$

Thus, $S_{3D}$ is determined univocally by $P_1$ and $P_2$. The maximum $S_{3D} = \log_3 3 = 1$ corresponds to $P_1 = P_2 = P_{3D} = 0$, while the minimum $S_{3D} = 0$ is reached when $P_1 = P_2 = P_{3D} = 1$.

Due to the relevant nature of the concepts related with entropy, Réfrégier et al. [108,114,115] have dealt with the Shannon entropy and with the Kullback relative entropy, including their relations with the *n*-dimensional degrees of polarization for optical waves with Gaussian and non-Gaussian probability density functions. In these works, the degree of polarization is related to the measure of proximity between probability density functions and to the measure of disorder provided by the Shannon entropy.

Taking into account the expression of the von Neumann entropy associated with 2x2 coherency matrices given by Eq. (34), as well as that in the case of 3x3 coherency matrices there exist two indices of purity defined as relative differences of the eigenvalues, it is possible to introduce a proper definition of two respective *partial entropies*, namely $S_{2D}(P_2)$ and $S_{2D}(P_1)$

Condition $0 \leq P_1 \leq P_2 \leq 1$ on the indices of purity has its counterpart in the partial entropies $0 \leq S_{2D}(P_2) \leq S_{2D}(P_1) \leq 1$.

Furthermore, following the definition of Réfrégier, Roche and Goudail [107] of the polarimetric contrast, we find that $I, P_1, P_2$ can be considered an appropriate set of characteristic parameters of the polarimetric contrast of the corresponding $3\times 3$ coherency matrix $\mathbf{R}$.

## 4. Mathematical representation of the polarimetric effects of material media

The polarimetric effects of a material sample depend on its intrinsic properties, on the spectral characteristics of interacting light, and on the particular type of interaction under study. Thus, given a direction of incidence of the light probe, it is possible to study the effects in the refracted, reflected, diffracted or scattered light in different directions. The transformation of the polarization state of the incoming light into the outgoing one due to the interaction with a passive linear medium can be represented through a linear transformation of the corresponding Stokes parameters [10-13].

The main problem of theoretical polarimetry is the physical interpretation of the information that provides the sixteen elements of the transformation matrix (Mueller matrix). To address this problem a suitable mathematical characterization of these matrices is required, which is obtained through the construction of a 4×4 coherency matrix whose properties are analogous to those of the 2×2 and 3×3 coherency matrices studied previously.

In the case of deterministic nondepolarizing linear material systems (pure systems), pure states are transformed into pure states, so that these basic interactions involve birefringence and diattenuation properties. In the general case, the incoherent superposition of emerging light pencils results in depolarization phenomena.

In this section, a general characterization of physical Mueller matrices is obtained from the representation of a passive, linear optical system as a parallel combination of pure elements. This model is justified because of the essential nature of scattering phenomena: any polarimetric behavior of a material medium is derived from the secondary radiation of the accelerated atomic and molecular charges of the medium.

This general characterization is obtained by means of a complete set of explicit necessary and sufficient conditions for a real 4×4 matrix to be a physical Mueller matrix. A subset of four conditions is derived from the mathematical properties of the covariance matrix associated with a Mueller matrix (covariance conditions). Another subset of conditions, composed of two *passivity conditions* is based on the physical restriction that the elements of the parallel combination do not amplify the intensity of light.

### 4.1. Basic polarimetric interaction. The Jones matrix

In order to represent the effects of a deterministic-nondepolarizing optical system (or, briefly, *pure system*) on the polarization properties of an electromagnetic wave that interacts with it, let us first consider the Jones formalism. It should be noted that we use the term *deterministic nondepolarizing* in the sense that totally polarized incident light always emerges totally polarized. This observation is important because, as Simon has pointed out [116], the degree of polarization of a partially polarized light beam can decrease when it interacts with some kinds of deterministic optical systems (for example partial polarizers). Moreover, there exists the possibility of depolarizing effects when polychromatic light passes through some deterministic systems [117-120].

For passive pure systems, the linear transformation of the electric field components of the light beam interacting with them is given by the corresponding Jones matrix **T**

$$\boldsymbol{\varepsilon}'(t) = \mathbf{T}\boldsymbol{\varepsilon}(t), \tag{97}$$

where $\boldsymbol{\varepsilon}'$, $\boldsymbol{\varepsilon}$, are the emerging and incident Jones vectors (Fig. 6).

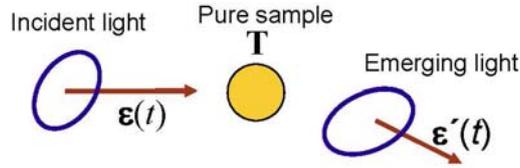

Fig. 6. Basic polarimetric interaction.

Thus, we can say that any polarimetrically-pure system (hereafter *pure*) can be represented by its corresponding Jones matrix and vice-versa.

In general both emerging and incident fields can fluctuate, so that, to describe the properties of partially polarized light, it is necessary to use the 2x2 coherency matrices rather than the Jones vectors. The transformation of the coherency matrix is given by

$$\boldsymbol{\Phi}' = \left\langle \boldsymbol{\varepsilon}' \otimes \boldsymbol{\varepsilon}'^\dagger \right\rangle = \left\langle \mathbf{T}\boldsymbol{\varepsilon} \otimes (\mathbf{T}\boldsymbol{\varepsilon})^\dagger \right\rangle = \left\langle \mathbf{T}\boldsymbol{\varepsilon} \otimes \boldsymbol{\varepsilon}^\dagger \mathbf{T}^\dagger \right\rangle = \mathbf{T}\left\langle \boldsymbol{\varepsilon} \otimes \boldsymbol{\varepsilon}^\dagger \right\rangle \mathbf{T}^\dagger = \mathbf{T}\boldsymbol{\Phi}\mathbf{T}^\dagger. \tag{98}$$

Like other authors [13,121], and to make easier subsequent calculations, we introduce the following alternative notation for the elements of $\boldsymbol{\Phi}$

$$\varphi_0 \equiv \varphi_{11},\ \varphi_1 \equiv \varphi_{12},\ \varphi_2 \equiv \varphi_{21},\ \varphi_3 \equiv \varphi_{22}. \tag{99}$$

The elements of the polarization matrix can also be written as the column-vector $\boldsymbol{\varphi}$ defined as

$$\boldsymbol{\varphi} \equiv (\varphi_0, \varphi_1, \varphi_2, \varphi_3)^T = \left(\left\langle \varepsilon_1 \varepsilon_1^* \right\rangle, \left\langle \varepsilon_1 \varepsilon_2^* \right\rangle, \left\langle \varepsilon_2 \varepsilon_1^* \right\rangle, \left\langle \varepsilon_2 \varepsilon_2^* \right\rangle\right)^T = \left\langle \boldsymbol{\varepsilon} \otimes \boldsymbol{\varepsilon}^* \right\rangle. \tag{100}$$

Vector $\boldsymbol{\varphi}$ and the corresponding Stokes vector $\mathbf{s}$ are related by means of the following expression [13,47]

$$\mathbf{s} = \mathbf{L}\boldsymbol{\varphi}, \tag{101}$$

where

$$\mathbf{L} = \begin{pmatrix} 1 & 0 & 0 & 1 \\ 1 & 0 & 0 & -1 \\ 0 & 1 & 1 & 0 \\ 0 & i & -i & 0 \end{pmatrix}. \tag{102}$$

This matrix $\mathbf{L}$ that connects both vector representations of the polarization states has the following property

$$\mathbf{L}^{-1} = \frac{1}{2}\mathbf{L}^\dagger. \tag{103}$$

Taking into account some properties of the Kronecker product, it is straightforward to show that the Stokes vector $\mathbf{s}'$ of the emerging beam is given by

$$\begin{aligned}\mathbf{s}' &= \mathbf{L}\left\langle \boldsymbol{\varepsilon}' \otimes \boldsymbol{\varepsilon}'^* \right\rangle = \mathbf{L}\left\langle (\mathbf{T}\boldsymbol{\varepsilon}) \otimes (\mathbf{T}\boldsymbol{\varepsilon})^* \right\rangle = \mathbf{L}\left\langle (\mathbf{T} \otimes \mathbf{T}^*)(\boldsymbol{\varepsilon} \otimes \boldsymbol{\varepsilon}^*) \right\rangle = \\ &= \mathbf{L}(\mathbf{T} \otimes \mathbf{T}^*)\left\langle (\boldsymbol{\varepsilon} \otimes \boldsymbol{\varepsilon}^*) \right\rangle = \mathbf{L}(\mathbf{T} \otimes \mathbf{T}^*)\boldsymbol{\varphi} = \mathbf{L}(\mathbf{T} \otimes \mathbf{T}^*)\mathbf{L}^{-1}\mathbf{s}.\end{aligned} \tag{104}$$

Therefore, these linear transformations of Stokes vectors are given by the *Mueller-Jones matrix* $\mathbf{M}_J$ [13]

$$\mathbf{M}_J = \mathbf{L}\left(\mathbf{T} \otimes \mathbf{T}^*\right)\mathbf{L}^{-1}, \tag{105}$$

or, in components,

$$m_{kl} = \frac{1}{2}\mathrm{tr}\left(\boldsymbol{\sigma}_k \mathbf{T} \boldsymbol{\sigma}_l \mathbf{T}^\dagger\right). \tag{106}$$

In the next section we will deal with optical systems whose effect on polarized light cannot be represented by means of Jones matrices but can be represented by Mueller matrices. Therefore, like other authors, we will distinguish between Mueller-Jones matrices (or *pure* Mueller matrices [122]), which correspond to pure systems (i.e. characterized by Jones matrices) and Mueller matrices in general.

For the sake of clarity, hereafter we will use the following notation for the different types of matrices: $\mathbf{M}_J$, Mueller-Jones matrices and $\mathbf{M}$, general Mueller matrices.

As a number of authors have pointed out [17,18,45,123-126,], the condition for the elements of the Jones matrix to represent a pure *physically realizable* (or physical) system arises from the restriction that the gain (intensity transmittance) $g$ of the optical system, defined as the ratio between the intensities of the emerging and incident light beams, should be limited to the interval $0 \leq g \leq 1$. This condition, called the *passivity condition* (also *transmittance condition*) can be written as a function of the elements of $\mathbf{T}$ as follows [124,126]

$$\frac{1}{2}\left\{\mathrm{tr}\left(\mathbf{T}^\dagger \mathbf{T}\right) + \sqrt{\mathrm{tr}^2\left(\mathbf{T}^\dagger \mathbf{T}\right) - 4\det\left(\mathbf{T}^\dagger \mathbf{T}\right)}\right\} \leq 1. \tag{107}$$

It should be noted that the condition $0 \leq g$, is directly satisfied [124].

Moreover, it is also worth to consider the reversibility properties of the matrices that represent optical systems, i.e. the operation of interchanging the incident and emerging light beams. It was Jones who first formulated a reciprocity theorem in terms of Jones matrices [9]. Given a linear optical system characterized by a Jones matrix $\mathbf{T}$, the relation between the input $\boldsymbol{\eta}$ and output $\boldsymbol{\eta}'$ Jones vectors when the light passes through the same system in the opposite direction is given by

$$\boldsymbol{\eta}' = \mathbf{T}^r \boldsymbol{\eta}, \quad \mathbf{T}^r \equiv \mathbf{D}(1,-1)\mathbf{T}^T \mathbf{D}(1,-1), \quad \left[\mathbf{D}(1,-1) \equiv \mathrm{diag}(1,-1)\right]. \tag{108}$$

That is to say, the Jones matrix representing the system when incident and emerging beams are interchanged is $\mathbf{T}^r$ (to avoid confusion it is worth recall that in the seminal works of Jones [9] he used a peculiar convention and stated that $\mathbf{T}^r = \mathbf{T}^T$, and this is the reason why $\mathbf{T}^r = \mathbf{T}^T$ has been reported in some works as the rule for reciprocity [127-129]). The above result is not applicable when the system exhibits magneto-optic effects. For example, for a Faraday cell, the forward-propagating and counter-propagating Jones matrices are equal [129].

Moreover, the symmetry of Eq. (107) shows that if it is valid for $\mathbf{T}$, it is also valid for $\mathbf{T}^r$ and for $\mathbf{T}^T$. Therefore, if a Jones matrix $\mathbf{T}$ is physically realizable, $\mathbf{T}^r$ and $\mathbf{T}^T$ are also physically realizable, and vice-versa (note that the only requirement for a 2x2 complex matrix to be a physically realizable Jones matrix is its larger singular value is smaller or equal than 1).

For a Mueller-Jones matrix $\mathbf{M}_J \equiv \mathbf{M}(\mathbf{T})$, the corresponding *reciprocal Mueller-Jones matrix* $\mathbf{M}_J^r$ is given by

$$\mathbf{M}_J^r \equiv \mathbf{M}_J\left(\mathbf{T}^r\right) = \mathbf{D}(1,1,-1,1)\, \mathbf{M}_J^T\, \mathbf{D}(1,1,-1,1), \tag{109}$$

where the diagonal matrix $\mathbf{D}(1,1,-1,1)$ performs a change of the sign in $s_2$ in the reverse interaction. As in the case of Jones matrices, a particular analysis of the form of $\mathbf{M}'_J$ is required when the system includes magneto-optic effects.

The passivity condition, expressed as a function of the elements of $\mathbf{M}_J$, is the following [124]

$$g_f \leq 1, \quad g_f \equiv m_{00} + \sqrt{m_{01}^2 + m_{02}^2 + m_{03}^2}, \tag{110}$$

where $g_f$ is the maximum transmittance of $\mathbf{M}_J$.

In the case of Mueller-Jones matrices, the following equality is satisfied [130,131]

$$m_{01}^2 + m_{02}^2 + m_{03}^2 = m_{10}^2 + m_{20}^2 + m_{30}^2, \tag{111}$$

so that the passivity condition can also be expressed as

$$g_r \leq 1, \quad g_r \equiv m_{00} + \sqrt{m_{10}^2 + m_{20}^2 + m_{30}^2}, \tag{112}$$

which shows that

$$g_r(\mathbf{M}_J) = g_f(\mathbf{M}'_J). \tag{113}$$

In consequence, $g_r$ can be properly considered as the maximum reverse transmittance. We will see that, although for Mueller-Jones matrices both *gains* are equal $g_r(\mathbf{M}_J) = g_f(\mathbf{M}_J)$, this equality fails in general, $g_r(\mathbf{M}) \neq g_f(\mathbf{M})$.

The above conclusions are also valid when the medium exhibits magneto-optic effects because changes of the sign in $m_{0i}, m_{j0}$ $(i, j = 1, 2, 3)$ do not affect the value of the maximum gain.

In a further section we will show that any system is polarimetrically equivalent to a certain parallel combination of pure systems, and that any pure system is polarimetrically equivalent to a serial combination of retarders and diattenuators (partial or total polarizers) [132]. The following sections are devoted to the properties of pure systems, prior to dealing with nonpure systems.

### 4.2. Retarders

Retarders are non-absorbing materials that exhibit different refraction indices (birefringence) for respective orthogonal eigenstates of polarization, and are represented by unitary Jones matrices

$$\mathbf{T}_R(\alpha,\delta,\Delta) \equiv \begin{pmatrix} c_\alpha^2 e^{i\Delta/2} + s_\alpha^2 e^{-i\Delta/2} & s_\alpha c_\alpha \left(e^{i\Delta/2} - e^{-i\Delta/2}\right) e^{-i\delta} \\ s_\alpha c_\alpha \left(e^{i\Delta/2} - e^{-i\Delta/2}\right) e^{i\delta} & s_\alpha^2 e^{i\Delta/2} + c_\alpha^2 e^{-i\Delta/2} \end{pmatrix}, \tag{114}$$

$$s_\alpha \equiv \sin\alpha,\ c_\alpha \equiv \cos\alpha,$$

where $\Delta$ stands for the retardance caused between the two orthogonal elliptically-polarized eigenstates (Fig. 7). Their respective azimuths, (namely $\varphi, \varphi + \pi/2$) and ellipticities ($\chi, -\chi$) are given by

$$\tan 2\varphi = \tan 2\alpha \cos\delta,\ \sin 2\chi = \sin 2\alpha \sin\delta,\ (0 \leq \alpha \leq \pi/2, -\pi < \delta \leq \pi). \tag{115}$$

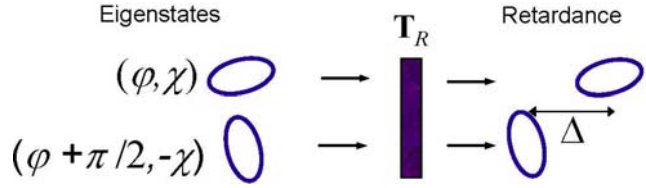

Fig. 7. Retarder.

The operational notation used for the Jones matrices is similar to that used by Priebe [133] for Mueller-Jones matrices.

It is important to point out that birefringent systems can produce depolarization in quasimonochromatic light. For instance, this is the case with long fiber optics, where the shift between the modes is larger than the coherence length. Thus, as we have indicated previously, these cases are not covered by the Jones model, which can only be applied when the accumulated shift is smaller than the coherence length of the light beam. The treatment of linear interactions involving depolarizing effects requires the use of the Mueller formalism.

Moreover, the chromatic dependence of the effective retardance of retarders is applied to the design of Lyot and Solc filters, which are constituted by appropriate serial combinations of retarders placed between total polarizers. The spectral calculations can be performed through the Jones formalism, whereas the study of the integrated depolarization effects requires the Mueller formalism [134,117]. All these considerations also have important applications to liquid crystal optical filters, electro-optical polarization controllers, polarization maintaining fiber ring lasers, birefringent tuning in dye lasers, thin film filters for optical sensing of gas concentration, active filters for color imaging, optical flip-flop systems and many other industrial and scientific devices.

Particular cases of retarders are the linear retarder $\mathbf{T}_R(\alpha,0,\Delta)$ and the circular retarder $\mathbf{T}_R(\pi/4,\pi/2,\Delta)$.

Given a Jones matrix $\mathbf{T}$ in a coordinate system $XY$ ($Z$ is the propagation direction of the light beam), the corresponding Jones matrix $\mathbf{T}'$ with respect to a rotated coordinate system $X'Y'$ ($\theta$: angle between $X$ and $X'$) is given by

$$\mathbf{T}'_R \equiv \mathbf{T}_G(-\theta)\,\mathbf{T}_R\,\mathbf{T}_G(\theta); \quad \mathbf{T}_G(\theta) \equiv \begin{pmatrix} \cos\theta & \sin\theta \\ -\sin\theta & \cos\theta \end{pmatrix}. \tag{116}$$

With this operational notation for the unitary Jones matrices it is easy to show the following equivalences [127,130,132,135-137]:

*a)* $\mathbf{T}_R(\alpha,0,\Delta) = \mathbf{T}_G(-\alpha)\,\mathbf{T}_R(0,0,\Delta)\,\mathbf{T}_G(\alpha)$

*b)* $\mathbf{T}_R^{-1}(\alpha,\delta,\Delta) = \mathbf{T}_R^\Delta(\alpha,\delta,\Delta) = \mathbf{T}_R(\alpha,\delta,-\Delta)$.

*c)* $\mathbf{T}_R(0,0,0) = \mathbf{D}(1,1)$.

*d)* $\mathbf{T}_R(0,0,\pi) = \mathbf{D}(1,-1)$.

*e)* $\mathbf{T}_R(0,\pi/2,\Delta) = \mathbf{T}_R(0,0,-\Delta)$.

*f)* $\mathbf{T}_G(\theta) = \mathbf{T}_R(\pi/4, \pi/2, 2\theta)$: A rotation matrix is equivalent to the matrix of a circular retarder with $\Delta = 2\theta$.

*g)* A system composed of any serial combination of retarders is equivalent to an elliptical retarder.

*h)* An elliptical retarder is equivalent to a serial combination of a linear retarder and a circular retarder (rotator).

*i)* An elliptical retarder is equivalent to a serial combination of two linear retarders.

*j)* $\mathbf{T}_R(\alpha, \delta, \Delta) = \mathbf{T}_R(0, 0, -\delta) \mathbf{T}_R(\alpha, 0, \Delta) \mathbf{T}_R(0, 0, \delta)$

Any retarder is optically equivalent to a linear retarder placed between two linear retarders whose fast axes coincide with the reference axes *X* and *Y* respectively.

*k)* $\mathbf{T}_R(\alpha, 0, \Delta) = \mathbf{T}_R(0, 0, \delta_1) \mathbf{T}_R(\theta, 0, \delta_2) \mathbf{T}_R(0, 0, \delta_1)$:

A linear retarder placed between two equal linear retarders whose fast axes have equal orientations is equivalent to a linear retarder [138]. The parameters $\alpha, \Delta$ of the resulting equivalent linear retarder are

$$\cos(\Delta/2) = \cos\delta_1 \cos(\delta_2/2) - \sin\delta_1 \sin(\delta_2/2) \cos 2\theta$$

$$\tan 2\alpha = \frac{\sin 2\theta}{\sin\delta_1 \cot(\delta_2/2) + \cos\delta_1 \cos 2\theta}$$

This serial combination of three linear retarders provides a simple method for designing tunable compensators by adjusting the angle of the intermediate retarder.

In agreement with theorem *(j)*, any unitary Jones matrix $\mathbf{T}_R$ can be expressed as $\mathbf{T}_R = e^{i\mathbf{K}}$, where $\mathbf{K}$ is the Hermitian matrix given by $\mathbf{K} = i[\mathbf{D}(1,1)) - \mathbf{T}_R][\mathbf{D}(1,1)) + \mathbf{T}_R]^{-1}$ [103]. Thus, from this point of view, any retarder has an associated Hermitian matrix $\mathbf{K}$, in such a manner that its Jones matrix has the above exponential form. This representation is useful for calculations concerning the evolution of polarization states along the direction of propagation of the wave inside thick media [139-141]. In these problems, differential forms of Jones matrices and Mueller-Jones matrices are used [139,142-144].

An interesting case of retarder is a twisted optical fiber with local linear polarized eigenstates. The twist results in an overall behavior equivalent to an elliptical retarder. Nevertheless, the appropriate physical model for the study of polarization mode dispersion is given by the *principal states of polarization* [145,146]. Despite the overall equivalence to an elliptical retarder, the system is also equivalent to a linear retarder followed by a circular retarder (rotator). Thus, the incoming eigenstates of the equivalent linear retarder result in a pair of orthogonal linear polarized outgoing states, which are rotated with respect to the incoming ones. As we will see, this model of *singular states of polarization* is also related with the singular value decomposition of Jones and Mueller matrices, and can be very useful for modeling serial and parallel combinations of pure systems
[See J. J. Gil, R. Ossikovski, Polarized light and the Mueller matrix approach, CRC Press, 2016].

The Jones formulation presented for retarders can easily be translated to the Mueller formalism. We have seen that any matrix of retarder can be expressed as a product of simple matrices corresponding to linear retarders and rotation matrices. The respective Mueller matrices of linear retarders and rotation matrices have the following forms

$$\mathbf{M}_R(0,0,\Delta) = \begin{pmatrix} 1 & 0 & 0 & 0 \\ 0 & 1 & 0 & 0 \\ 0 & 0 & c_\Delta & s_\Delta \\ 0 & 0 & -s_\Delta & c_\Delta \end{pmatrix}, \quad \mathbf{M}_G(\theta) = \mathbf{M}_R(\pi/4,\pi/2,2\theta) = \begin{pmatrix} 1 & 0 & 0 & 0 \\ 0 & c_{2\theta} & s_{2\theta} & 0 \\ 0 & -s_{2\theta} & c_{2\theta} & 0 \\ 0 & 0 & 0 & 1 \end{pmatrix}, \quad (117)$$

so that, in the same way as in the Jones model,

$$\begin{aligned}\mathbf{M}_R(\alpha,\delta,\Delta) &= \mathbf{M}_R(0,0,-\delta)\,\mathbf{M}_R(\alpha,0,\Delta)\,\mathbf{M}_R(0,0,\delta) = \\ &= \mathbf{M}_R(0,0,-\delta)\,\mathbf{M}_G(-\alpha)\,\mathbf{M}_R(0,0,\Delta)\,\mathbf{M}_G(\alpha)\,\mathbf{M}_R(0,0,\delta).\end{aligned} \quad (118)$$

These orthogonal matrices have the general form

$$\mathbf{M}_R = \begin{pmatrix} 1 & \mathbf{0}^T \\ \mathbf{0} & \mathbf{m}_R \end{pmatrix}, \quad (119)$$

where $\mathbf{m}_R$ represents a proper rotation in the abstract space defined by the three Stokes parameters $s_1, s_2, s_2$, i.e. it corresponds to a rotation in the Poincaré sphere. It should be noted that, in accordance with the definition of the Stokes parameters, these axes do not correspond to those of the spatial reference system *XYZ*.

Orthogonal Mueller matrices (which always correspond to retarders) can be diagonalized through the following similarity transformation

$$\mathbf{D}(1,1,e^{i\Delta},e^{-i\Delta}) = \mathbf{M}_R(0,0,\delta)\,\mathbf{M}_G(\alpha)\,\mathbf{C}^\dagger\,\mathbf{M}_R(\alpha,\delta,\Delta)\,\mathbf{C}\,\mathbf{M}_G(-\alpha)\,\mathbf{M}_R(0,0,-\delta), \quad (120)$$

where the non-Mueller unitary matrix $\mathbf{C}$ is [147]

$$\mathbf{C} = \frac{1}{\sqrt{2}}\begin{pmatrix} 1 & 1 & 0 & 0 \\ 1 & -1 & 0 & 0 \\ 0 & 0 & 1 & 1 \\ 0 & 0 & i & -i \end{pmatrix}. \quad (121)$$

Although this *modal matrix* is not the only one that satisfies Eq. (120), it has been chosen because it is the same as that we will use to diagonalize the Mueller matrices of diattenuators.

Thus, any retarder has two physical eigenstates, with eigenvalues equal to one, which are the following [148]

$$\mathbf{t}^{(0)} = \frac{1}{\sqrt{2}}(1,c_{2\alpha},s_{2\alpha}c_\delta,s_{2\alpha}s_\delta)^T, \quad \mathbf{t}^{(1)} = \frac{1}{\sqrt{2}}(1,-c_{2\alpha},-s_{2\alpha}c_\delta,-s_{2\alpha}s_\delta)^T, \quad (122)$$

and the Mueller matrix also has the following two non-physical eigenvectors (with respective eigenvalues $e^{i\Delta}, e^{-i\Delta}$)

$$\frac{1}{\sqrt{2}}(1,-s_{2\alpha},c_{2\alpha}c_\delta+is_\delta,c_{2\alpha}s_\delta-ic_\delta)^T, \quad \frac{1}{\sqrt{2}}(1,-s_{2\alpha},c_{2\alpha}c_\delta-is_\delta,c_{2\alpha}s_\delta+ic_\delta)^T. \quad (123)$$

The action of a retarder can be represented in the Poincaré sphere as a rotation of an angle $\Delta$ about the axis defined by the two antipodal points corresponding to the two mutually-orthogonal physical eigenstates [127].

The submatrix $\mathbf{m}_R$ characteristic of a retarder can be expressed in terms of the normalized Stokes eigenvector for the fast axis $\hat{\mathbf{t}}^{(0)} \equiv \mathbf{t}^{(0)}/t_0^{(0)}$ and the retardance $\Delta$ as follows [149]

$$m_{ij} = \delta_{ij} \cos\Delta + \hat{t}_i^{(0)} \hat{t}_j^{(0)} (1 - \cos\Delta) + \sum_{k=1}^{3} \left( \epsilon_{ijk} \, \hat{t}_k^{(0)} \right) \sin\Delta; \quad i,j = 1,2,3. \tag{124}$$

where $\delta_{ij}$ is the Kronecker delta and $\epsilon_{ijk}$ is the Lecy-Cività permutation symbol.

Conversely, the retardance and the elements of the eigenvectors of $\mathbf{M}_R$ can be obtained as [16,149]

$$\cos\Delta = \frac{\operatorname{tr}\mathbf{M}_R}{2} - 1, \quad \hat{t}_i^{(0)} = \frac{1}{2\sin\Delta} \sum_{j,k=1}^{3} \epsilon_{ijk} \, m_{jk}. \tag{125}$$

It should also be noted that the transmittance of a retarder is independent of the incoming Stokes vector and $g(\mathbf{M}_R) = 1$.

Any serial combination of retarders is equivalent to a certain elliptical retarder represented by a unitary Jones matrix (orthogonal Mueller matrix). Therefore, any kind of retarder has two orthogonal eigenstates.

Hereafter we will distinguish between serial combinations, where the light passes through successive elements arranged along the direction of propagation, and parallel combinations, where the incoming light beam falls simultaneously on different parts of the material target and the light pencils emerging from the different components are recombined into a whole emerging beam. The Mueller (or Jones) matrix of a serial combination is given by the ordered product of the Mueller (or Jones) matrices corresponding to the different components. The Jones matrix of a coherent parallel combination is given by a linear combination of the Jones matrices corresponding to the different components [13] where the absolute values of the squares of the coefficients sum to one. The Mueller matrix of an incoherent parallel combination is given by a convex linear combination of the Mueller matrices corresponding to the different components [13].

An original method for the graphic representation of retarders has been introduced on the basis of a four-dimensional spherical parameterization of the Jones matrix. This representation takes the form of a solid cylinder in such a manner that the projection of the point representing the retarder on the cylinder base gives the corresponding Jones eigenvectors [150].

Eq. (119) shows the possibility of a particular treatment of Mueller-Jones matrices representing devices composed of retarders by means of 3x3 matrices applicable directly to the *vectorial part* of the Stokes vectors $(s_1, s_2, s_3)^T$ [137]. Moreover, the use of a complex number to represent the polarization ellipse of light passing through a thick retarder allows us to obtain the evolution of the state of polarization along the direction of propagation. This evolution is given by a second-degree ordinary differential equation [151]. As we will see in next section, this model cannot be applied to pure systems exhibiting diattenuation.

### 4.3. Diattenuators

Diattenuators are material systems exhibiting selective transmittances for two incoming states of polarization. A diattenuator is called *homogeneous* (or *normal*) when it has two orthogonal eigenstates and is represented by a positive semidefinite Hermitian Jones matrix

$$\mathbf{T}_D(\beta, \gamma, k_1, k_2) \equiv \begin{pmatrix} k_1 c_\beta^2 + k_2 s_\beta^2 & s_\beta c_\beta (k_1 - k_2) e^{-i\gamma} \\ s_\beta c_\beta (k_1 - k_2) e^{i\gamma} & k_1 s_\beta^2 + k_2 c_\beta^2 \end{pmatrix},$$

$$s_\beta \equiv \sin\beta, \; c_\beta \equiv \cos\beta, \tag{126}$$

where $k_1, k_2 \ (0 \leq k_2 \leq k_1 \leq 1)$ are the coefficients of amplitude transmittance for the two orthogonal eigenstates (Fig. 8). Their respective azimuths, (namely $\psi, \psi + \pi/2$) and ellipticities ($\nu, -\nu$) are given by

$$\tan 2\psi = \tan 2\beta \cos\gamma, \ \sin 2\nu = \sin 2\beta \sin\gamma, \ (0 \leq \beta \leq \pi/2, -\pi < \gamma \leq \pi). \tag{127}$$

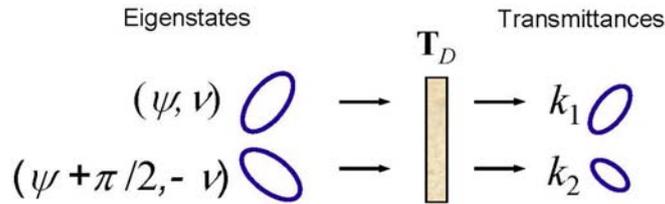

Fig. 8. Diattenuator.

Some interesting specific cases are the following: total diattenuator (or *polarizer*) $\mathbf{T}_D(\beta, \gamma, k_1, 0)$; linear diattenuator $\mathbf{T}_D(\beta, 0, k_1, k_2)$, and circular diattenuator $\mathbf{T}_D(\pi/4, \pi/2, k_1, k_2)$.

The inverse matrix of a nonsingular Hermitian Jones matrix is

$$\mathbf{T}_D^{-1}(\beta, \gamma, k_1, k_2) = \mathbf{T}_D(\beta, \gamma, 1/k_1, 1/k_2), \ (0 < k_2 \leq k_1 \leq 1). \tag{128}$$

We observe that this is a non-Jones matrix because it violates the passivity condition. For singular Hermitian Jones matrices $(k_2 = 0)$, the corresponding pseudoinverse is

$$\mathbf{T}_D^-(\beta, \gamma, k_1, k_2) = \mathbf{T}_D(\beta, \gamma, 1/k_1, 0), \ (0 < k_1 \leq 1), \tag{129}$$

so that

$$\left(\mathbf{T}_D^- \mathbf{T}_D\right) \mathbf{T}_D = \mathbf{T}_D. \tag{130}$$

When $k_2 = k_1$ the diattenuator degenerates into a partially-transparent isotropic medium

$$\mathbf{T}_D(\beta, \gamma, k_1, k_1) = k_1 \mathbf{D}(1,1). \tag{131}$$

The matrix of a generic normal elliptical diattenuator can be written as

$$\begin{aligned}\mathbf{T}_D(\beta, \gamma, k_1, k_2) &= \mathbf{T}_R(0,0,-\gamma) \mathbf{T}_D(\beta, 0, k_1, k_2) \mathbf{T}_R(0,0,\gamma) \\ &= \mathbf{T}_R(0,0,-\gamma) \mathbf{T}_G(-\beta) \mathbf{T}_D(0,0,k_1,k_2) \mathbf{T}_G(\beta) \mathbf{T}_R(0,0,\gamma),\end{aligned} \tag{132}$$

i.e. any elliptical diattenuator is optically equivalent to a linear diattenuator placed between two linear retarders whose fast axes coincide with the reference axes *X* and *Y* respectively.

It should be noted that the limits established for $k_1, k_2$, which are just the singular values of $\mathbf{T}_D$, ensure the fulfillment of the passivity condition

$$g(\mathbf{T}_D) = k_1^2 \leq 1. \tag{133}$$

As with the case of retarders, the Mueller-Jones matrices allow for the equivalences shown for the Jones model to be replicated. Nevertheless, it is important to consider the Mueller-Jones matrices because they can be used in the general Stokes-Mueller model.

We have seen that any matrix of a normal diattenuator can be expressed as a product of simple matrices corresponding to linear diattenuators, linear retarders (aligned with the reference axes) and rotation matrices. The Mueller matrix of a linear diattenuator with $\beta = 0$ is

$$\mathbf{M}_D(0, 0, k_1, k_2) = \frac{1}{2\,k_1^2} \begin{pmatrix} k_1^2 + k_2^2 & k_1^2 - k_2^2 & 0 & 0 \\ k_1^2 - k_2^2 & k_1^2 + k_2^2 & 0 & 0 \\ 0 & 0 & 2k_1 k_2 & 0 \\ 0 & 0 & 0 & 2k_1 k_2 \end{pmatrix}, \tag{134}$$

so that, as in the Jones model,

$$\begin{aligned} \mathbf{M}_D(\beta, \gamma, k_1, k_2) &= \mathbf{M}_R(0, 0, -\gamma)\, \mathbf{M}_D(\beta, 0, k_1, k_2)\, \mathbf{M}_R(0, 0, \gamma) = \\ &= \mathbf{M}_R(0, 0, -\gamma)\, \mathbf{M}_G(-\beta)\, \mathbf{M}_D(0, 0, k_1, k_2)\, \mathbf{M}_G(\beta)\, \mathbf{M}_R(0, 0, \gamma). \end{aligned} \tag{135}$$

Mueller-Jones matrices corresponding to normal diattenuators are symmetric and can be diagonalized through the following similarity transformation

$$\mathbf{D}(1, 1, k_1^2, k_2^2) = \mathbf{M}_R(0, 0, \gamma)\, \mathbf{M}_G(\beta)\, \mathbf{C}^\dagger\, \mathbf{M}_D(0, 0, k_1, k_2)\, \mathbf{C}\, \mathbf{M}_G(-\beta)\, \mathbf{M}_R(0, 0, -\gamma), \tag{136}$$

where $\mathbf{C}$ is the modal matrix in Eq. (121).

Therefore, the eigenvalues of a diattenuator are $\left(k_1^2, k_2^2, k_1 k_2, k_1 k_2\right)$, which correspond to the same eigenvectors as those obtained for the Mueller-Jones matrices of retarders (but replacing $\alpha$, $\delta$ by $\beta$, $\gamma$) [148,147]. $k_1^2, k_2^2$ are the intensity transmittances for to the two physical eigenstates, whereas the duplicated eigenvalue $k_1 k_2$ corresponds to the other two unphysical eigenvectors.

The Mueller matrix of a *normal* (or *homogeneous*) diattenuator can be written as [149]

$$\mathbf{M}_D = m_{00} \begin{pmatrix} 1 & \mathbf{D}^T \\ \mathbf{D} & \mathbf{m}_D \end{pmatrix}, \tag{137}$$

where

$$\mathbf{D} \equiv \frac{(m_{10}, m_{20}, m_{30})^T}{m_{00}} = \frac{(m_{01}, m_{02}, m_{03})^T}{m_{00}},$$

$$\mathbf{m}_D = (1 - D^2)^{1/2}\, \mathbf{I} + \left[1 - (1 - D^2)^{1/2}\right] \frac{\mathbf{D} \otimes \mathbf{D}^T}{D^2}; \quad D \equiv |\mathbf{D}|. \tag{138}$$

where $\mathbf{I}$ is the identity matrix.

These expressions show that the Mueller matrix $\mathbf{M}_D$ is fully determined by its transmittance for unpolarized light $m_{00}$ and its polarizance/diattenuation vector $\mathbf{D}$ [149], that is to say, $\mathbf{M}_D$ is completely characterized by the Stokes vector $m_{00}(1, \mathbf{D}^T)^T$ [152].

The parameters $(\beta, \gamma, k_1, k_2)$ of the normal diattenuator can be calculated from $\mathbf{M}_D$ as [16]

$$\tan 2\beta = \frac{(m_{20}^2 + m_{30}^2)^{1/2}}{m_{10}} = \frac{(m_{02}^2 + m_{03}^2)^{1/2}}{m_{01}}, \quad \tan \gamma = \frac{m_{30}}{m_{20}} = \frac{m_{03}}{m_{02}}, \tag{139}$$

$$k_1^2 = m_{00}(1 + D); \quad k_2^2 = m_{00}(1 - D).$$

Any serial combination of diattenuators is equivalent to a diattenuator whose Jones matrix (Mueller matrix) is a product of Hermitian (symmetric) matrices. Unlike the case of unitary (orthogonal) matrices, the product of Hermitian (symmetric) matrices is not a Hermitian (symmetric) matrix. This means that the most general case of diattenuator has non-orthogonal eigenstates (nonnormal diattenuator). Moreover, general conditions for pure systems to be dichroic (diattenuators) and birefringent (retarders) have been studied by Savenkov, Sydoruk and Muttiah [153].

The effects of diattenuators on the input field can also be represented in quantum theoretical terms by replacing the classical amplitude transmittances with their respective annihilation operators. This subject has been studied by Agarwal [154], showing that the Wigner function of the field is transformed in a simple manner by polarizing devices (squeezing devices) and finding the relation of the Berry phase with this kind of transformation.

### 4.4. Polar decomposition of pure systems

A general matrix description of pure systems is given by the polar decomposition of its corresponding matrices [135,16]. This decomposition can be carried out in both Jones and Mueller-Jones formalisms.

Any 2x2 complex matrix $\mathbf{T}$ that satisfies the passivity condition given by Eq. (107) is a Jones matrix and can be written as the product of a Hermitian matrix (diattenuator) and a unitary matrix (retarder) in either of the two possible relative positions (Fig. 9):

$$\mathbf{T} = \mathbf{T}_D(\beta,\gamma,k_1,k_2)\mathbf{T}_R(\alpha,\delta,\Delta); \quad \mathbf{T} = \mathbf{T}_R(\alpha',\delta',\Delta')\mathbf{T}_D(\beta',\gamma',k'_1,k'_2). \tag{140}$$

Obviously, these expressions are directly translatable to the corresponding Mueller-Jones matrices

$$\mathbf{M}_J = \mathbf{M}_D(\beta,\gamma,k_1,k_2)\,\mathbf{M}_R(\alpha,\delta,\Delta); \quad \mathbf{M}_J = \mathbf{M}_R(\alpha',\delta',\Delta')\,\mathbf{M}_D(\beta',\gamma',k'_1,k'_2), \tag{141}$$

where $\mathbf{M}_D$ is a symmetric Mueller-Jones matrix and $\mathbf{M}_R$ is an orthogonal Mueller-Jones matrix.

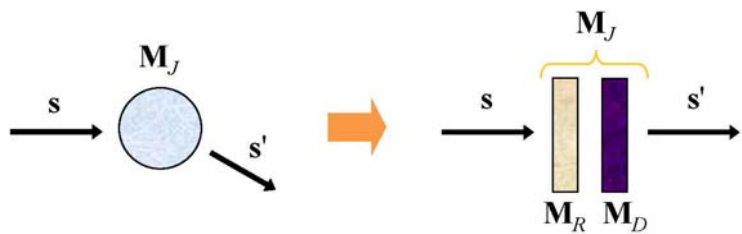

Fig. 9. Retarder-diattenuator equivalent system based on the polar decomposition of pure matrices.

The set of physical parameters $(\alpha,\delta,\Delta,\beta,\gamma,k_1,k_2)$, which are restricted to the limits

$$0 \le \alpha \le \pi/2,\ -\pi < \delta \le \pi,\ 0 \le \Delta \le \pi,\ 0 \le \beta \le \pi/2,\ -\pi < \gamma \le \pi,\ 0 \le k_2 \le k_1 \le 1, \tag{142}$$

provides all the information obtainable through polarimetric measurements. Alternatively, $(\alpha',\delta',\Delta',\beta',\gamma',k'_1,k'_2)$ can be used to characterize the pure system.

When $\mathbf{M}_J$ is singular [$\det \mathbf{M}_J = \det^4 \mathbf{T} = 0$] the polar decomposition is not unique and arbitrary values of $\delta,\delta'$ can be chosen.

Regardless of whether $\mathbf{M}_J$ is singular or not, the parameters corresponding to the equivalent diattenuators are [16]

$$k_1^2 = m_{00}(1+D); \quad k_2^2 = m_{00}(1-D),$$

$$\tan 2\beta = \frac{(m_{20}^2 + m_{30}^2)^{1/2}}{m_{10}}, \quad \tan \gamma = \frac{m_{30}}{m_{20}}, \tag{143}$$

$$\tan 2\beta' = \frac{(m_{02}^2 + m_{03}^2)^{1/2}}{m_{01}}, \quad \tan \gamma' = \frac{m_{03}}{m_{02}},$$

where

$$D \equiv \frac{1}{m_{00}}\sqrt{m_{01}^2 + m_{02}^2 + m_{03}^2} = \frac{1}{m_{00}}\sqrt{m_{10}^2 + m_{20}^2 + m_{30}^2}. \tag{144}$$

This last equality expresses a particular property of pure Mueller matrices.

It should be noted that $k_1^2, k_2^2, k_1 k_2, k_1 k_2$ are the singular values of $\mathbf{M}_J \equiv \mathbf{M}(\mathbf{T})$ and that, obviously, the value of the maximum transmittance does not depend on the mathematical formalism used, i.e.

$$g(\mathbf{M}_J) = g(\mathbf{T}) = k_1^2 = \frac{1}{2}\left\{\mathrm{tr}(\mathbf{T}^\dagger \mathbf{T}) + \sqrt{\mathrm{tr}^2(\mathbf{T}^\dagger \mathbf{T}) - 4\det(\mathbf{T}^\dagger \mathbf{T})}\right\} \leq 1. \tag{145}$$

Moreover, the minimum transmittance $k_2^2$ can be expressed as

$$k_2^2 = \frac{1}{2}\left\{\mathrm{tr}(\mathbf{T}^\dagger \mathbf{T}) - \sqrt{\mathrm{tr}^2(\mathbf{T}^\dagger \mathbf{T}) - 4\det(\mathbf{T}^\dagger \mathbf{T})}\right\} \leq 1. \tag{146}$$

Now, for the obtainment of the parameters of the equivalent retarder, we distinguish two cases [16]:

1) $\mathbf{M}_J$ nonsingular ($0 < k_2$)

As a first step, by applying the inverse matrices of the respective diattenuators, the matrices of the retarders are calculated

$$\begin{aligned}\mathbf{E} &\equiv \mathbf{M}_D(\beta, \gamma, 1/k_1, 1/k_2)\mathbf{M}_J = \mathbf{M}_R(\alpha, \delta, \Delta),\\ \mathbf{E}' &\equiv \mathbf{M}_J \mathbf{M}_D(\beta', \gamma', 1/k_1, 1/k_2) = \mathbf{M}_R(\alpha', \delta', \Delta').\end{aligned} \tag{147}$$

Furthermore, the angular parameters $\varphi, \chi, \Delta$ of the equivalent retarder are obtained through the following relations

$$\cos \Delta = \frac{\mathrm{tr}\,\mathbf{E}}{2} - 1, \quad \cos \Delta' = \frac{\mathrm{tr}\,\mathbf{E}'}{2} - 1,$$

$$\sin 2\chi = \frac{e_{12} - e_{21}}{2\sin(\Delta/2)}, \quad \sin 2\chi' = \frac{e'_{12} - e'_{21}}{2\sin(\Delta'/2)}, \tag{148}$$

$$\sin 2\varphi = \frac{e_{31} - e_{13}}{2\cos 2\chi \sin(\Delta/2)}, \quad \sin 2\varphi' = \frac{e'_{31} - e'_{13}}{2\cos 2\chi' \sin(\Delta'/2)},$$

which lead to

$$\cos 2\alpha = \cos 2\varphi \cos 2\chi, \quad \cos 2\alpha' = \cos 2\varphi' \cos 2\chi',$$

$$\tan \delta = \frac{\tan 2\chi}{\sin 2\varphi}, \quad \tan \delta' = \frac{\tan 2\chi'}{\sin 2\varphi'}, \qquad (149)$$

2) $\mathbf{M}_J$ singular ($0 = k_2$)

In this case, the decomposition is not unique. An arbitrary value can be chosen for $\delta$, (or $\delta'$) for example $\delta = 0$ ($\delta' = 0$), so that the equivalent retarder is a linear retarder. Then $\alpha, \alpha'$ are given by

$$\tan 2\alpha = \tan 2\alpha' = \frac{m_{10} - m_{01}}{m_{02} - m_{20}}, \qquad (150)$$

and for $\Delta(\Delta')$, two different values $\Delta_1, \Delta_2$ ($\Delta'_1, \Delta'_2$) are possible

$$\tan \Delta_1 = \tan \Delta'_1 = \frac{b\, m_{30} - a\, m_{03}}{b^2 - m_{03}^2},$$

$$\tan \Delta_2 = \tan(\pi - \Delta'_2) = \frac{b\, m_{30} + a\, m_{03}}{b^2 - m_{03}^2}, \qquad (151)$$

where

$$a \equiv m_{01} \sin \alpha - m_{02} \cos \alpha, \quad b \equiv m_{10} \sin \alpha - m_{20} \cos \alpha. \qquad (152)$$

In the case that $m_{03} = 0$, these results should be replaced by $\Delta = \Delta' = \pi/2$

Each one of the pure components of the equivalent system given by the polar decomposition is normal (or homogeneous) [17], i.e. has orthogonal eigenstates. In particular, the fact that the equivalent diattenuator has orthogonal eigenstates (and, hence, has not the most general form of a diattenuator) does not represent a lack of validity, but it is a peculiarity of the polar decomposition.

From the general expressions of the Jones matrices (or the corresponding Mueller-Jones matrices) of retarders, diattenuators and pure systems, a number of theorems can be stated. All of them can be deduced from those mentioned in this section [132,127, 130,135-137,155].

The equivalent model (retarder-diattenuator) of a pure system, which has been obtained by the polar decomposition of its corresponding matrix $\mathbf{T}$, results in the singular value decomposition of $\mathbf{T}$. In fact, Eq. (140) can be expressed as

$$\mathbf{T} = \left[ \mathbf{T}_R(0,0,-\gamma)\, \mathbf{T}_G(-\beta) \right] \mathbf{T}_D(0,0,k_1,k_2) \left[ \mathbf{T}_G(\beta)\, \mathbf{T}_R(0,0,\gamma)\, \mathbf{T}_R(\alpha,\delta,\Delta) \right]. \qquad (153)$$

This Jones algebra allows us to easily distinguish between *normal* (or *homogeneous*) Jones matrices (i.e. the eigenvalues coincide with the singular values) and *nonnormal* (*inhomogeneous*) Jones matrices (i.e. the eigenvalues are different from the singular values). Normal Jones matrices correspond to systems in which the eigenvectors of the equivalent retarder coincide with those of the equivalent diattenuator. A particular case of normal Jones matrices are those that represent retarders. This subject has been studied by Lu and Chipman [17], who have defined an *inhomogeneity parameter* $\eta$ as the scalar product of the eigenvectors of the system, so that for normal systems $\eta = 0$, whereas the value $\eta = 1$ represents the maximum possible value of the inhomogeneity. An example of system with $\eta = 1$ is a serial combination composed of a retarder placed between two crossed total polarizers (*black sandwich*) [156].

Obviously, given the direct relation between Jones and Mueller-Jones matrices, the above arguments are also valid for Muller-Jones matrices.

The above-mentioned authors, Lu and Chipman, have analyzed the retardation and diattenuation properties for both normal and nonnormal cases through the polar decomposition of their Jones matrices [17,149]. In terms of the elements of the Mueller-Jones matrix $\mathbf{M}_J$, the polarizance and diattenuation vectors are defined as

$$\mathbf{D} \equiv \frac{1}{m_{00}}(m_{10}, m_{20}, m_{30})^T, \quad \mathbf{P} \equiv \frac{1}{m_{00}}(m_{01}, m_{02}, m_{03})^T, \tag{154}$$

so that $P$ contains all the information on the polarizing power for incident unpolarized radiation, whereas $D$ describes completely the diattenuation power of $\mathbf{M}_J$.

Taking into account the reciprocity conditions for Mueller matrices, we see that

$$\mathbf{D}(\mathbf{M}_J) = \mathbf{P}(\mathbf{M}_J^r), \quad \mathbf{P}(\mathbf{M}_J) = \mathbf{D}(\mathbf{M}_J^r), \tag{155}$$

and, from Eq. (144), we observe that for pure systems the magnitudes of polarizance and diattenuation vectors are equal.

The properties of nonsingular Jones and Mueller-Jones matrices can be studied by means of their representation in the SL(2C) group or in the proper orthochronous Lorentz group respectively [157-160]. In the case of diattenuators, this requires a normalization that violates the passivity condition. As we will see, a powerful group-theoretic framework for the representations of the polarimetric properties of material media arises from the concept of the coherency matrix associated with a Mueller matrix [161-163,41].

## 5. Interaction of polarized light with non-deterministic passive optical systems

Many scientific and industrial applications of polarimetry involve depolarization phenomena. The study and characterization of the general polarimetric behavior of material samples is of considerable importance in order to take maximum advantage of the potential possibilities of these non-destructive techniques.

In general, for an incident totally polarized light beam, a material system produces a superposition of totally polarized outgoing light pencils with different polarizations. Depending on the spectral distribution of the incident light, on the nature of the material target and on the nature of the interaction, these pencils can present different relative degrees of mutual coherence. Thus, the superposition of some pencils can be coherent (so that their respective Jones vectors should be added in order to obtain the corresponding resultant Jones vector), the superposition of some pencils can be incoherent (so that their respective Stokes vectors should be added in order to obtain the corresponding resultant Stokes vector) and the superposition of other outgoing pencils can be partially coherent (so that their coherent parts are "Jones-added" and their incoherent parts are "Stokes-added"). Taking into account the principle of optical equivalence of polarization states [33], the above observation implies that the polarimetric effect of a material medium is equivalent to that of a system composed of a parallel combination of a number of pure optical systems. Therefore, the state of polarization of the outgoing light beam is physically equivalent to the state of polarization of the superposition of the light beams emerging from each one of the pure components of the equivalent parallel combination.

In this paper, we deal only with physical (or physically realizable) Mueller matrices, i.e. 4x4 real matrices that represent the linear polarimetric behavior of material systems. In accordance with the above considerations, we will show that the Mueller matrix of the whole material system is given by a convex sum of the Mueller-Jones matrices of the incoherent pure elements included in the parallel combination. This definition of general Mueller matrices is equivalent to saying that the Mueller matrix of the system can be considered an ensemble average of pure Mueller matrices [47,161,164-168].

Although in the literature the properties of matrices transforming Stokes vectors into Stokes vectors (i.e. satisfying the so-called *Stokes criterion*) have been studied [150, 160,167,169-175], this kind of matrix, hereafter named *Stokes matrices*, does not coincide with the set of *Mueller matrices*, which is derived from the *ensemble criterion*. Some properties of Stokes matrices have been obtained from the Stokes criterion, which means that the degree of polarization of outgoing light must be smaller than, or equal to, unity (i.e. the Stokes matrix does not *overpolarize*) [169,170,173]. These and other relevant results have been derived from the properties of the eigenvectors of $\mathbf{GM}^T\mathbf{GM}$, or $\mathbf{M}^T\mathbf{GM}$ and from their spectral decomposition, where $\mathbf{G}$ is the diagonal matrix $\mathbf{G} \equiv \mathbf{D}(1,-1,-1,-1)$ [160,172,174].

Obviously, any physical Mueller matrix is a Stokes matrix but, in general, the converse is not true. No method has been quoted to physically realize a Stokes matrix being non derivable from the ensemble criterion. It is obvious that only physical Mueller matrices (from now on, Mueller matrices) are the subject of physical interest and, as some authors have pointed out [165,167,168], the ensemble-based model represents a suitable theoretical framework for the study of polarimetric phenomena. In the following sections we formulate mathematically the conditions for a $4 \times 4$ real matrix to be a Mueller matrix.

### 5.1. Construction of a Mueller matrix

In general, an optical system can exhibit spatial inhomogeneity over the area illuminated by the incident light beam, as well as dispersive effects, producing depolarization. The emerging light is consequently composed of a number of incoherent contributions, and the optical system cannot be represented by means of a Jones matrix. However, taking into account that, essentially, any linear effect is due to certain sort of scattering, the system can be considered as being composed of a number of deterministic-nondepolarizing elements, each one with a well-defined Jones matrix, in such a manner that the light beam is shared among these different elements. In other words, the system is a parallel combination of polarimetrically-pure components (Fig. 10).

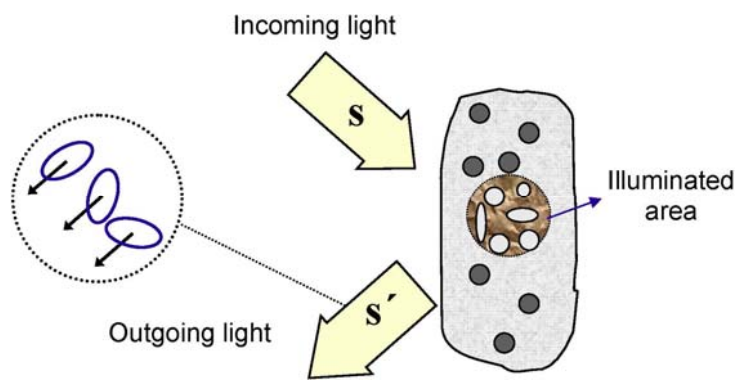

Fig. 10. Physical transformation of Stokes vectors.

Let $I_i$ be the intensity of the portion of light that interacts with the "$i$" element. The ratio between $I_i$ and the intensity $I$ of the whole beam is denoted by a respective coefficient $p_i \equiv I_i/I$ so that

$$p_i = \frac{I_i}{I}, \quad \sum_i p_i = 1. \tag{156}$$

Now we denote by $\mathbf{T}_i$ and $\mathbf{M}_{Ji}$, the respective Jones and Mueller-Jones matrices representing the "*i*" element. Thus, the Jones vector of the light pencil emerging from each element is given by

$$\boldsymbol{\varepsilon}'_i(t) = \mathbf{T}_i \sqrt{p_i} \boldsymbol{\varepsilon}(t), \tag{157}$$

and the corresponding Stokes vector is given by

$$\begin{aligned}
\mathbf{s}'_i &= \mathbf{L}\left\langle \boldsymbol{\varepsilon}'_i \otimes \boldsymbol{\varepsilon}'^*_i \right\rangle = \mathbf{L}\left\langle \left(\mathbf{T}_i \sqrt{p_i}\boldsymbol{\varepsilon}\right) \otimes \left(\mathbf{T}_i \sqrt{p_i}\boldsymbol{\varepsilon}\right)^* \right\rangle \\
&= \mathbf{L}\left\langle p_i \left(\mathbf{T}_i \otimes \mathbf{T}_i^*\right)\left(\boldsymbol{\varepsilon} \otimes \boldsymbol{\varepsilon}^*\right) \right\rangle = p_i \mathbf{L}\left(\mathbf{T}_i \otimes \mathbf{T}_i^*\right)\left\langle \boldsymbol{\varepsilon} \otimes \boldsymbol{\varepsilon}^* \right\rangle \\
&= p_i \left[ \mathbf{L}\left(\mathbf{T}_i \otimes \mathbf{T}_i^*\right)\mathbf{L}^{-1} \right] \mathbf{s}.
\end{aligned} \tag{158}$$

The polarization state of the complete emerging beam is obtained through the incoherent superposition of the beams emerging from the different elements, resulting in the following Stokes vector

$$\mathbf{s}' = \sum_i \mathbf{s}'_i = \left(\sum_i p_i \mathbf{M}_{Ji}\right) \mathbf{s} = \mathbf{M}\mathbf{s}, \tag{159.a}$$

where

$$\mathbf{M} \equiv \left(\sum_i p_i \mathbf{M}_{Ji}\right), \quad \mathbf{M}_{Ji} \equiv \mathbf{L}\left(\mathbf{T}_i \otimes \mathbf{T}_i^*\right)\mathbf{L}^{-1}, \quad p_i \geq 0, \quad \sum_i p_i = 1. \tag{159.a}$$

We have obtained this result by considering the optical system as composed of a set of parallel elements, but it is worth pointing out that the same result is obtained by considering the system as an ensemble, so that each realization "*i*", characterized by a well-defined Jones matrix $\mathbf{T}_i$, occurs with a probability $p_i$. Therefore, we can consider the optical system as composed of such an ensemble [165]. We will refer to ensemble averages by means of

$$\langle x \rangle_e \equiv \sum_i p_i x^{(i)}; \quad p_i \geq 0, \quad \sum_i p_i = 1, \tag{160}$$

so that

$$\mathbf{M} = \langle \mathbf{M}_J \rangle_e = \mathbf{L}\langle \mathbf{T} \otimes \mathbf{T}^* \rangle_e \mathbf{L}^{-1}, \tag{161}$$

or, in components [165],

$$m_{kl} = \langle m_{kl} \rangle_e = \frac{1}{2}\left\langle \mathrm{tr}\left(\boldsymbol{\sigma}_k \mathbf{T} \boldsymbol{\sigma}_l \mathbf{T}^\dagger\right) \right\rangle_e, \quad k,l = 0,1,2,3. \tag{162}$$

### 5.2. Coherency matrix associated with a Mueller matrix and covariance conditions

A high degree of knowledge of the relationships between the elements of a Mueller matrix is important for analyzing the information contained in such matrices obtained from experimental measurements. As we have indicated above, contributions of various authors who have dealt with this subject provide specific sets of constraining conditions for the elements of Mueller matrices [130,164,167,170,172,174,176,177]. As we will see, the nonnegativity of the four eigenvalues of the Hermitian matrix $\mathbf{H}$ associated with a Mueller matrix $\mathbf{M}$ plays a fundamental role in the general characterization of Mueller matrices. This criterion was introduced by Cloude [161] and derives directly from the ensemble model. Another criterion that should be added in order to complete the mathematical characterization of Mueller matrices, is that the Mueller matrix of any passive optical system should satisfy the *passivity condition*, i.e. $g_f \leq 1$, where $g_f$ is the maximum transmittance (gain) with regard to all possible incident polarization states [124].

As a step previous to the definition of the coherency matrix associated with a Mueller matrix, we introduce the following notation for the elements of $\mathbf{T}$ $t_0 \equiv t_{11}$, $t_1 \equiv t_{12}$, $t_2 \equiv t_{21}$, $t_3 \equiv t_{22}$, which allows the matrix $\langle \mathbf{T} \otimes \mathbf{T}^* \rangle_e$ to be written in the following manner

$$\langle \mathbf{T} \otimes \mathbf{T}^* \rangle_e = \begin{pmatrix} \langle t_0 t_0^* \rangle_e & \langle t_0 t_1^* \rangle_e & \langle t_1 t_0^* \rangle_e & \langle t_1 t_1^* \rangle_e \\ \langle t_0 t_2^* \rangle_e & \langle t_0 t_3^* \rangle_e & \langle t_1 t_2^* \rangle_e & \langle t_1 t_3^* \rangle_e \\ \langle t_2 t_0^* \rangle_e & \langle t_2 t_1^* \rangle_e & \langle t_3 t_0^* \rangle_e & \langle t_3 t_1^* \rangle_e \\ \langle t_2 t_2^* \rangle_e & \langle t_2 t_3^* \rangle_e & \langle t_3 t_2^* \rangle_e & \langle t_3 t_3^* \rangle_e \end{pmatrix}, \qquad (163)$$

where the elements can be suitably reordered to construct the matrix $\mathbf{H}$ [121,168,178] whose elements are defined as

$$h_{kl} \equiv \frac{1}{2} \langle t_k t_l^* \rangle_e, \quad k,l = 0,1,2,3. \qquad (164)$$

This matrix was first introduced by Simon [121], who distinguished between two cases, (1) $\mathbf{H}$ positive semidefinite and (2) $\mathbf{H}$ negative definite [178]. As we will see in next paragraph this last case never corresponds to physical Mueller matrices.

This definition of the Hermitian matrix $\mathbf{H}$ shows that its elements are, in fact, the second-order moments of the variables $t_k/\sqrt{2}$. A necessary and sufficient condition for a Hermitian matrix $\mathbf{H}$ to be a matrix of second-order moments is that $\mathbf{H}$ be positive semidefinite [88]. Moreover, a Hermitian matrix $\mathbf{H}$ is positive semidefinite if, and only if, its eigenvalues $\lambda_i$ are nonnegative [103].

Through symbolic calculus it is possible to obtain the very complicated algebraic expressions for the four eigenvalues of $\mathbf{H}$ and, hence, to express the conditions of their nonnegativity.

Another equivalent set of conditions, whose algebraic expressions are simpler, is derived from the nonnegativity of four nested principal minors of $\mathbf{H}$ [168]. This explicit formulation of the conditions could be useful in order to show some properties of the structure of Mueller matrices, to analyze polarimetric measurements and to study physical models.

To obtain convenient expressions of these constraining conditions, it is useful to write the $h_{kl}$ elements in terms of statistical parameters

$$h_{kl} \equiv \mu_{kl} \sigma_k \sigma_l, \qquad (165.a)$$

where

$$\sigma_k^2 = h_{kk} = \left\langle \left(t_k/\sqrt{2}\right)\left(t_k/\sqrt{2}\right)^* \right\rangle_e = \frac{1}{2} \langle |t_k|^2 \rangle_e, \quad \mu_{kl} = \frac{h_{kl}}{\sigma_k \sigma_l} \quad \left(\mu_{kl} \equiv \rho_{kl} e^{i\beta_{kl}}\right). \qquad (165.b)$$

Thus, $\mathbf{H}$ can be written as follows

$$\mathbf{H} = \begin{pmatrix} \sigma_0^2 & \mu_{01}\sigma_0\sigma_1 & \mu_{02}\sigma_0\sigma_2 & \mu_{03}\sigma_0\sigma_3 \\ \mu_{01}^*\sigma_0\sigma_1 & \sigma_1^2 & \mu_{12}\sigma_1\sigma_2 & \mu_{13}\sigma_1\sigma_3 \\ \mu_{02}^*\sigma_0\sigma_2 & \mu_{12}^*\sigma_1\sigma_2 & \sigma_2^2 & \mu_{23}\sigma_2\sigma_3 \\ \mu_{03}^*\sigma_0\sigma_3 & \mu_{13}^*\sigma_1\sigma_3 & \mu_{23}^*\sigma_2\sigma_3 & \sigma_3^2 \end{pmatrix}. \tag{166}$$

From the nonnegativity of four nested principal minors, we can write the set of necessary and sufficient conditions (hereafter called the *covariance conditions*) for a Hermitian matrix $\mathbf{H}$ to be a matrix of second-order moments [168]

$$h_{00} \geq 0;$$

$$1 \geq \rho_{01};$$

$$1 + 2\rho_{01}\rho_{12}\rho_{02}\cos(\beta_{01} + \beta_{12} - \beta_{02}) \geq \rho_{01}^2 + \rho_{12}^2 + \rho_{02}^2;$$

$$\begin{aligned} \det \mathbf{H} &= 1 + 2\rho_{12}\rho_{23}\rho_{13}\cos(\beta_{12} + \beta_{23} - \beta_{13}) + 2\rho_{02}\rho_{23}\rho_{03}\cos(\beta_{02} + \beta_{23} - \beta_{03}) \\ &\quad + 2\rho_{01}\rho_{13}\rho_{03}\cos(\beta_{01} + \beta_{13} - \beta_{03}) + 2\rho_{01}\rho_{12}\rho_{02}\cos\beta(\beta_{01} + \beta_{12} - \beta_{02}) \\ &\quad - 2\rho_{01}\rho_{02}\rho_{13}\rho_{23}\cos(\beta_{01} - \beta_{02} + \beta_{13} - \beta_{23}) \\ &\quad - 2\rho_{01}\rho_{03}\rho_{12}\rho_{23}\cos(\beta_{01} - \beta_{03} + \beta_{12} + \beta_{23}) \\ &\quad - 2\rho_{02}\rho_{03}\rho_{12}\rho_{13}\cos(\beta_{02} - \beta_{03} - \beta_{12} + \beta_{13}) \\ &\quad + \rho_{01}^2\rho_{23}^2 + \rho_{02}^2\rho_{13}^2 + \rho_{03}^2\rho_{12}^2 - \rho_{01}^2 - \rho_{02}^2 - \rho_{03}^2 - \rho_{12}^2 - \rho_{13}^2 - \rho_{23}^2 \geq 0. \end{aligned} \tag{167}$$

*For updated details, see*

J. J Gil, I San José, "Explicit algebraic characterization of Mueller matrices," Opt. Lett. Optics Letters 39, 4041-4044 (2014).

It should be noted that, regardless of the fact that the above four inequalities provide a set of sufficient conditions; all the fifteen principal minors of $\mathbf{H}$ are nonnegative.

The pure case occurs when the following nine independent conditions are satisfied

$$\rho_{ij} = 1 \ (i, j = 0,1,2,3),$$
$$\beta_{01} + \beta_{12} = \beta_{02}, \ \beta_{12} + \beta_{23} = \beta_{13}, \ \beta_{02} + \beta_{23} = \beta_{03}. \tag{168}$$

This *deterministic* case corresponds to a Mueller-Jones matrix $\mathbf{M}(\mathbf{H})$. [168]

As a basis for the expansion of $\mathbf{H}$, let us now consider the following set of Hermitian trace-orthogonal matrices [179,161]

$$\mathbf{E}_{ij} = \boldsymbol{\sigma}_i \otimes \boldsymbol{\sigma}_j \ (i, j = 0,1,2,3), \tag{169}$$

which are defined as direct products of the $\boldsymbol{\sigma}_i$ matrices used for the expansion of the 2D coherency matrix (i.e. the Pauli matrices plus the identity matrix).

As occurs with the basis used for the expansions of the 2x2 and 3x3 coherency matrices, these matrices $\mathbf{E}_{ij}$ are Hermitian $\mathbf{E}_{ij} = \mathbf{E}_{ij}^\dagger$; trace-orthogonal $\mathrm{tr}(\mathbf{E}_{ij}\mathbf{E}_{kl}) = 4\delta_{ik}\delta_{jl}$, and satisfy $\mathbf{E}_{ij}^2 = \mathbf{D}(1,1,1,1)$. Thus, these matrices are unitary and, except for $\mathbf{E}_{00} = \mathbf{D}(1,1,1,1)$, are traceless. This basis allows $\mathbf{H}$ to be expressed as a linear combination of $\mathbf{E}_{ij}$

$$\mathbf{H} = \frac{1}{4} \sum_{i,j=0}^{3} m_{ij} \mathbf{E}_{ij} \; , \; m_{ij} = \mathrm{tr}\left(\mathbf{E}_{ij}\mathbf{H}\right), \tag{170}$$

where the sixteen real coefficients $m_{ij}$ are just the elements of the Mueller matrix associated with the *coherency matrix* $\mathbf{H}$ (commonly called the *covariance matrix*). It should be noted that the term *coherency matrix* is used by Cloude [161], van der Mee [169] and other authors to refer to other Hermitian matrices defined from $\mathbf{M}$ in a different way.

The formulation given by Eq. (170), together with the nonnegativity of $\mathbf{H}$ has been considered by Aiello, Puentes and Woerdman [180], who have analyzed the connection between classical polarization optics and quantum mechanics of two level systems.

The explicit relations between the coherency matrix $\mathbf{H}$ and its corresponding Mueller matrix $\mathbf{M}$ are [121,178,168]

$$\mathbf{H} = \frac{1}{4}\begin{pmatrix} \begin{matrix} m_{00}+m_{01} \\ +m_{10}+m_{11} \end{matrix} & \begin{matrix} m_{02}+m_{12} \\ +i(m_{03}+m_{13}) \end{matrix} & \begin{matrix} m_{20}+m_{21} \\ -i(m_{30}+m_{31}) \end{matrix} & \begin{matrix} m_{22}+m_{33} \\ +i(m_{23}-m_{32}) \end{matrix} \\ \begin{matrix} m_{02}+m_{12} \\ -i(m_{03}+m_{13}) \end{matrix} & \begin{matrix} m_{00}-m_{01} \\ +m_{10}-m_{11} \end{matrix} & \begin{matrix} m_{22}-m_{33} \\ -i(m_{23}+m_{32}) \end{matrix} & \begin{matrix} m_{20}-m_{21} \\ -i(m_{30}-m_{31}) \end{matrix} \\ \begin{matrix} m_{20}+m_{21} \\ +i(m_{30}+m_{31}) \end{matrix} & \begin{matrix} m_{22}-m_{33} \\ +i(m_{23}+m_{32}) \end{matrix} & \begin{matrix} m_{00}+m_{01} \\ -m_{10}-m_{11} \end{matrix} & \begin{matrix} m_{02}-m_{12} \\ +i(m_{03}-m_{13}) \end{matrix} \\ \begin{matrix} m_{22}+m_{33} \\ -i(m_{23}-m_{32}) \end{matrix} & \begin{matrix} m_{20}-m_{21} \\ +i(m_{30}-m_{31}) \end{matrix} & \begin{matrix} m_{02}-m_{12} \\ -i(m_{03}-m_{31}) \end{matrix} & \begin{matrix} m_{00}-m_{01} \\ -m_{10}+m_{11} \end{matrix} \end{pmatrix}, \tag{171}$$

$$\mathbf{M} = \begin{pmatrix} \begin{matrix} h_{00}+h_{11} \\ +h_{22}+h_{33} \end{matrix} & \begin{matrix} h_{00}-h_{11} \\ +h_{22}-h_{33} \end{matrix} & \begin{matrix} h_{01}+h_{10} \\ +h_{23}+h_{32} \end{matrix} & \begin{matrix} -i(h_{01}-h_{10}) \\ -i(h_{23}-h_{32}) \end{matrix} \\ \begin{matrix} h_{00}+h_{11} \\ -h_{22}-h_{33} \end{matrix} & \begin{matrix} h_{00}-h_{11} \\ -h_{22}+h_{33} \end{matrix} & \begin{matrix} h_{02}+h_{20} \\ -h_{13}-h_{32} \end{matrix} & \begin{matrix} -i(h_{01}-h_{10}) \\ +i(h_{23}-h_{32}) \end{matrix} \\ \begin{matrix} h_{02}+h_{20} \\ +h_{13}+h_{31} \end{matrix} & \begin{matrix} h_{02}+h_{20} \\ -h_{13}-h_{31} \end{matrix} & \begin{matrix} h_{03}+h_{30} \\ +h_{12}+h_{21} \end{matrix} & \begin{matrix} -i(h_{03}-h_{30}) \\ +i(h_{12}-h_{21}) \end{matrix} \\ \begin{matrix} i(h_{02}-h_{20}) \\ +i(h_{13}-h_{31}) \end{matrix} & \begin{matrix} i(h_{02}-h_{20}) \\ -i(h_{13}-h_{31}) \end{matrix} & \begin{matrix} i(h_{03}-h_{30}) \\ +i(h_{12}-h_{21}) \end{matrix} & \begin{matrix} h_{03}+h_{30} \\ -h_{12}-h_{21} \end{matrix} \end{pmatrix}. \tag{172}$$

The constraining inequalities between $m_{ij}$ (derived from the nonnegativity of the eigenvalues of $\mathbf{H}$) are of higher degree of complexity than in the 2D and 3D Stokes parameters, and involve the invariants $\mathrm{tr}\,\mathbf{H}$, $\mathrm{tr}\,\mathbf{H}^2$, $\mathrm{tr}\,\mathbf{H}^3$ and $\mathrm{tr}\,\mathbf{H}^4$. Leaving aside the fact that the 4x4 coherency matrix $\mathbf{H}$ represents the polarimetric properties of material media, from $\mathbf{H}$ we can reproduce easily the coherency matrices $\boldsymbol{\Phi}$ and $\mathbf{R}$ representing the polarimetric properties of light in the respective 2D and 3D models. Therefore, by eliminating the corresponding extra components of the variables $t_i$, the 2D model arises as the upper-left 2x2 matrix by setting

$$s_0 = 2(m_{00}+m_{01}), \; s_1 = 2(m_{10}+m_{11}), \; s_2 = 2(m_{20}+m_{21}), \; s_3 = 2(m_{30}+m_{31}). \tag{173}$$

The restriction to the 3D model can be straightforward performed but, as occurs in the restriction 3D-2D, the explicit expressions for the 3D Stokes parameters in terms of $m_{ij}$ are more complicated because of the mathematical asymmetry of the Gell-Mann matrices with respect to the $\mathbf{E}_{ij}$ matrices.

The expansion given by Eq. (170) provides a fundamental significance for the elements of the Mueller matrix, deeper than their simple phenomenological significance as coefficients of the linear transformation of Stokes vectors. We see that the relation between the complex covariance matrix $\mathbf{H}$ and its corresponding Mueller matrix is, in fact, analogous to the relation between the 2×2 coherency matrix and its corresponding Stokes parameters. Thus, leaving aside the physical meaning of the elements $m_{ij}$ of the Mueller matrix, they can be mathematically considered as 4D Stokes parameters. This shows the symmetry of the operators representing polarization quantities for both light and material media.

Sudha Shenoy, Gopala Rao, Usha Devi and Rajagopal [181] emphasized the mathematical analogy between the formulation of the coherency matrix $\mathbf{H}$, derived from the ensemble criterion, and the positive operator valued measures (POVM) of the quantum density matrix and have proposed a way of realizing different types of Mueller devices.

### 5.3. Passivity conditions (transmittance conditions)

*For updated details, see*
I. San José and J. J. Gil, "Characterization of passivity in Mueller matrices," J. Opt. Soc. Am. A **37**, 199-208 (2020).

The covariance conditions have been obtained from the construction of the Mueller matrix as the average given by Eq. (159.a) -or Eq. (161)-, where $\mathbf{T}_i$ are 2×2 complex matrices, but without taking into account the conditions derived from the fact that $\mathbf{T}_i$ are passive Jones matrices. Thus, in order to complete the set of conditions for a 4×4 real matrix to be a Mueller matrix, it is necessary to consider the conditions derived from the fact that all the pure Mueller matrices of the elements involved in the average satisfy the passivity condition.

The direct application to a Mueller matrix of the integral passivity condition (i.e. its transmittance never exceeds the value 1) leads to the well known condition [124,126]

$$m_{00} + \sqrt{m_{01}^2 + m_{02}^2 + m_{03}^2} \leq 1. \tag{174}$$

Nevertheless, through calculations based on two main hypotheses: 1) the system is considered an ensemble (the Mueller matrix is given by an ensemble average of pure Mueller matrices) and 2) each single realization is a passive system (i.e. its pure Mueller matrix satisfies the passivity condition), the following pair of conditions can be demonstrated [168]

$$g_f \leq 1, \; g_r \leq 1; \quad g_f \equiv m_{00} + \sqrt{m_{01}^2 + m_{02}^2 + m_{03}^2}, \; g_r \equiv m_{00} + \sqrt{m_{10}^2 + m_{20}^2 + m_{30}^2}. \tag{175}$$

Thus, the procedure followed to construct the general Mueller matrices as a convex sum (or as an ensemble average) of passive pure Mueller matrices, leads to the fact that any Mueller matrix must satisfy, not only the *forward passivity condition*

$$m_{00} + \sqrt{m_{01}^2 + m_{02}^2 + m_{03}^2} \leq 1, \tag{176}$$

but also the *reverse passivity condition*

$$m_{00} + \sqrt{m_{10}^2 + m_{20}^2 + m_{30}^2} \leq 1. \tag{177}$$

The origin of these names comes from the fact that $g_r$ represents the maximal transmittance for light passing through the system in the reverse direction.

A good example for analyzing these properties of Mueller matrices is the following matrix

$$\begin{pmatrix} 1 & 0 & 0 & 0 \\ 0 & 0 & 0 & 0 \\ 0 & 0 & 0 & 0 \\ 1 & 0 & 0 & 0 \end{pmatrix}. \tag{178}$$

It is easy to check that this matrix satisfies the covariance conditions as well as the forward passivity condition, but it does not satisfy the reverse passivity condition. This is consistent with the fact that this matrix cannot be obtained as an average of pure Mueller matrices and, hence, it is not possible to realize it, which implies that such a matrix is not a Mueller matrix.

### 5.4. Characterization theorem for Mueller matrices

Let us now consider the starting premises for obtaining the characterization of Mueller matrices:

1) A 2×2 complex matrix is a Jones matrix if and only if it satisfies the passivity condition given by Eq. (107).

2) A Mueller-Jones matrix is defined as a 4×4 real matrix that can be derived from a Jones matrix. Therefore, a Mueller-Jones matrix can be expressed by means of Eq. (105) as a function of its corresponding Jones matrix. In consequence, a Mueller-Jones matrix satisfies the passivity condition given by Eq. (110), which can also be expressed by means of Eq. (112).

3) The set of Mueller matrices is defined by all the 4×4 real matrices that can be obtained by means of a convex sum of Mueller-Jones matrices.

4) *Characterization theorem*. The results of the previous sections lead to the following statement: *A real 4×4 matrix $M$ is a Mueller matrix if and only if the four eigenvalues of $\mathbf{H}(\mathbf{M})$ are nonnegative and $M$ satisfies the passivity conditions given by Eq. (176) and Eq. (177)*.

   In terms of simpler algebraically explicit expressions, this statement can be formulated as: *A real 4×4 matrix is a Mueller matrix if and only if it satisfies the six inequalities constituted by 1) the four covariance conditions given by Eq. (167), and 2) the two passivity conditions given by Eq. (176) and Eq. (177)*.

Obviously, in all the previous paragraphs we use the terms *Jones matrices* and *Mueller matrices* to refer to physically realizable Jones and Mueller matrices respectively. Despite the reiteration, we consider it appropriate to emphasize this in order to avoid any confusion due to the different terms and definitions used in the literature [182].

When a Mueller matrix is obtained from an experimental measurement, we of course expect that, leaving aside experimental errors, it corresponds to a real optical system. However, it is of interest to consider the above general constraints in order to know the range of possible expectations as well as to check that the results obtained are consistent. Furthermore, the knowledge of a general characterization of Mueller matrices allows for interpreting the results in terms of parameters with specific physical meaning.

A useful way to explicitly show most of the covariance conditions (not all) in the expression of the Mueller matrix $\mathbf{M}$, is to write it as a function of the statistical parameters, so that we get

$$\mathbf{M} = \frac{1}{2}\left(\begin{array}{cc|cc|cc}
\sigma_0^2+\sigma_1^2+\sigma_2^2+\sigma_3^2 & \sigma_0^2+\sigma_1^2-\sigma_2^2-\sigma_3^2 & 2\rho_{02}\sigma_0\sigma_2 c_{02} & 2\rho_{02}\sigma_0\sigma_2 s_{02} \\
 & & +2\rho_{13}\sigma_1\sigma_3 c_{13} & -2\rho_{13}\sigma_1\sigma_3 s_{13} \\
\hline
\sigma_0^2-\sigma_1^2+\sigma_2^2-\sigma_3^2 & \sigma_0^2-\sigma_1^2-\sigma_2^2+\sigma_3^2 & 2\rho_{02}\sigma_0\sigma_2 c_{02} & 2\rho_{02}\sigma_0\sigma_2 s_{02} \\
 & & -2\rho_{13}\sigma_1\sigma_3 c_{13} & +2\rho_{13}\sigma_1\sigma_3 s_{13} \\
\hline
2\rho_{01}\sigma_0\sigma_1 c_{01} & 2\rho_{01}\sigma_0\sigma_1 c_{01} & 2\rho_{03}\sigma_0\sigma_3 c_{03} & -2\rho_{03}\sigma_0\sigma_3 s_{03} \\
+2\rho_{23}\sigma_2\sigma_3 c_{23} & -2\rho_{23}\sigma_2\sigma_3 c_{23} & +2\rho_{12}\sigma_1\sigma_2 c_{12} & -2\rho_{12}\sigma_1\sigma_2 s_{12} \\
\hline
-2\rho_{01}\sigma_0\sigma_1 s_{01} & -2\rho_{01}\sigma_0\sigma_1 s_{01} & -2\rho_{03}\sigma_0\sigma_3 s_{03} & -2\rho_{03}\sigma_0\sigma_3 c_{03} \\
-2\rho_{23}\sigma_2\sigma_3 s_{23} & +2\rho_{23}\sigma_2\sigma_3 s_{23} & +2\rho_{12}\sigma_1\sigma_2 s_{12} & +2\rho_{12}\sigma_1\sigma_2 c_{12}
\end{array}\right) \quad (179)$$

where $c_{ij} = \cos\beta_{ij}$, $s_{ij} = \sin\beta_{ij}$.

From Eq. (179) we observe that the following Stokes vectors can be built in terms of the elements of **M**

$$\mathbf{s}^{(02)} \equiv \begin{pmatrix} m_{00}+m_{10} \\ m_{01}+m_{11} \\ m_{02}+m_{12} \\ m_{03}+m_{13} \end{pmatrix} = \begin{pmatrix} \sigma_0^2+\sigma_2^2 \\ \sigma_0^2-\sigma_2^2 \\ 2\rho_{02}\sigma_0\sigma_2 c_{02} \\ 2\rho_{02}\sigma_0\sigma_2 s_{02} \end{pmatrix}, \quad \mathbf{s}^{(13)} \equiv \begin{pmatrix} m_{00}-m_{10} \\ m_{01}-m_{11} \\ m_{02}-m_{12} \\ m_{03}-m_{13} \end{pmatrix} = \begin{pmatrix} \sigma_1^2+\sigma_3^2 \\ \sigma_1^2-\sigma_3^2 \\ 2\rho_{13}\sigma_1\sigma_3 c_{13} \\ -2\rho_{13}\sigma_1\sigma_3 s_{13} \end{pmatrix},$$

$$\mathbf{s}^{(01)} \equiv \begin{pmatrix} m_{00}+m_{01} \\ m_{10}+m_{11} \\ m_{20}+m_{21} \\ m_{30}+m_{31} \end{pmatrix} = \begin{pmatrix} \sigma_0^2+\sigma_1^2 \\ \sigma_0^2-\sigma_1^2 \\ 2\rho_{01}\sigma_0\sigma_1 c_{01} \\ -2\rho_{01}\sigma_0\sigma_1 s_{01} \end{pmatrix}, \quad \mathbf{s}^{(23)} \equiv \begin{pmatrix} m_{00}-m_{01} \\ m_{10}-m_{11} \\ m_{20}-m_{21} \\ m_{30}-m_{31} \end{pmatrix} = \begin{pmatrix} \sigma_2^2+\sigma_3^2 \\ \sigma_2^2-\sigma_3^2 \\ 2\rho_{23}\sigma_2\sigma_3 c_{23} \\ -2\rho_{23}\sigma_2\sigma_3 s_{23} \end{pmatrix}, \quad (180)$$

$$\mathbf{s}^{(03)} \equiv \begin{pmatrix} m_{00}+m_{11} \\ m_{10}+m_{01} \\ m_{22}-m_{33} \\ m_{23}+m_{32} \end{pmatrix} = \begin{pmatrix} \sigma_0^2+\sigma_3^2 \\ \sigma_0^2-\sigma_3^2 \\ 2\rho_{03}\sigma_0\sigma_3 c_{03} \\ -2\rho_{03}\sigma_0\sigma_3 s_{03} \end{pmatrix}, \quad \mathbf{s}^{(12)} \equiv \begin{pmatrix} m_{00}-m_{11} \\ m_{01}-m_{10} \\ m_{22}+m_{33} \\ m_{23}-m_{32} \end{pmatrix} = \begin{pmatrix} \sigma_1^2+\sigma_2^2 \\ \sigma_1^2-\sigma_2^2 \\ 2\rho_{12}\sigma_1\sigma_2 c_{12} \\ -2\rho_{12}\sigma_1\sigma_2 s_{12} \end{pmatrix}.$$

These expressions are formally identical to those corresponding to certain Stokes vectors where $\sigma_k^2$ represent the mean quadratic amplitudes and $\rho_{kl}, \beta_{kl}$ represent the respective magnitudes and arguments of the degrees of coherence.

It is important to point out that the first four Stokes vectors in Eq. (180) can be obtained as

$$\mathbf{s}^{(02)} = \mathbf{M}^T (1,1,0,0)^T, \quad \mathbf{s}^{(13)} = \mathbf{M}^T (1,-1,0,0)^T,$$
$$\mathbf{s}^{(01)} = \mathbf{M} (1,1,0,0)^T, \quad \mathbf{s}^{(23)} = \mathbf{M} (1,-1,0,0)^T. \quad (181)$$

(recall that, for any given Mueller matrix **M**, $\mathbf{M}^T$ is also a Mueller matrix). Nevertheless, the last two $\mathbf{s}^{(03)}$ and $\mathbf{s}^{(12)}$, cannot be obtained by means of direct transformations of Stokes vectors performed through **M** or $\mathbf{M}^T$. This property arises from the peculiar structure of Mueller matrices and not from the fact that they transform Stokes vectors into Stokes vectors.

Obviously, any Mueller matrix is a Stokes matrix. The converse statement is not true because there are Stokes matrices such as $\mathbf{X} \equiv \mathbf{D}(1,0,1,-1)$, whose corresponding vector $\mathbf{s}^{(03)}(\mathbf{X}) = (1,0,2,0)^T$ is not a Stokes vector. Another counterexample is the Stokes matrix $\mathbf{Y} \equiv \mathbf{D}(1,1,1,0)$, whose corresponding vector $\mathbf{s}^{(12)}(\mathbf{Y}) = (0,0,1,0)^T$ is not a Stokes vector.

Since any Mueller matrix is a Stokes matrix, we conclude that the numerous inequalities quoted in the literature for Stokes matrices are all deducible, as necessary conditions, from the above general characterization of Mueller matrices.

### 5.5. The purity criterion for Mueller matrices and the degree of polarimetric purity of material media

The study of the necessary and sufficient conditions for a Mueller matrix to be a pure Mueller matrix is an important subject dealt with by some authors [121,122,125,164-178,182-188]. We can now analyze this subject in the light of the properties of the coherency matrix $\mathbf{H}$.

As with 2x2 and 3x3 coherency matrices, we consider here the Euclidean norms of $\mathbf{H}$ and $\mathbf{M}$

$$\|\mathbf{H}\|_2 \equiv \sqrt{\sum_{i,j=0}^{3} |h_{ij}|^2} = \sqrt{\operatorname{tr}(\mathbf{H}^\dagger \mathbf{H})} = \sqrt{\operatorname{tr}(\mathbf{H}^2)},$$

$$\|\mathbf{M}\|_2 \equiv \sqrt{\sum_{i,j=0}^{3} m_{ij}^2} = \sqrt{\operatorname{tr}(\mathbf{M}^T \mathbf{M})},$$

(182)

and we define the norm

$$\|\mathbf{H}\|_0 \equiv \operatorname{tr}\mathbf{H} = \left\|\sqrt{\mathbf{H}}\right\|_2^2.$$ (183)

It is easy to show that these norms satisfy the following relations

$$\|\mathbf{H}\|_2^2 = \frac{1}{4}\|\mathbf{M}\|_2^2,$$ (184)

$$\|\mathbf{H}\|_0 = m_{00},$$ (185)

$$\frac{1}{4}\|\mathbf{H}\|_0^2 \le \|\mathbf{H}\|_2^2 \le \|\mathbf{H}\|_0^2.$$ (186)

The last relation shows that $\|\mathbf{H}\|_2 = \|\mathbf{H}\|_0$, or equivalently $\sum_{i,j=0}^{3} m_{ij}^2 = 4m_{00}^2$, is a necessary and sufficient condition for $\mathbf{H}$ to have only a nonzero eigenvalue and, in consequence, it is a necessary and sufficient condition for a 4x4 coherency matrix to correspond to a Mueller-Jones matrix [168]. The other limit $\|\mathbf{H}\|_2 = \|\mathbf{H}\|_0 / 2$ is reached when the system is composed of an equiprobable mixture of pure elements.

The inequality $\|\mathbf{H}\|_2 \le \|\mathbf{H}\|_0$ was first pointed out by Fry and Kattawar [164] in terms of the elements of the Mueller matrix, and they found that the equality

$$\operatorname{tr}(\mathbf{M}^T \mathbf{M}) = 4m_{00}^2,$$ (187)

is satisfied by any Mueller-Jones matrix. However, the question on whether, given a Mueller matrix $\mathbf{M}$, this is a sufficient condition for $\mathbf{M}$ to be a Mueller-Jones matrix has been a subject discussed in a number of works [121,122,165,173,176,178,182,186,187,189, 190]. It is clear that, under the premise that a Mueller matrix is expressible as a convex sum of Mueller-Jones matrices, we can state: *given a Mueller matrix* $\mathbf{M}$, *Eq. (187) is a sufficient condition for* $\mathbf{M}$ *to be a Mueller-Jones matrix.* All the arguments contrary to this statement arise from considering real 4x4 matrices which do not satisfy the eigenvalue conditions or the passivity conditions [173,178,182,187,189,190].

The degree of polarimetric purity $P_\Delta \equiv P_{4D}$ of a material sample is defined as [131,168,41]

$$P_\Delta = \sqrt{\frac{1}{3}\left(\frac{4\,\mathrm{tr}\mathbf{H}^2}{\mathrm{tr}^2\mathbf{H}} - 1\right)}. \tag{188}$$

This invariant nondimensional parameter is restricted to the interval $0 \leq P_\Delta \leq 1$. The minimum $P_\Delta = 0$ corresponds to an ideal total depolarizer, characterized by the fact that all eigenvalues of $\mathbf{H}$ are equal, i.e. the medium is composed of an equiprobable mixture of elements, and does not exhibit any polarimetric preference ($m_{ij} = 0$ except for $m_{00}$). The maximum, $P_\Delta = 1$, corresponds to a pure system ($\lambda_0 > 0$, $\lambda_1 = \lambda_2 = \lambda_3 = 0$).

This fundamental physical quantity $P_\Delta$ was first introduced by Gil and Bernabéu in the form of a *depolarization index* in terms of the elements of $\mathbf{M}$ [131]

$$P_\Delta = \sqrt{\frac{1}{3}\left(\frac{\sum_{i,j=0}^{3} m_{ij}^2}{m_{00}^2} - 1\right)} = \sqrt{\frac{1}{3}\left(\frac{\mathrm{tr}(\mathbf{M}^T\mathbf{M})}{m_{00}^2} - 1\right)}. \tag{189}$$

In terms of both norms of $\mathbf{H}$, $P_\Delta$ is expressed as

$$P_\Delta = \sqrt{\frac{1}{3}\left(\frac{4\|\mathbf{H}\|_2^2}{\|\mathbf{H}\|_0^2} - 1\right)}. \tag{190}$$

$P_\Delta$ gives an objective measure of the global polarimetric purity of the system, as well as of its depolarizing power, and provides criteria for the analysis of measured Mueller matrices [58,191-193]. Moreover, the *depolarizance* $D_\Delta$ of an optical system can be defined as

$$D_\Delta \equiv \sqrt{1 - P_\Delta^2}. \tag{191}$$

This definition is proper because $D_\Delta$ can be obtained as an average measure of the depolarization produced by the system over all incident pure states [131]. A comparison between $P_\Delta$ and another scalar measure of the depolarizing power of a material sample, the *average degree of polarization A* was performed by Chipman [194]. Both measures are defined from averages of the degree of polarization of the exiting states. The difference between these two quantities is particularly significant for systems involving an asymmetric serial arrangement of diattenuators and depolarizers. This is a direct consequence of the fact that, whereas $P_\Delta$ is an invariant quantity defined from a symmetric average involving exiting states for both direct ($\mathbf{M}$) and reciprocal ($\mathbf{M}^r$) Mueller matrices, *A* only takes into account the effects of direct propagation. Although the definition of *A* seems to be more natural, the symmetric nature of the characteristic properties of the Mueller matrices leads to a simple geometric interpretation of $D_\Delta$ as a normalized distance from $\mathbf{M}$ to an ideal depolarizer [194].

### 5.6. Parallel decompositions of matrices representing material media

As done with 3×3 coherency matrices, we consider here the possible parallel decompositions of the coherency matrix $\mathbf{H}(\mathbf{M})$ representing the properties of a material sample. The physical realizability of parallel decompositions of $\mathbf{H}$ requires that it has the form of a convex linear combination. That is to say, the incident light is shared among the components, so that the emerging pencils are recombined into the whole emerging beam. The convex linear combination of passive components ensures the passive behavior of the whole system.

5.6.1. The spectral decomposition of the Mueller matrix of a material medium

Since $\mathbf{H}$ is a positive semidefinite Hermitian matrix, it can be diagonalized through a unitary transformation

$$\mathbf{H} = \mathbf{U}\mathbf{D}(\lambda_1, \lambda_2, \lambda_3, \lambda_4)\mathbf{U}^\dagger, \tag{192}$$

where $\mathbf{D}(\lambda_1, \lambda_2, \lambda_3, \lambda_4)$ represents the diagonal matrix composed of the four nonnegative eigenvalues $0 \leq \lambda_4 \leq \lambda_3 \leq \lambda_2 \leq \lambda_1$. The columns $\mathbf{u}_k$ $(k=1,2,3,4)$ of the 4x4 unitary matrix $\mathbf{U}$ are the respective unitary, and mutually orthogonal, eigenvectors.

Therefore, $\mathbf{H}$ can be expressed as the following convex linear combination of four rank-one coherency matrices that represent respective pure systems

$$\begin{aligned}\mathbf{H} = {}&\lambda_1 \mathbf{U}\mathbf{D}(1,0,0,0)\mathbf{U}^\dagger + \lambda_2 \mathbf{U}\mathbf{D}(0,1,0,0)\mathbf{U}^\dagger \\ &+ \lambda_3 \mathbf{U}\mathbf{D}(0,0,1,0)\mathbf{U}^\dagger + \lambda_4 \mathbf{U}\mathbf{D}(0,0,0,1)\mathbf{U}^\dagger,\end{aligned} \tag{193}$$

where each term in the sum is affected by its corresponding eigenvector $\mathbf{u}_i$, so that

$$\mathbf{H} = \sum_{k=1}^{4} \hat{\lambda}_k \mathbf{H}_{Jk}, \quad \hat{\lambda}_k \equiv \lambda_k / m_{00},$$

$$\left[\mathbf{H}_{Jk} \equiv m_{00}\left(\mathbf{u}_k \otimes \mathbf{u}_k^\dagger\right), \quad \sum_{k=1}^{4} \hat{\lambda}_k = 1, \quad m_{00} = \mathrm{tr}\,\mathbf{H}\right]. \tag{194}$$

This *spectral decomposition* can be expressed as follows in terms of the corresponding pure Mueller matrices

$$\mathbf{M} = \sum_{k=1}^{4} \hat{\lambda}_k \mathbf{M}_{Jk},$$

$$\left[\mathbf{M}_{Jk} \equiv \mathbf{M}_{Jk}(\mathbf{H}_{Jk}), \quad m_{00k} = m_{00} = \mathrm{tr}\,\mathbf{H}, \quad \sum_{k=1}^{4} \hat{\lambda}_k = 1\right],$$

showing that any linear system can be considered a parallel combination of up to four pure systems with weights proportional to the eigenvalues of $\mathbf{H}$.

Due to the practical importance of this kind of decomposition, it should be noted that when an eigenvalue $\lambda_k$ has a multiplicity $m$ $(1 < m \leq 4)$, the eigenvectors of the corresponding invariant $m$-dimensional subspace are not unique and can be chosen arbitrarily as a set of orthonormal vectors covering the corresponding subspace.

Moreover, the statistical nature of $\mathbf{H}$ leads to a probabilistic interpretation of its eigenvalues. This fact has direct consequences in the interpretation of quantities such as the von Neumann entropy and the indices of polarimetric purity, which will be considered in further sections.

5.6.2. The characteristic decomposition of the Mueller matrix of a material medium

Let us now consider the *4D characteristic decomposition* (also called the *trivial decomposition*) of the coherency matrix **H** as a convex linear combination of four systems with equal mean transmittances

$$\mathbf{H} = \mathbf{U}\,\mathbf{D}(\lambda_1, \lambda_2, \lambda_3, \lambda_4)\,\mathbf{U}^\dagger = \frac{\lambda_1 - \lambda_2}{\mathrm{tr}\,\mathbf{H}} \mathbf{H}_J + 2\frac{\lambda_2 - \lambda_3}{\mathrm{tr}\,\mathbf{H}} \mathbf{H}_2 + 3\frac{\lambda_3 - \lambda_4}{\mathrm{tr}\,\mathbf{H}} \mathbf{H}_3 + 4\frac{\lambda_4}{\mathrm{tr}\,\mathbf{H}} \mathbf{H}_4,$$

$$\mathbf{H}_J \equiv \mathrm{tr}\,\mathbf{H}\left[\mathbf{U}\mathbf{D}(1,0,0,0)\mathbf{U}^\dagger\right],$$

$$\mathbf{H}_2 \equiv \frac{1}{2}\mathrm{tr}\,\mathbf{H}\left[\mathbf{U}\mathbf{D}(1,1,0,0)\mathbf{U}^\dagger\right],$$

$$\mathbf{H}_3 \equiv \frac{1}{3}\mathrm{tr}\,\mathbf{H}\left[\mathbf{U}\mathbf{D}(1,1,1,0)\mathbf{U}^\dagger\right],$$

$$\mathbf{H}_4 \equiv \frac{1}{4}(\mathrm{tr}\,\mathbf{H})\,\mathbf{I},$$

$$\left[\mathrm{rank}\,\mathbf{H}_i = 1 \ (i = J, 2, 3, 4), \ \ \mathbf{I} \equiv \mathbf{D}(1,1,1,1)\right]$$

(195)

We observe that any state of the material system is polarimetrically equivalent to a convex linear combination of up to four components with equal mean transmittances: a *pure component*, a *2D unpolarized system*, a *3D unpolarized system* and a *4D unpolarized system*.

This result can also be expressed in terms of Mueller matrices as follows

$$\mathbf{M} = \frac{\lambda_1 - \lambda_2}{m_{00}} \mathbf{M}_J + 2\frac{\lambda_2 - \lambda_3}{m_{00}} \mathbf{M}_2 + 3\frac{\lambda_3 - \lambda_4}{m_{00}} \mathbf{M}_3 + 4\frac{\lambda_4}{m_{00}} \mathbf{M}_4,$$

$$\mathbf{M}_i = \mathbf{M}_i(\mathbf{H}_i) \ (i = J, 2, 3, 4).$$

(196)

As expected, it is straightforward to reproduce the characteristic decomposition for 3D and 2D coherency matrices.

Moreover, it is clear that, in general, a nonpure system cannot be decomposed into sum of a pure component and a 4D unpolarized component. As we have observed previously, a $n \times n$ coherency matrix of a mixed state depends, in general, on $n^2$ independent real parameters, whereas a $n \times n$ coherency matrix of a pure state depends, in general, on $2n - 1$ independent real parameters. It is obvious that the characteristic decomposition of a $n \times n$ coherency matrix ($n^2$ parameters) into a convex sum of the coherency matrix of a pure state ($2n - 1$ independent real parameters) and a fully random state (one parameter) is only possible for $n = 2$.

The characteristic decomposition constitutes the appropriate framework for distinguishing the pure component from the random or noise component of the material sample. It also allows for a proper treatment of the experimental errors when it is known that the target under measurement is pure.
[See J. J. Gil, "On optimal filtering of measured Mueller matrices," *Appl. Opt.* **55**, 5449–5455 (2016).]

5.6.3. The arbitrary decomposition of the Mueller matrix of a material medium
*For updated details, see*
J. J. Gil and I. San José, "Arbitrary decomposition of a Mueller matrix," Opt. Lett. **23**, 5715-5718 (2019),
J. J. Gil and I. San José, "Polarimetric subtraction of Mueller matrices," J. Opt. Soc. Am. A **30**, 1078-1088 (2013)

Given the simple linear relation between a coherency matrix and its corresponding Mueller matrix, it is possible to classify Mueller Matrices according to the rank of **H**, leading to the possible *target* *decompositions* [113,191,192]. The methods for decomposing measured Mueller matrices play an important role in many applications of polarimetry because they can be used for different purposes as, for example, to identify elements in the sample and to improve the contrast of images obtained by radar polarimetry [59,195].

As we have demonstrated for 3×3 coherency matrices, the spectral decomposition is not the only possibility for decomposing a mixed state. It can also be decomposed through the arbitrary decomposition. This decomposition can be applied to $n \times n$ coherency matrices [102] and, in particular, to $\mathbf{H}$. Moreover, the physical possibility of synthesizing experimentally nonpure systems by means of different parallel combinations of pure systems suggests the existence of decompositions other than the spectral decomposition.

Let us consider a nonpure 4x4 coherency matrix $\mathbf{H}$ ($1 < \operatorname{rank} \mathbf{H} \leq 4$). Through the same arguments as those considered for the arbitrary decomposition of 3x3 coherency matrices, we obtain the following arbitrary decomposition of the coherency matrix into a linear convex combination of pure coherency matrices

$$\mathbf{H} = \sum_{i=1}^{4} p_i \mathbf{H}_{Ji}, \quad \sum_{i=1}^{4} p_i = 1, \quad \operatorname{rank} \mathbf{H}_{Ji} = 1, \quad \operatorname{tr} \mathbf{H}_{Ji} = \operatorname{tr} \mathbf{H}. \tag{197}$$

This *homogeneous arbitrary decomposition* can also be expressed as

$$\mathbf{H} = (\operatorname{tr} \mathbf{H}) \sum_{i=1}^{r} p_i \left( \hat{\mathbf{w}}_i \otimes \hat{\mathbf{w}}_i^\dagger \right), \quad p_i = \frac{1}{\sum_{j=1}^{r} \frac{1}{\hat{\lambda}_j} \left| \left( \mathbf{U}^\dagger \mathbf{w}_i \right)_j \right|^2},$$

$$\left[ \left| \hat{\mathbf{w}}_i \right| = 1, \ r \equiv \mathbf{H}, \ \hat{\lambda}_j \equiv \lambda_j / \operatorname{tr} \mathbf{H} \right], \tag{198}$$

where $r \equiv \operatorname{rank} \mathbf{H}$ and $\hat{\mathbf{w}}_i$ is a set of $r$ arbitrary linearly independent vectors constituting a generalized basis of the subspace generated by the eigenvectors of $\mathbf{H}$ with nonzero eigenvalues. Obviously, the spectral decomposition is a particular case of the arbitrary decomposition. Each pure component can be expressed as a function of the corresponding unit vector $\hat{\mathbf{w}}_i$. The coherency matrices of the components have been chosen so as to satisfy $\operatorname{tr} \mathbf{H}_{Ji} = \operatorname{tr} \mathbf{H} = m_{00}$ (the term *homogeneous* refer to this feature), but arbitrary values for the mean transmittances $m_{00i}$ of the components can be easily considered by an appropriate recalculation of the coefficients $p_i$ of the new corresponding convex sum. In particular, the arbitrary decomposition can be performed in terms of passive components (i.e., pure Mueller matrices satisfying the passivity criterion) [see I. San José and J. J. Gil, "Characterization of passivity in Mueller matrices," J. Opt. Soc. Am A **37,** 199-208 (2020)].

When $\operatorname{rank} \mathbf{H} = 4$, any pure system can be considered as a component and, once chosen, the successive choices of the second and third components are restricted by the exigency that the eigenvector with a nonzero eigenvalue of $\mathbf{H}_{Ji}$ belongs to the subspace generated by the eigenvectors with nonzero eigenvalues of the coherency matrix resulting from the previous subtractions. The fourth component is fully determined by the previous choices of the other pure components.

Obviously, as occurs in the spectral decomposition, the number of pure components of the equivalent system is equal to the rank of $\mathbf{H}$.

We see that any nonpure system is polarimetrically equivalent to a parallel combination of one to four pure components with the same mean transmittance $m_{00}$. As in other sections of this paper, *parallel combination* here means that the set of pure elements of the equivalent system are illuminated by exclusive portions of the incident light, so that their respective emerging beams are incoherently superposed.

It is very important to point out that, among these mathematically possible decompositions, only those where all the components satisfy the passivity condition are physically realizable, i.e. the corresponding maximum gains satisfy $k_{1i}(\mathbf{H}_{Ji}) \leq 1$. These inequalities are not ensured by the conditions $\operatorname{tr} \mathbf{H}_{Ji} = \operatorname{tr} \mathbf{H}$ and, given a component $\mathbf{H}_{Ji}$, different procedures can be followed in order to check the passivity condition. For instance, by observing the condition in terms of the elements $h_{kl(i)}$ of $\mathbf{H}_{Ji}$ [168]

$$g_{f(i)} = g_{r(i)} = k_{1(i)}^2$$
$$= \operatorname{tr}(\mathbf{H}) + \sqrt{\left(h_{00(i)} - h_{11(i)} + h_{22(i)} - h_{33(i)}\right)^2 + 4\left(h_{01(i)} + h_{23(i)}\right)\left(h_{01(i)}^* + h_{23(i)}^*\right)} \leq 1. \quad (199)$$

or by checking Eq. (110), which expresses the passivity condition in terms of the elements of the pure Mueller matrix $\mathbf{M}_{Ji}(\mathbf{H}_{Ji})$.

For many experimental cases where the target under measurement contains a known (or suspected to be) pure component, the arbitrary decomposition provides a procedure for a proper subtraction of the known component and, i.e., to isolate the unknown part of the sample. The procedure can be iterated if there exists more than one known pure component.

Moreover, the arbitrary decomposition has important consequences because it allows for the identification of all the possible physically realizable target decompositions. In fact, it provides a method for analyzing measured samples and adjusting the target decomposition in order to obtain improvements in identifying unknown components. When applied to imaging polarimetry, this polarimetric subtraction can be applied to improve strongly the contrast of particular elements of the target with respect to the homogeneous substrate.

The *homogeneous arbitrary decomposition* can be expressed in terms of the corresponding Mueller matrices as follows

$$\mathbf{M} = \sum_{i=1}^{r} p_i \mathbf{M}_{Ji}, \quad p_i = \frac{1}{\sum_{j=1}^{r} \frac{1}{\hat{\lambda}_j} \left|\left(\mathbf{U}^\dagger \mathbf{w}_i\right)_j\right|^2},$$

$$p_i \geq 0, \quad \sum_{i=1}^{r} p_i = 1, \quad m_{00i} = m_{00}, \quad (200)$$

where the Mueller matrix $\mathbf{M}$ of the system is obtained as a convex linear combination of, up to four, Mueller-Jones matrices $\mathbf{M}_{Ji}$ with equal mean intensity coefficients $m_{00i} = m_{00}$. The general *arbitrary decomposition* (without the constraint $m_{00i} = m_{00}$) has been formulated as follows in [J. J. Gil, I. San José, "Arbitrary decomposition of a Mueller matrix," Opt. Lett. **44**, 5715-5718 (2019)]

$$\mathbf{M} = \sum_{i=1}^{r} k_i \mathbf{M}_{Ji}, \quad k_i = \frac{1}{m_{00i} \sum_{j=1}^{r} \frac{1}{\hat{\lambda}_j} \left|\left(\mathbf{U}^\dagger \mathbf{w}_i\right)_j\right|^2} \quad (k_i \geq 0), \quad \sum_{i=1}^{r} k_i = 1.$$

The general arbitrary decomposition of passive Mueller matrices in terms of passive components has been formulated in [I. San José and J. J. Gil, "Characterization of passivity in Mueller matrices," J. Opt. Soc. Am A **37,** 199-208 (2020)]. As an example of an arbitrary decomposition of a passive Mueller matrix whose components are non-passive, let us consider the Mueller matrix $\mathbf{M}_{\Delta 0}$ of an ideal depolarizer ($m_{ij}=0$, except $m_{00}=1$). The canonical spectral decomposition of $\mathbf{H}(\mathbf{M}_{\Delta 0})$ leads to the following decomposition of $\mathbf{M}_{\Delta 0}$

$$\mathbf{M}_{\Delta 0} \equiv \begin{pmatrix} 1 & 0 & 0 & 0 \\ 0 & 0 & 0 & 0 \\ 0 & 0 & 0 & 0 \\ 0 & 0 & 0 & 0 \end{pmatrix} = \frac{1}{4}\begin{pmatrix} 1 & 1 & 0 & 0 \\ 1 & 1 & 0 & 0 \\ 0 & 0 & 0 & 0 \\ 0 & 0 & 0 & 0 \end{pmatrix} + \frac{1}{4}\begin{pmatrix} 1 & -1 & 0 & 0 \\ 1 & -1 & 0 & 0 \\ 0 & 0 & 0 & 0 \\ 0 & 0 & 0 & 0 \end{pmatrix} + \frac{1}{4}\begin{pmatrix} 1 & 1 & 0 & 0 \\ -1 & -1 & 0 & 0 \\ 0 & 0 & 0 & 0 \\ 0 & 0 & 0 & 0 \end{pmatrix} + \frac{1}{4}\begin{pmatrix} 1 & -1 & 0 & 0 \\ -1 & 1 & 0 & 0 \\ 0 & 0 & 0 & 0 \\ 0 & 0 & 0 & 0 \end{pmatrix}. \qquad (201)$$

Obviously, the components of this linear convex combination do not satisfy the passivity condition (the factors $1/4$ are the coefficients of the convex sum). This example demonstrates that not all the mathematically possible arbitrary decompositions are physically realizable.

Nevertheless, the passivity condition is satisfied by all the elements of the following alternative decomposition

$$\mathbf{M}_{\Delta 0} = \frac{1}{4}\mathbf{M}_R(0,0,\pi/2) + \frac{1}{4}\mathbf{M}_R(0,0,-\pi/2) + \frac{1}{4}\mathbf{M}_R(\pi/4,0,\pi) + \frac{1}{4}\mathbf{M}_R(\pi/4,\pi/2,\pi), \qquad (202)$$

whose explicit expression is

$$\mathbf{M}_{\Delta 0} = \frac{1}{4}\begin{pmatrix} 1 & 0 & 0 & 0 \\ 0 & 1 & 0 & 0 \\ 0 & 0 & 0 & -1 \\ 0 & 0 & 1 & 0 \end{pmatrix} + \frac{1}{4}\begin{pmatrix} 1 & 0 & 0 & 0 \\ 0 & 1 & 0 & 0 \\ 0 & 0 & 0 & 1 \\ 0 & 0 & -1 & 0 \end{pmatrix} + \frac{1}{4}\begin{pmatrix} 1 & 0 & 0 & 0 \\ 0 & -1 & 0 & 0 \\ 0 & 0 & -1 & 0 \\ 0 & 0 & 0 & 1 \end{pmatrix} + \frac{1}{4}\begin{pmatrix} 1 & 0 & 0 & 0 \\ 0 & -1 & 0 & 0 \\ 0 & 0 & 1 & 0 \\ 0 & 0 & 0 & -1 \end{pmatrix}. \qquad (203)$$

### 5.7. Geometric maps of the degree of polarization

*For a comprehensive formulation of the Poincaré sphere mapping by Mueller matrices, see*

R. Ossikovski, J. J. Gil, I. San José, "Poincaré sphere mapping by Mueller matrices," J. Opt. Soc. Am. A **30**, 2291-2305 (2013)

Let us consider now some possible graphical representations of the polarimetric effects of material media. As it is well known, the Poincaré sphere maps states of polarization with intensity equal to one, so that the distance to the origin is the degree of polarization. Points on the surface represent pure states, whereas points inside represent nonpure states. The linear interaction of polarized light with material targets produce changes in both intensity and degree of polarization and, therefore, the Poincaré sphere representation (defined for states with intensity equal to 1) is not directly applicable for mapping the output states.

As a geometrical and visual tool for analyzing and classifying material samples, based on their effects on the degree of polarization, some authors have studied the $P$-image defined by the points [2,196-199]

$$\frac{1}{s_0'}(s_1', s_2', s_3'), \qquad (204)$$

which correspond to the output Stokes parameters

$$(s_0', s_1', s_2', s_3')^T = \mathbf{M}(s_0, s_1, s_2, s_3)^T \quad \left(1 = s_0^2 \geq s_1^2 + s_2^2 + s_3^2\right). \qquad (205)$$

The input Stokes vectors map all the points of the solid Poincaré sphere. The distance of each point of the $P$-image to the origin is just its degree of polarization and, in general, varies depending on the corresponding input Stokes vector. The surface of this solid object is called the *DoP surface* [198] and corresponds to totally polarized input states.

Moreover, given $\mathbf{M}$, the $I$-image is defined as the surface [198,199]

$$\frac{s'_0}{\sqrt{s'^2_1 + s'^2_2 + s'^2_3}} \left( s'_1, s'_2, s'_3 \right)^T. \tag{206}$$

These representations allow one to classify the different polarimetric behaviors,

*a) Normal diattenuators*: The $I$-image is not spherical, whereas the surface shape of the $P$-image remains being a sphere of radius 1. Notice that the Mueller matrix is symmetric. Topologically the points are more concentrated around two poles on the surface. The position of these poles is given by the intersection of the maximum and minimum polarizance vectors with the surface. The straight segment connecting these points intersects the image point of the origin of the Poincaré sphere.

*b) Retarders*: None of the objects is deformed (neither morphologically nor topologically). The result is a rotation of both objects with respect to the axis defined by the two antipodal eigenstates of the retarder.

*c) Parallel combination (or statistical mixture) of diattenuators*: The surface of the $P$-image is deformed and its origin is displaced. The $I$-image is also deformed. When the diattenuators are homogeneous, some symmetries of the objects correspond to the property $\mathbf{M}^T = \mathbf{M}$.

*d) Parallel combination (or statistical mixture) of retarders*: The surface of the $P$-image is deformed and the origin remains unchanged. The $I$-image is not deformed.

*e) Parallel combination (or statistical mixture) of diattenuators and retarders*: The surface of the $P$-image is deformed and its origin is displaced. The $I$-image is also deformed.

A detailed study of these kinds of representations, including plane maps, has been developed by DeBoo, Sasian, and Chipman [198].

### 5.8. Indices of polarimetric purity

*For updated details, see*

I. San José and J. J. Gil, "Invariant *indices of polarimetric purity*: generalized *indices* of *purity* for n× n covariance matrices," Opt. Commun. **284**, 38-47, (2011).

Note that, after the publication of the original version of this Review, the indices of polarimetric purity associated with a Mueller matrix have been redefined in a more appropriate manner (see the above reference). This does not prevent the validity of the former formulation.

Given the statistical nature of the coherency matrices $\mathbf{H}$ representing material media, we emphasize the importance of obtaining parameters that provide measures of their polarimetric purity. Usually the *degree of polarimetric purity* (or *depolarization index*) $P_{4D}$ [131,200] or, alternatively, the polarization entropy [57], is used as a quantity characterizing the overall purity.

Thus, in a similar manner that for the cases of 2x2 and 3x3 coherency matrices representing states of light, the study of the structure of purity of $\mathbf{H}$ requires the use of a set of relative differences between the four eigenvalues of $\mathbf{H}$. Thus, in addition to $P_{4D}$, three new invariant *indices of polarimetric purity* (IPP) can be defined from the eigenvalues of $\mathbf{H}$. This set of three quantities contains all the information concerning the polarimetric purity. It should be noted that neither $P_{4D}$ nor the polarization entropy cover all the information mentioned, but they can be calculated from the indices of purity.

At this point, it is important to consider the concept of *polarimetric contrast* introduced by Réfrégier, Roche and Goudail [107] for 3D coherency matrices. These authors have pointed out that for 3D polarization states with Gaussian fluctuations, three invariant quantities are relevant to characterize the contrast in polarimetric imagery. There are certain particular cases where a $3 \times 3$ representation of the coherency matrix is enough for the characterization of the measured polarimetric quantities (e.g. monostatic radar polarimetry). Nevertheless, complete Mueller polarimetry requires the consideration of the $4 \times 4$ coherency matrix $\mathbf{H}$ and, in consequence, at least four invariant quantities should be considered for a complete characterization of the polarimetric contrast. A proper set of such quantities is constituted by the three indices of purity (defined below) together with the transmittance for unpolarized light $m_{00} = \mathrm{tr}\mathbf{H} = \lambda_1 + \lambda_2 + \lambda_3 + \lambda_4$, $\lambda_i$ being the eigenvalues of $\mathbf{H}$.

An examination of the expressions of the eigenvalues of $\mathbf{H}$ obtained by algebraic computation shows that it is possible to write them in terms of three nonnegative parameters

$$\lambda_1 = \frac{\mathrm{tr}\mathbf{H}}{4}(1 + P_2 + 2P_1), \quad \lambda_2 = \frac{\mathrm{tr}\mathbf{H}}{4}(1 + P_2 - 2P_1),$$
$$\lambda_3 = \frac{\mathrm{tr}\mathbf{H}}{4}(1 - P_2 + 2P_3), \quad \lambda_4 = \frac{\mathrm{tr}\mathbf{H}}{4}(1 - P_2 - 2P_3), \tag{207}$$

where the *indices of polarimetric purity* (IPP) $P_1, P_2, P_3$, are defined as

$$P_1 \equiv \frac{\lambda_1 - \lambda_2}{\mathrm{tr}\mathbf{H}}, \quad P_2 \equiv \frac{(\lambda_1 + \lambda_2) - (\lambda_3 + \lambda_4)}{\mathrm{tr}\mathbf{H}}, \quad P_3 \equiv \frac{\lambda_3 - \lambda_4}{\mathrm{tr}\mathbf{H}}. \tag{208}$$

Note that, despite the fact that the above former formulation is considered in this work, the IPP where redefined in the following more appropriate manner, which is that is currently used in related works

$$P_1 \equiv \frac{\lambda_1 - \lambda_2}{\mathrm{tr}\mathbf{H}}, \quad P_2 \equiv \frac{\lambda_1 + \lambda_2 - 2\lambda_3}{\mathrm{tr}\mathbf{H}}, \quad P_3 \equiv \frac{\lambda_1 + \lambda_2 + \lambda_3 - 3\lambda_4}{\mathrm{tr}\mathbf{H}} = 1 - 4\lambda_4. \tag{IPP}$$

From Eq. (208), we see that they the IPP are expressed in a similar way to that a degree of polarization. In fact:

- $P_1$ is a relative measure of the difference between the weights of the two more significant pure components of the system.

- $P_2$ is a relative measure of the difference between the combined weight of the two more significant pure components and the combined weight of the two less significant pure components of the system,

- $P_3$ is a relative measure of the difference between the weights of the two less significant pure components.

We see that the indices of purity can be interpreted as probabilistic relative measures, which provide complete information on the relative amounts of the *equivalent pure components* of the target.

From the above equations, the following quadratic relation between $P_\Delta$ and the three indices of purity $P_1, P_2, P_3$ is obtained

$$P_\Delta^2 = \frac{1}{3}\left(2P_1^2 + P_2^2 + 2P_3^2\right). \tag{209}$$

Another interesting expression of $P_\Delta$ as a symmetric quadratic average of all the relative differences between pairs of eigenvalues is given by

$$P_\Delta^2 = \frac{1}{3}\sum_{\substack{i,j=0 \\ i<j}}^{3} p_{ij}^2, \quad p_{ij} \equiv \frac{\lambda_i - \lambda_j}{\mathrm{tr}\mathbf{H}}. \tag{210}$$

Pure systems are characterized by $P_\Delta = P_1 = P_2 = 1$, $P_3 = 0$. Moreover, the values $P_\Delta = P_1 = P_2 = P_3 = 0$ correspond to certain equiprobable mixtures of four (or more) incoherent elements, resulting in a Mueller matrix $\mathbf{M}_{\Delta 0}$ whose elements are zero except for $m_{00}$.

The degree of polarimetric purity provides a global measure of the purity of the system, whereas a detailed analysis requires a consideration of the three indices of purity.

By applying the starting conditions for the eigenvalues $0 \leq \lambda_4 \leq \lambda_3 \leq \lambda_2 \leq \lambda_1$, we find that the indices of purity are restricted by the following conditions

$$P_1 \geq 0, \ P_3 \geq 0, \ P_1 + P_3 \leq P_2, \ P_3 \leq \frac{1}{2}(1 - P_2). \tag{211}$$

Some useful inequalities derived from these conditions, but less restrictive and hence insufficient for re-obtaining the conditions given by Eq. (211) are the following

$$0 \leq P_1 \leq P_2 \leq 1, \ 0 \leq P_3 \leq P_2 \leq 1, \ 0 \leq P_3 \leq \frac{1}{3}(1 - P_1). \tag{212}$$

Fig.11 shows the feasible region of the purity indices in the *purity space*. The restriction to the plane $P_3 = 0$ reproduces the feasible region for the indices of purity $P_1, P_2$ corresponding to 3x3 coherency matrices, and the feasible region for the degree of polarization of 2x2 coherency matrices corresponds to the segment $P_2 = 1$, $0 \leq P_1 \leq 1$.

Fig.11 summarizes the different physically realizable possibilities in terms of the values of the indices of purity. Nevertheless, it is worth considering some particular cases in order to understand the physical meaning of $P_i$, as well as to reproduce the feasible regions for 3D and 2D representations of the indices of purity.

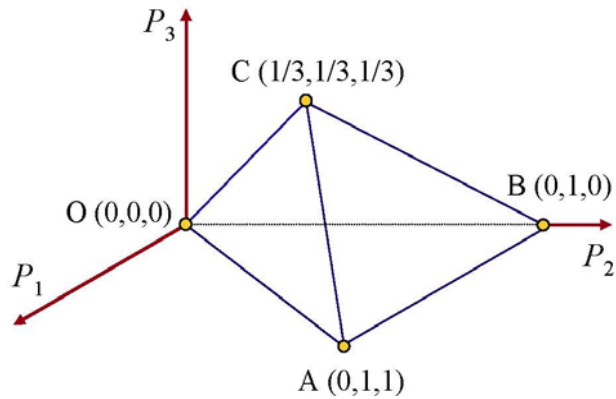

Fig. 11. Feasible region for $P_1, P_2, P_3$ in the purity space.

Given a fixed value of $P_1$, the maximum of $P_3$ is $P_3 = (1 - P_1)/3$ which is reached when $P_2 = (1 + 2P_1)/3$.

Given fixed values for $P_1$ and $P_2$, the maximum of $P_3$ should be analyzed through the two following cases

i) $P_2 \leq \frac{1}{3}(1+2P_1) \;\Rightarrow\; P_3 \leq P_2 - P_1$

ii) $P_2 \geq \frac{1}{3}(1+2P_1) \;\Rightarrow\; P_3 \leq \frac{1}{2}(1-P_2)$

Next we consider some interesting cases based on the values of the indices of purity.

*a)* The face OBA corresponds to states with $P_3 = 0$. Here we analyze the intervals $0 < P_1 \leq P_2 \leq 1$ $(0 < P_\Delta \leq 1)$.

The system is composed of two or four pure elements (if there are four, the two less significant have equal cross sections $\lambda_3 = \lambda_4$). The feasible region is determined by the points located on the triangle OBA (vertex O excluded). This case is mathematically equivalent to that obtained for 3x3 coherency matrices (3D polarization states).

*a.1)* $P_3 = 0$, $P_1 = 0$, $0 < P_2 < 1$ $\left(0 < P_\Delta < 1/\sqrt{3}\right)$.

The system is composed of four pure elements. The two more significant have equal cross sections $\lambda_1 = \lambda_2$ and the two less significant have equal cross sections $\lambda_3 = \lambda_4$. The feasible region is the edge OB (vertices O and B excluded).

*a.2)* $P_3 = 0$, $0 \leq P_1 \leq 1$, $P_2 = 1$ $\left(1/\sqrt{3} \leq P_\Delta \leq 1\right)$.

The system is composed of two pure elements. The feasible region is the edge BA.

*a.2.1)* $P_3 = 0$, $P_1 = P_2 = 1$ $(P_\Delta = 1)$.

Pure system (nondepolarizing deterministic system) characterized by a Mueller-Jones matrix. The vertex A represents this state.

*a.2.2)* $P_3 = 0$, $P_1 = 0$, $P_2 = 1$ $\left(P_\Delta = 1/\sqrt{3}\right)$.

The system is composed of two pure elements with equal cross sections. Point B represents this state.

*b)* The face OBC corresponds to states with $P_1 = 0$. Here we consider the ranges $0 < P_2 \leq 1$, $0 < P_3$ $\left(0 < P_\Delta \leq 1/\sqrt{3}\right)$

The system is composed of three or four pure elements. The two more significant have equal cross sections $\lambda_1 = \lambda_2$ whereas the two less significant have different cross sections $\lambda_3 \neq \lambda_4$. The face OBC (excluding the segment OB) determines the feasible region. These states are exclusive of the 4D representation because the reproduction of 3D states (and, hence, possible 2D states) requires the equality $\lambda_3 = \lambda_4$.

*b.1)* $P_1 = 0$, $0 < P_2 = P_3$ $\left(P_\Delta \leq 1/\sqrt{3}\right)$.

The system is composed of three pure elements with equal cross sections ($0 < \lambda_1 = \lambda_2 = \lambda_3, \lambda_4 = 0$) and is represented by the edge OC (vertex O excluded).

*b,1,1)* $P_1 = 0$, $P_2 = P_3 = 1/3$ $\left(P_\Delta = 1/\sqrt{3}\right)$.

The system is composed of three pure elements with equal cross sections. The vertex C represents this state.

*c)* The face CBA (excluded the edge BA) corresponds to states with $1/3 \leq P_2 < 1$, $P_3 = (1-P_2)/2$, $P_1 \leq P_2$ $\left(1/\sqrt{3} \leq P_\Delta < 1\right)$.

The system is composed of three or four elements. If there are four, the two less significant have different cross sections. These states are exclusive of the 4D representation.

d) The face OCA (vertex O excluded) corresponds to states with $P_2 = P_1 + P_3$, $P_1 < P_2$ $(0 \leq P_\Delta \leq 1)$.

The system is composed of four pure elements. The three more significant have equal cross-sections $\lambda_1 = \lambda_2 = \lambda_3$ and different from the least significant. These states are exclusive of the 4D representation. The edge CA corresponds to $P_2 = (1 + 2P_1)/3$.

e) $P_2 = 0 \Rightarrow P_1 = P_2 = P_3 = P_\Delta = 0$.

This system, equivalent to an ideal depolarizer, is composed of four pure elements with equal cross sections and is represented by the vertex A. All the elements of the Mueller matrix are zero except for $m_{00}$.

While the indices of polarimetric purity (IPP) [either in its original definition or in its current form, see Eq. (IPP) above] provide quantitative information on the structure of polarimetric purity of the medium represented by **M**, the components of purity, namely the diattenuation $D$, the polarizance $P$ and the degree of spherical purity $P_S = \|\mathbf{m}\|_2 / 3$ have been introduced in after the original publication of this work as a meaningful complementary approach that provides qualitative information on the polarimetric purity of **M**. Both sets provide respective qualitative and quantitative views of the contributions to purity and are linked by the interesting relation

$$P_\Delta^2 = \frac{1}{3}\left(2P_1^2 + \frac{2}{3}P_2^2 + \frac{1}{3}P_3^2\right) = \frac{1}{3}\left(D^2 + P^2 + 3P_s^2\right),$$

where the updated expression in Eq. (IPP) for the IPP has been used.

### 5.9. Polarization entropy

Both experimental and industrial polarimetry involve the measurement of up to 16 physical parameters that are constrained by certain complicated nonlinear relations. Thus, considerable efforts have been made to address this problem in order to take advantage of the information provided by polarimetric measurements. Polarization entropy $S$ [57] is a concept related with the lack of polarimetric purity of the material samples. The measurement of $S$ is useful for some purposes and is commonly used in problems concerning light scattering, SAR polarimetry and, in general, situations where depolarization is a relevant subject. This parameter, which is related to the indices of polarimetric purity (IPP) of the material sample, is defined as

$$S(\mathbf{H}) = -\mathrm{tr}\left(\hat{\mathbf{H}} \log_4 \hat{\mathbf{H}}\right) = -\sum_{i=1}^{4}\left(\hat{\lambda}_i \log_4 \hat{\lambda}_i\right). \tag{213}$$

where $\hat{\mathbf{H}} = \mathbf{H}/\mathrm{tr}\,\mathbf{H}$ is the coherency density matrix and $\hat{\lambda}_1, \hat{\lambda}_2, \hat{\lambda}_3, \hat{\lambda}_4$ are the eigenvalues of $\hat{\mathbf{H}}$.

As we have seen, the IPP provide complete information on the depolarization properties and about the polarimetric purity of material samples, in such a manner that they cover a scope of information wider than $S$ and $P_\Delta$. In fact, $S$ and $P_\Delta$ can be directly obtained from the indices of purity.

The study of the relation between the relevant parameters $P_\Delta$ and $S$ leads to the result that all scattering media must satisfy some universal constraints [193,201]. These constraints apply to both classical and quantum scattering processes and have important applications in fields like quantum communication where depolarization is related with decoherence.

Moreover, given the practical importance of having at hand suitable techniques for improving the analysis of polarimetric images, as we have indicated previously, some recent works [108,115] have dealt with the concept of Shannon entropy and Kullback relative entropy for optical waves, and their relations with the $n$-dimensional degrees of polarization. We see that these concepts can be applied not only to $n$-dimensional partially polarized light, but also to the analysis of the polarimetric purity of material systems.

Following the idea introduced in the section devoted to the entropy of 3x3 coherency matrices, it is possible to extend to 4x4 coherency matrices the concept of partial entropies by defining the following quantities: $S_{2D}(P_2)$, $S_{2D}(P_1)$, and $S_{2D}(P_3)$. The latter is exclusive of $n \times n$ coherency matrices with $n \geq 4$.

Note that, by writing the eigenvalues of **H** as functions of the IPP, an explicit expression of $S(\mathbf{H})$ in terms of the IPP is obtained. This can be made either with the former definition of the IPP in Eq. (208) or, better, by using the definitive definition of the IPP for 4D coherency matrices, namely $P_1 \equiv \hat{\lambda}_1 - \hat{\lambda}_2$, $P_2 \equiv \hat{\lambda}_1 + \hat{\lambda}_2 - 2\hat{\lambda}_3$, $P_3 \equiv 1 - 4\lambda_4$.

### 5.10. Unified polarization algebra

*For updated details, see*

I. San José and J. J. Gil, "Invariant *indices of polarimetric purity*: generalized *indices* of *purity* for n× n covariance matrices," Opt. Commun. **284**, 38-47, (2011).

The intimate relation between $n \times n$ coherency matrices and the Lie groups SU($n$) has been pointed out and studied by Cloude [161-163] who has shown that this is a suitable and fruitful framework for polarization algebra.

The polarization states are characterized by means of their respective coherency matrices, which contain all the physical measurable information [202]. In fact, throughout the sections devoted to the mathematical representation of the polarimetric properties of light and optical systems, we have seen that there exists a strong and clear symmetry in the description of these properties. The expansion of the $n \times n$ coherency matrix in a basis of $n$ trace-orthogonal Hermitian matrices leads to the corresponding $n$D Stokes parameters. The overall polarimetric purity is defined by means of the degree of purity $P_{nD}$, whereas the detailed information on polarimetric purity is given by the corresponding ($n$-1) set of indices of purity. The relation between the norms associated with the coherency matrix provides an appropriate purity criterion.

Thus, if $\mathbf{\Omega}$ is a $n \times n$ coherency matrix and $\mathbf{\Theta}_{ij}\ (i, j = 1, n)$ constitute a basis of Hermitian trace-orthogonal $n \times n$ matrices composed of the $n-1$ traceless generators of the SU($n$) group plus the identity matrix, all of them satisfying $\mathbf{\Theta}_{ij}^2 = \mathbf{1}$ and $\mathrm{tr}(\mathbf{\Theta}_{ij}\mathbf{\Theta}_{kl}) = n\,\delta_{ik}\delta_{jl}$, the real coefficients $c_{ij} = \mathrm{tr}(\mathbf{\Theta}_{ij}\mathbf{\Omega})$ of the expansion

$$\mathbf{\Omega} = \frac{1}{n}\sum_{i,j=1}^{n} c_{ij}\mathbf{\Theta}_{ij}, \tag{214}$$

are the measurable quantities ($n$D Stokes parameters) characterizing completely the polarimetric properties of the system.

The following relations are satisfied

$$\|\mathbf{\Omega}\|_2^2 = \frac{1}{n}\|\mathbf{C}\|_2^2, \quad \|\mathbf{\Omega}\|_0 = c_{11}, \quad \frac{1}{n}\|\mathbf{\Omega}\|_0^2 \leq \|\mathbf{\Omega}\|_2^2 \leq \|\mathbf{\Omega}\|_0^2, \tag{215}$$

where $\|\ \|_2$ stands for Euclidean norm; the norm $\|\ \|_0$ is defined as $\|\mathbf{\Omega}\|_0 \equiv \mathrm{tr}\mathbf{\Omega} = \left\|\sqrt{\mathbf{\Omega}}\right\|_2^2$, and **C** is the matrix whose elements are $c_{ij}$. Therefore we can state the following purity criterion: given a

coherency matrix $\mathbf{\Omega}$, $\|\mathbf{\Omega}\|_2 = \|\mathbf{\Omega}\|_0$ is a necessary and sufficient condition for $\mathbf{\Omega}$ to represent a pure system.

The degree of purity is defined as the following invariant nondimensional quantity

$$P_{nD} = \sqrt{\frac{1}{n-1}\left(\frac{n\|\mathbf{\Omega}\|_2^2}{\|\mathbf{\Omega}\|_0^2} - 1\right)}, \tag{216}$$

which gives a global measure of the purity of the system, so that $0 \leq P_{nD} \leq 1$. The minimum value of the degree of purity $P_{nD} = 0$ corresponds to a fully random system, whereas the maximum value $P_{nD} = 1$ corresponds to a pure system.

The degree of purity can also be expressed as

$$P_{nD}^2 = \frac{1}{n-1}\sum_{\substack{i,j=1\\i<j}}^{n} p_{ij}^2, \quad p_{ij} \equiv \frac{\lambda_i - \lambda_j}{\mathrm{tr}\,\mathbf{H}}. \tag{217}$$

This has a particular relevance because it provides information similar to the entropy, but in a simpler and meaningful way. Through the indices of purity, the invariant information is structured in an optimum manner, so that these results can be useful in other fields of Physics. For example, in the same way that the Bloch equations lead to a set of parameters with the same mathematical properties as the Stokes parameters [46,203-205], the degree of purity and the indices of purity can be applied to $n$-dimensional density matrices characteristics of $n$-level systems [180].

In general, the above-mentioned invariant quantities can be applied to any problem susceptible to be modeled by means of $n \times n$ positive semidefinite Hermitian matrices. Therefore, $n-1$ indices of purity $P_i$ are necessary for the complete characterization of the polarimetric purity of the system.

Concerning the different decompositions of the coherency matrix $\mathbf{\Omega}$, it is straightforward to generalize the spectral, characteristic, and arbitrary decompositions. Unlike the spectral decomposition, which is widely known and used, the two others are generally not so known and we emphasize the high potential of their applications in polarimetry and in other fields.

General expressions for the IPP for $n$D coherency matrices can be found in [I. San José and J. J. Gil, "Invariant *indices of polarimetric purity*: generalized *indices* of *purity* for n× n covariance matrices," Opt. Commun. **284**, 38-47, (2011)]

$$P_k = \sum_{i=1}^{k} \hat{\lambda}_i - k\hat{\lambda}_k,$$

so that, with this definitive definition, the IPP are constrained to the nested limits

$$0 \leq P_1 \leq ... \leq P_{n-1} \leq 1.$$

### 5.11. Macroscopic polarimetric behaviors of material media.

In previous sections, we have analyzed the properties of retarders and diattenuators. Now we consider the overall properties of systems composed of an incoherent mixture of pure elements (incoherent parallel combination).

An alternative view of the Stokes-Mueller model is based on a consideration of the spectral Stokes vector $\mathbf{s}(\nu)$ (related to the corresponding spectral coherency matrix). Under the assumption of cross-spectral purity [7,118], $\mathbf{s}(\nu)$ can be expressed as

$$\mathbf{s}(\nu) = g(\nu)\overline{\mathbf{s}}, \tag{218}$$

where $g(\nu)$ is the normalized spectral distribution of the light beam and $\overline{\mathbf{s}}$ is a frequency-independent Stokes vector. The medium is represented by its corresponding *spectral Mueller-Jones matrix* $\mathbf{M}_J(\nu)$ so that the spectral stokes vector of the outgoing light is given by

$$\mathbf{s}'(\nu) = \mathbf{M}_J(\nu)\mathbf{s}(\nu), \tag{219}$$

and, by integrating over all frequencies,

$$\overline{\mathbf{s}}' = \mathbf{M}_J \overline{\mathbf{s}}, \tag{220}$$

where $\overline{\mathbf{s}}$ is the resulting integrated Stokes vector and $\mathbf{M}$ is the integrated Mueller matrix defined by Dlugnikov [118]

$$\overline{\mathbf{s}}' \equiv \int_0^\infty \mathbf{s}'(\nu)\, d\nu, \quad \mathbf{M} \equiv \int_0^\infty g(\nu)\mathbf{M}_J(\nu)\, d\nu. \tag{221}$$

By means of this formulation, some authors have studied the effects of dispersive media on the degree of polarization of the transmitted light [206,117,119].

Let us now analyze the polarizance/diattenuation [131,149,200,207] properties of $\mathbf{M}$. All the information on diattenuation is included in its first row, whereas all information on polarizance is included in its first column. Taking into account the reciprocity properties of Mueller matrices, when appropriate we will refer to the polarizance vector $\mathbf{P}$ as the *forward polarizance vector* and to the diattenuation vector $\mathbf{D}$ as the *reverse polarizance vector*

$$\mathbf{P} \equiv \frac{1}{m_{00}}(m_{10}, m_{20}, m_{30})^T, \quad \mathbf{D} \equiv \frac{1}{m_{00}}(m_{01}, m_{02}, m_{03})^T, \tag{222}$$

whose magnitudes are the *forward polarizance* (polarizance) $P$ and the *reverse polarizance* (diattenuation) $D$.

The polarizances are restricted to the following values

$$0 \leq P \leq 1, \quad 0 \leq D \leq 1, \tag{223}$$

Following the notation suggested by Xing [152] and Lu and Chipman [149,208], a Mueller matrix can be written as

$$\mathbf{M} = m_{00}\begin{pmatrix} 1 & \mathbf{D}^T \\ \mathbf{P} & \mathbf{m} \end{pmatrix}; \quad \mathbf{m} \equiv \frac{1}{m_{00}}\begin{pmatrix} m_{11} & m_{12} & m_{13} \\ m_{21} & m_{22} & m_{23} \\ m_{31} & m_{32} & m_{33} \end{pmatrix}. \tag{224}$$

Taking into account the expressions for the extreme values of the transmittance

$$m_{00}(1+P) \leq 1, \quad m_{00}(1+D) \leq 1, \tag{225}$$

it turns out that $\mathbf{D}$ and $\mathbf{P}$ provide all the information on forward and reverse diattenuation respectively.

In general, depolarization and diattenuation appear in a combined form. Pure systems are characterized by their respective Mueller-Jones matrices and do not produce depolarization effects on incoming totally polarized light. Mueller-Jones matrices are Mueller matrices that satisfy $P_\Delta(\mathbf{M}) = 1$ and, consequently, they satisfy the property $P = D$. Incoherent mixtures (parallel combinations) of pure systems produce depolarizing effects, which can appear combined with retardation and polarization properties. It is important to observe that $P = D$ is a necessary but not a sufficient condition for a Mueller matrix to be a Mueller-Jones matrix. Moreover, $P \neq D$ implies that the system is nonpure, while $P = D = 1$ implies total purity.

It is interesting to consider some cases that are representative of different physical behaviors of material media.

a) $P > 0$, $D = 0$. The system exhibits forward polarizance, zero reverse polarizance and depolarization. An example of this is a system composed of an ideal depolarizer followed by a total polarizer.

$$\mathbf{M}_D(\beta,0,1,1)\begin{pmatrix} 1 & 0 & 0 & 0 \\ 0 & 0 & 0 & 0 \\ 0 & 0 & 0 & 0 \\ 0 & 0 & 0 & 0 \end{pmatrix} = \begin{pmatrix} m_{00} & 0 & 0 & 0 \\ m_{10} & 0 & 0 & 0 \\ m_{20} & 0 & 0 & 0 \\ m_{30} & 0 & 0 & 0 \end{pmatrix}.$$

b) $D > 0$, $P = 0$. The system exhibits reverse polarizance, zero forward polarizance and depolarization. An example is a system composed of a total polarizer followed by an ideal depolarizer. The resulting matrix has the form of the transposed matrix of the above matrix.

c) $P = 1$, $0 < D < 1$. The system (denoted as a *depolarizing polarizer*) exhibits maximum forward polarizance, reverse polarizance and depolarization. An example is a serial combination constituted by a partial polarizer, an ideal depolarizer and a total polarizer, in this order. Although the system produce totally polarized light for all incoming states, the "internal" depolarizing effect affects the corresponding Mueller matrix, so that this nonpure system cannot be represented by a Mueller-Jones matrix.

d) $0 < P < 1$, $D = 1$. The system (denoted as a *depolarizing analyzer*) exhibits maximum reverse polarizance, forward polarizance and depolarization. An example is a serial combination constituted by a total polarizer, an ideal depolarizer and a partial polarizer, in that order. Obviously, this case is the *reciprocal* of the previous one.

e) $0 < P < 1$, $0 < D < 1$, $P \neq D$. The system exhibits forward polarizance, reverse polarizance and depolarization. An example is a serial combination constituted by a diattenuator (partial polarizer), an ideal depolarizer and a diattenuator (partial polarizer), in that order.

f) $0 < P = D < 1$,. The system exhibits equal polarizances. This condition does not determine if the system exhibits depolarization or not. The global purity of the system is given by $P_\Delta$.

g) $P = D = 1$. The system (denoted as a *depolarizer*) exhibits maximum forward polarizance and maximum reverse polarizance and is equivalent to a serial combination of pure or nonpure elements placed between two total polarizers. It is important to stress that the system behaves as pure regardless of the possible existence of depolarizers placed between the total polarizers. An interesting example is an ideal depolarizer placed between two polarizers

$$\mathbf{M} = \mathbf{M}'_D(\beta,1,1)\begin{pmatrix} 1 & 0 & 0 & 0 \\ 0 & 0 & 0 & 0 \\ 0 & 0 & 0 & 0 \\ 0 & 0 & 0 & 0 \end{pmatrix} \mathbf{M}_D(\theta,1,1).$$

It is easy to show that in this case $P_\Delta = 1$, despite the effect of the intermediate depolarizer, which affects only to the transmittance of the complete system represented by $\mathbf{M}$.

### 5.12. The generalized polar decomposition

Some models for the serial decomposition of Stokes matrices have been introduced by Xing [152] and Sridhar and Simon [160] on the basis of the Stokes criterion. As van der Mee has pointed out [167], these decompositions are not totally general. Moreover, Lu and Chipman have emphasized the necessity of serial decompositions where the basic physical behaviors (diattenuation/polarization, retardation and depolarization) appear explicitly in the different components of the equivalent serial system and have presented an algorithm to decompose a Mueller matrix into a serial combination of a diattenuator, a retarder and a partial polarizer-depolarizer [149]. This *generalized polar decomposition* of a Mueller matrix is currently used for many purposes concerning the physical interpretation of experimental measures as well as for studying physical models of a great variety of material targets [209-211].

5.12.1. The Lu-Chipman decomposition

In this sub-section we summarize the main results of the work of Lu and Chipman [149].

Taking into account the expression of a Mueller matrix $\mathbf{M}$ given by Eq. (224) these authors have shown that $\mathbf{M}$ can be written as [149]

$$\mathbf{M} = m_{00} \begin{pmatrix} 1 & \mathbf{D}^T \\ \mathbf{P} & \mathbf{m} \end{pmatrix} = m_{00}\, \hat{\mathbf{M}}_{\Delta P}\, \mathbf{M}_R\, \hat{\mathbf{M}}_D,$$

$$\hat{\mathbf{M}}_{\Delta P} \equiv \begin{pmatrix} 1 & \mathbf{0}^T \\ \mathbf{P}_\Delta & \mathbf{m}_\Delta \end{pmatrix},\quad \mathbf{M}_R \equiv \begin{pmatrix} 1 & \mathbf{0}^T \\ \mathbf{0} & \mathbf{m}_R \end{pmatrix},\quad \hat{\mathbf{M}}_D \equiv \begin{pmatrix} 1 & \mathbf{D}^T \\ \mathbf{D} & \mathbf{m}_D \end{pmatrix},$$
(226)

where $\mathbf{0}^T \equiv (0,0,0)$ and:

- Matrix $\hat{\mathbf{M}}_D$ is proportional to a diattenuator.

- Matrix $\mathbf{M}_R$ represents a retarder.

- Matrix $\hat{\mathbf{M}}_{\Delta P}$ is proportional to a polarizer-depolarizer with zero diattenuation. This class of system exhibits simultaneously both depolarizing and polarizing properties.

As indicated in Eq. (226), the matrix $\hat{\mathbf{M}}_D$ can be calculated from the diattenuation vector $\mathbf{D}$ of $\mathbf{M}$. Moreover,

$$\mathbf{P}_\Delta = \frac{1}{1-D^2}(\mathbf{P} - \mathbf{m}\mathbf{D}).$$
(227)

For the calculation of $\hat{\mathbf{M}}_{\Delta P}$ and $\mathbf{M}_R$ the cases of $\mathbf{M}$ nonsingular and $\mathbf{M}$ singular should be distinguished

*a)* $\mathbf{M}$ nonsingular

In this case $\hat{\mathbf{M}}_D$ is nonsingular, so that it is possible to define the matrix $\mathbf{M}' = \mathbf{M}\,\mathbf{M}_D^{-1}$. This matrix lacks diattenuation and can be written as

$$\mathbf{M}' = \frac{1}{m_{00}}\hat{\mathbf{M}}_{\Delta P}\mathbf{M}_R = \frac{1}{m_{00}}\begin{pmatrix} 1 & \mathbf{0}^T \\ \mathbf{P}_\Delta & \mathbf{m}' \end{pmatrix},$$
(228)

and the calculation of the matrices $\hat{\mathbf{M}}_{\Delta P}$ and $\mathbf{M}_R$ can be performed through singular value decomposition of the matrix $\mathbf{m}'$

$$\mathbf{m}' = \mathbf{U}\left[\mathbf{D}(d_1, d_2, \varepsilon d_3)\right]\mathbf{V}, \quad d_1 \geq d_2 \geq d_3 \geq, \quad \varepsilon \equiv \frac{\det \mathbf{M}}{|\det \mathbf{M}|}, \tag{229}$$

where $\mathbf{U}, \mathbf{V}$ are $3 \times 3$ proper orthogonal matrices and $d_i$ are the eigenvalues of the positive semi-definite symmetric matrix $\mathbf{m}'\mathbf{m}'^T$.

Finally the matrices $\mathbf{m}_{\Delta P}$ and $\mathbf{m}_R$ are given by

$$\mathbf{m}_{\Delta P} = \mathbf{U}\,\mathbf{D}(d_1, d_2, \varepsilon d_3)\,\mathbf{U}^T, \quad \mathbf{m}_R = \mathbf{U}\mathbf{V}. \tag{230}$$

We observe that $\mathbf{m}_{\Delta P}$ is a symmetric matrix and $\mathbf{m}_R$ is a proper rotation matrix ($\det \mathbf{m}_R = 1$). Moreover, $\mathbf{m}_{\Delta P}$ and $\mathbf{m}_R$ are the symmetric and orthogonal components corresponding to the polar decomposition of $\mathbf{m}'$. This is the origin of the name *generalized polar decomposition* given by Lu and Chipman to the decomposition (226) despite the fact that, as we will see in the next section, the direct application of the polar decomposition to $\mathbf{M}$ results in an unphysical orthogonal component.

It should be noted that, in general, matrices $\hat{\mathbf{M}}_D$ and $\hat{\mathbf{M}}_{\Delta P}$ do not satisfy the passivity conditions [see J. J. Gil, "Transmittance constraints in serial decompositions of Mueller matrices. The arrow form of a Mueller matrix," J. Opt. Soc. Am. A **30**, 701-707 (2013)].

A physical decomposition in terms of passive components should be expressed in the form

$$\mathbf{M} = (c\mathbf{M}_{\Delta P})(b\mathbf{M}_R)(a\hat{\mathbf{M}}_D); \quad a,b,c \geq 0, \quad abc = m_{00}, \tag{231}$$

where $b \leq 1$ and $a$, $c$ must be small enough to satisfy

$$a(1+D) \leq 1, \quad c(1+P_\Delta) \leq 1. \tag{232}$$

Thus, the following condition is required for a passive realizable decomposition

$$(1+D)(1+P_\Delta) \leq 1/m_{00}. \tag{233}$$

It is clear that the components of the generalized polar decomposition are Stokes matrices. Nevertheless, the conventional form of the generalized polar decomposition assigns the coefficient $m_{00}$ to the equivalent diattenuator $\mathbf{M}_D$ so that the resulting polarizer-depolarizer component does not satisfy the reverse passivity condition. Thus, further work is required in clarifying the conditions under which all the components of the generalized polar decomposition are Mueller matrices, i.e. they satisfy the passivity conditions as well as the covariance conditions.

Provided Eq. (233) is satisfied, the Mueller matrix $\mathbf{M}$ can be decomposed as a product $\mathbf{M} = \mathbf{M}_{\Delta P} \mathbf{M}_R \mathbf{M}_D$ of a diattenuator $\mathbf{M}_D$, a retarder $\mathbf{M}_R$ and a polarizing-depolarizer $\mathbf{M}_{\Delta P}$

*b)* $\mathbf{M}$ singular

*For a comprehensive analysis of singular Mueller matrices, see*
J. J. Gil, R. Ossikovski and I. San José "Singular Mueller matrices," J. Opt. Soc. Am. A **33**, 600-609, (2016).

In this case, at least one of the following equalities is satisfied: $P = 1$, $D = 1$ and $\mathbf{M}$ can be decomposed as

$$\mathbf{M} = m_{00} \begin{pmatrix} 1 & \mathbf{D}^T \\ \mathbf{P} & \mathbf{P} \otimes \mathbf{D}^T \end{pmatrix} = m_{00} \hat{\mathbf{M}}_I \mathbf{M}_R \hat{\mathbf{M}}_D,$$

$$\hat{\mathbf{M}}_I \equiv \begin{pmatrix} 1 & \mathbf{0}^T \\ \mathbf{0} & P\,\mathbf{D}(1,1,1) \end{pmatrix}, \mathbf{M}_R \equiv \begin{pmatrix} 1 & \mathbf{0}^T \\ \mathbf{0} & \mathbf{m}_R \end{pmatrix}, \hat{\mathbf{M}}_D \equiv \begin{pmatrix} 1 & \mathbf{D}^T \\ \mathbf{D} & \mathbf{D} \otimes \mathbf{D}^T \end{pmatrix}, \tag{234}$$

where $m_{00}\hat{\mathbf{M}}_D$ is the Mueller matrix of a normal diattenuator and $\mathbf{M}_I$ represents a *diagonal depolarizer*.

5.12.2. The forward and reverse generalized polar decompositions

Let us now consider the possible different decompositions obtainable by modifying the order of the three components. Lu and Chipman [149], Morio and Goudail [212] and Ossikovski, De Martino and Guyot [213] have studied the six possible decompositions as well as some examples where certain decompositions lead to unphysical components. According with the results of Morio and Goudail [212], the following decompositions, where the diattenuator is placed before the polarizer-depolarizer (*forward generalized polar decompositions*), always lead to components represented by Stokes matrices

$$\hat{\mathbf{M}} = \hat{\mathbf{M}}_{\Delta P1}\mathbf{M}_{R1}\hat{\mathbf{M}}_{D1} = \hat{\mathbf{M}}_{\Delta P2}\hat{\mathbf{M}}_{D2}\mathbf{M}_{R2} = \mathbf{M}_{R5}\hat{\mathbf{M}}_{\Delta P5}\hat{\mathbf{M}}_{D5}, \tag{235}$$

whereas the other possible decompositions, where the polarizer-depolarizer is placed before the diattenuator (*reverse generalized polar decompositions*), are not always composed of Stokes matrices

$$\hat{\mathbf{M}} = \mathbf{M}_{R3}\hat{\mathbf{M}}_{D3}\hat{\mathbf{M}}_{\Delta D3} = \hat{\mathbf{M}}_{D4}\mathbf{M}_{R4}\hat{\mathbf{M}}_{\Delta D4} = \hat{\mathbf{M}}_{D6}\hat{\mathbf{M}}_{\Delta D6}\mathbf{M}_{R6}. \tag{236}$$

Concerning these kinds of decompositions and their physical validity it is necessary to check the passivity conditions for $\mathbf{M}$ and for the matrices of the different components. Some examples have been used in the literature where $m_{00}=1$ and $D \neq 0$ (or $D \neq 0$). These values never correspond to a Mueller matrix because for such cases at least one of the passivity conditions is not satisfied.

Prior to analyze other possible decompositions of $\mathbf{M}$, let us consider the Stokes matrix of a *diattenuator-depolarizer*

$$\mathbf{M}_{\Delta D} \equiv \begin{pmatrix} 1 & \mathbf{D}_\Delta^T \\ \mathbf{0} & \mathbf{m}_{\Delta D} \end{pmatrix}, \tag{237}$$

which exhibits zero polarizance, nonzero diattenuation and depolarization. We observe that to convert this Stokes matrix into a Mueller matrix, it should be multiplied by a coefficient $\alpha \leq 1/(1+D_\Delta)$.

In a previous section we have shown that, given a Mueller matrix $\mathbf{M}$, $\mathbf{M}^T$ is also a Mueller matrix. Thus, by applying the Lu-Chipman decomposition to $\mathbf{M}^T$ we obtain

$$\begin{aligned}\hat{\mathbf{M}} &= \left(\hat{\mathbf{M}}'_{\Delta P1}\mathbf{M}'_{R1}\hat{\mathbf{M}}'_{D1}\right)^T = \left(\hat{\mathbf{M}}'_{\Delta P2}\hat{\mathbf{M}}'_{D2}\mathbf{M}'_{R2}\right)^T = \left(\mathbf{M}'_{R5}\hat{\mathbf{M}}'_{\Delta P5}\hat{\mathbf{M}}'_{D5}\right)^T \\ &= \hat{\mathbf{M}}_{D4}\mathbf{M}_{R4}\hat{\mathbf{M}}_{\Delta D4} = \mathbf{M}_{R3}\hat{\mathbf{M}}_{D3}\hat{\mathbf{M}}_{\Delta D3} = \hat{\mathbf{M}}_{D6}\hat{\mathbf{M}}_{\Delta D6}\mathbf{M}_{R6}.\end{aligned} \tag{238}$$

We see that the operation of transposing does no affect the nature of the matrices of the pure components, whereas leads to replacing the polarizing-depolarizers by diattenuating-depolarizers.

Some relations between the pure components of each one of the three sets of decompositions can be derived from the polar decomposition of a pure Mueller matrix. Experimental validation of one of the forms of the reverse generalized polar decomposition has been presented recently by Anastasiadou, Ben Hatit, Ossikovski, Guyot and De Martino [214]

### 5.12.3. The normal decomposition of a Mueller matrix

*For updated details, see*

R. Ossikovski, "Analysis of depolarizing Mueller matrices through a symmetric decomposition," J. Opt. Soc. Am. A **26**, 1109 1118 (2009).

R. Ossikovski, "Canonical forms of depolarizing Mueller matrices," J. Opt. Soc. Am. A **27**, 123-130 (2010).

J. J. Gil, "Components of purity of a Mueller matrix," J. Opt. Soc. Am. A **28**, 1578-1585 (2011).

J. J. Gil, "Transmittance constraints in serial decompositions of Mueller matrices. The arrow form of a Mueller matrix," J. Opt. Soc. Am. A **30**, 701-707 (2013)

Although the Lu-Chipman decomposition has been demonstrated to be a powerful tool for analyzing measured Mueller matrices, the depolarizer component combines polarizing and depolarizing properties. Thus, an additional effort to separate the retarding, polarizing and depolarizing properties is desirable.

The general characterization of the Mueller matrix introduced in Sec 5.4 reveals a fundamental symmetry of Mueller matrices in the sense of any general property of a Muller matrix $\mathbf{M}$ can also be appropriately formulated for its transposed matrix $\mathbf{M}^T$. We now translate this basic symmetry to the Lu-Chipman decomposition in order to introduce a general decomposition of a Mueller matrix where the depolarizing properties are isolated from that related with retardation and polarization-diattenuation properties.

Let us consider a Mueller matrix $\mathbf{M}$ and apply one of the forward forms of the Lu Chipman decomposition, so that

$$\mathbf{M} = m_{00}\,\hat{\mathbf{M}}_{\Delta P2}\,\hat{\mathbf{M}}_{D2}\,\mathbf{M}_{R2}\,. \tag{239}$$

Moreover $\hat{\mathbf{M}}_{\Delta P2}$ can be expressed as the transposed matrix of the matrix of a diattenuator-depolarizer $\hat{\mathbf{M}}_{\Delta D2}$ with the form given by Eq. (237). Thus, the Lu-Chipman decomposition can be applied to $\hat{\mathbf{M}}_{\Delta D2}$ and obtain

$$\hat{\mathbf{M}}_{\Delta P2} = \hat{\mathbf{M}}_{\Delta D2}^T = \left[\hat{\mathbf{M}}'_{\Delta P}\,\hat{\mathbf{M}}'_D\,\mathbf{M}_R\right]^T = \mathbf{M}'^T_R\,\hat{\mathbf{M}}'^T_D\,\mathbf{M}'^T_{\Delta P}, \tag{240}$$

where $\mathbf{M}'^T_R$ represents a retarder, $\hat{\mathbf{M}}'^T_D$ represents a diattenuator and $\mathbf{M}'^T_{\Delta P}$ represents a diattenuating-depolarizer.

Contrary to what was stated in the original version of this work, it should be noted that, in general, $\mathbf{M}'^T_{\Delta P}$ exhibits nonzero forward and reverse polarizance, that is,

$$\mathbf{M} = \mathbf{M}'_{J2}\,\mathbf{M}_{\Delta PD}\,\mathbf{M}'_{J1},$$
$$\mathbf{M}_{\Delta PD} = \begin{pmatrix} 1 & \mathbf{D}'^T_\Delta \\ \mathbf{P}'_\Delta & \mathbf{m}_{\Delta PD} \end{pmatrix} \neq \begin{pmatrix} 1 & \mathbf{0}^T \\ \mathbf{0} & \mathbf{m}_{\Delta PD} \end{pmatrix}. \tag{241}$$

As demonstrated in

Y. Bolshakov, C. V. M. van der Mee, and A. C. M. Ran, "Polar decompositions in finite dimensional indefinite scalar product spaces: special cases and applications," In *Operator Theory: Advances and Applications, I*. Gohberg, ed. Birkhäuser Verlag, **87**, 61–94 (1996).

Y. Bolshakov, C. V. M. van der Mee, and A. C. M. Ran, "Errata for: Polar decompositions in finite dimensional indefinite scalar product spaces: special cases and applications," *Integral Equation Oper. Theory* **27**, 497–501 (1997).

A. V. Gopala Rao, S. Mallesh, and Sudha Shenoy, "On the algebraic characterization of a Mueller matrix in polarization optics. I. Identifying a Mueller matrix from its N-matrix," *J. Mod. Opt.* **45**, 955-987 (1998).

A. V. Gopala Rao, K. S. Mallesh, and Sudha Shenoy, "On the algebraic characterization of a Mueller matrix in polarization optics. II. Necessary and sufficient conditions for Jones derived Mueller matrices," *J. Mod. Opt.* **45**, 989-999 (1998).

R. Ossikovski, "Analysis of depolarizing Mueller matrices through a symmetric decomposition," J. Opt. Soc. Am. A **26**, 1109 1118 (2009).

R. Ossikovski, "Canonical forms of depolarizing Mueller matrices," J. Opt. Soc. Am. A **27**, 123-130 (2010).

any Mueller matrix admits the following *normal form decomposition*

$$\mathbf{M} = m_{00}\ \hat{\mathbf{M}}_{J2}\ \hat{\mathbf{M}}_{\Delta}\ \hat{\mathbf{M}}_{J1}, \tag{242}$$

which, by applying the polar decomposition to the pure serial components $\hat{\mathbf{M}}_{J1}$ and $\hat{\mathbf{M}}_{J2}$, leads to the *symmetric decomposition*

$$\mathbf{M} = m_{00}\ \mathbf{M}_{R2}\ \hat{\mathbf{M}}_{D2}\ \hat{\mathbf{M}}_{\Delta}\ \hat{\mathbf{M}}_{D1}\ \mathbf{M}_{R1},$$

where $\mathbf{M}_{R1}$ and $\mathbf{M}_{R2}$ represent respective retarders; $\hat{\mathbf{M}}_{D1}$ and $\hat{\mathbf{M}}_{D2}$ are normalized matrices of respective diattenuators, and $\hat{\mathbf{M}}_{\Delta}$ represents a canonical depolarizer [see R. Ossikovski, "Canonical forms of depolarizing Mueller matrices," J. Opt. Soc. Am. A **27**, 123-130 (2010)].

Thus, the previous result can be formulated by means of the following statement: *Any system is polarimetrically equivalent to a serial combination constituted by a canonical depolarizer placed between two pure components*. Or, in other words: *Any Mueller matrix can be expressed as the ordered product of the Mueller matrices of an "output pure system" a "canonical depolarizer" and an "input pure system"*.

As we have seen in the section devoted to the polar decomposition of a pure system, it is always possible to make the different alternative choices for the order of the diattenuator and the retarder of the pure equivalent components and, consequently, any Mueller matrix can be decomposed in the following different alternative ways

$$\begin{aligned}
\mathbf{M} &= m_{00}\mathbf{M}_{R2}\hat{\mathbf{M}}_{D2}\hat{\mathbf{M}}_{\Delta}\hat{\mathbf{M}}_{D1}\mathbf{M}_{R1} \\
&= m_{00}\hat{\mathbf{M}}'_{D2}\mathbf{M}'_{R2}\hat{\mathbf{M}}_{\Delta}\mathbf{M}'_{R1}\hat{\mathbf{M}}'_{D1} \\
&= m_{00}\hat{\mathbf{M}}'_{D2}\mathbf{M}'_{R2}\hat{\mathbf{M}}_{\Delta}\hat{\mathbf{M}}_{D1}\mathbf{M}_{R1} \\
&= m_{00}\mathbf{M}_{R2}\hat{\mathbf{M}}_{D2}\hat{\mathbf{M}}_{\Delta}\mathbf{M}'_{R1}\hat{\mathbf{M}}'_{D1}.
\end{aligned} \tag{243}$$

## 5.13. The Kernel and the arrow forms of a Mueller matrix

*For alternative invariant-equivalent forms of Mueller matrices, see*

J. J. Gil, "Components of purity of a Mueller matrix," J. Opt. Soc. Am. A **28**, 1578-1585 (2011).

J. J. Gil, "Transmittance constraints in serial decompositions of Mueller matrices. The arrow form of a Mueller matrix," J. Opt. Soc. Am. A **30**, 701-707 (2013).

J. J. Gil, "Review on Mueller matrix algebra for the analysis of polarimetric measurements," Journal of Applied Remote Sensing **8**, 081599 (2014).

J. J. Gil, "Invariant quantities of a Mueller matrix under rotation and retarder transformations," J. Opt. Soc. Am. A **33**, 52-58 (2016).

To complete the study of the general properties of Mueller matrices, along this section we consider the problem of identifying the Mueller serial component that is invariant from a physical point of view and, consequently, obtaining the corresponding set of physically invariant quantities involved in a Mueller matrix.

Although the polar decomposition is very useful for the study of pure systems, this kind of decomposition cannot be applied directly to the case of depolarizing systems because both the orthogonal and the symmetric components are not, in general, a Mueller matrix. Nevertheless, some algebraic rearrangements allow us to construct a new model where the Mueller matrix is decomposed into Mueller matrices (and not only Stokes matrices) and some invariant physical quantities appear grouped in the Kernel-matrix whose definition will be introduced below.

Let us consider a Mueller matrix $\mathbf{M}$ and its singular value decomposition (denoted from now on as SVD)

$$\mathbf{M} = \mathbf{X}\mathbf{D}_\Delta \mathbf{Y}, \tag{244}$$

where $\mathbf{X}, \mathbf{Y}$ are 4x4 orthogonal matrices and $\mathbf{D}_\Delta$ is the diagonal matrix whose elements are the nonnegative, ordered, singular values

$$d_0 \geq d_1 \geq d_2 \geq d_3. \tag{245}$$

Except for the particular case that $\mathbf{M}$ is a diagonal matrix, the orthogonal matrices $\mathbf{X}, \mathbf{Y}$ are not Mueller matrices. In fact, they can be written as products of six elementary 4x4 rotation Givens matrices

$$\mathbf{Q}_1(\varphi_1) = \begin{pmatrix} \cos\varphi_1 & 0 & 0 & \sin\varphi_1 \\ 0 & 1 & 0 & 0 \\ 0 & 0 & 1 & 0 \\ -\sin\varphi_1 & 0 & 0 & \cos\varphi_1 \end{pmatrix}, \; \mathbf{Q}_2(\varphi_2) = \begin{pmatrix} \cos\varphi_2 & 0 & \sin\varphi_2 & 0 \\ 0 & 1 & 0 & 0 \\ -\sin\varphi_2 & 0 & \cos\varphi_2 & 0 \\ 0 & 0 & 0 & 1 \end{pmatrix}, \; \mathbf{Q}_3(\varphi_3) = \begin{pmatrix} \cos\varphi_3 & \sin\varphi_3 & 0 & 0 \\ -\sin\varphi_3 & \cos\varphi_3 & 0 & 0 \\ 0 & 0 & 1 & 0 \\ 0 & 0 & 0 & 1 \end{pmatrix},$$

$$\mathbf{G}_1(\alpha_1) = \begin{pmatrix} 1 & 0 & 0 & 0 \\ 0 & 1 & 0 & 0 \\ 0 & 0 & \cos\alpha_1 & \sin\alpha_1 \\ 0 & 0 & -\sin\alpha_1 & \cos\alpha_1 \end{pmatrix}, \; \mathbf{G}_2(\alpha_2) = \begin{pmatrix} 1 & 0 & 0 & 0 \\ 0 & \cos\alpha_2 & 0 & \sin\alpha_2 \\ 0 & 0 & 1 & 0 \\ 0 & -\sin\alpha_2 & 0 & \cos\alpha_2 \end{pmatrix}, \; \mathbf{G}_3(\alpha_3) = \begin{pmatrix} 1 & 0 & 0 & 0 \\ 0 & \cos\alpha_3 & \sin\alpha_3 & 0 \\ 0 & -\sin\alpha_3 & \cos\alpha_3 & 0 \\ 0 & 0 & 0 & 1 \end{pmatrix}. \tag{246}$$

It should be noted that the order of the matrix factors can be chosen arbitrarily. Then $\mathbf{X}, \mathbf{Y}$ can be written as

$$\begin{aligned}\mathbf{X} &= \mathbf{G}_1(\alpha_1)\mathbf{G}_2(\alpha_2)\mathbf{G}_3(\alpha_3)\mathbf{Q}_1(\varphi_1)\mathbf{Q}_2(\varphi_2)\mathbf{Q}_3(\varphi_3), \\ \mathbf{Y} &= \mathbf{Q}_3^T(\varphi_6)\mathbf{Q}_2^T(\varphi_5)\mathbf{Q}_1^T(\varphi_4)\mathbf{G}_3^T(\alpha_4)\mathbf{G}_2^T(\alpha_5)\mathbf{G}_1^T(\alpha_6).\end{aligned} \tag{247}$$

or

$$\mathbf{X} = \mathbf{R}\mathbf{U}, \quad \mathbf{Y} = \mathbf{V}\mathbf{L}, \tag{248}$$

where

$$\begin{aligned}\mathbf{R} &\equiv \mathbf{G}_1(\alpha_1)\mathbf{G}_2(\alpha_2)\mathbf{G}_3(\alpha_3), \quad \mathbf{L} \equiv \mathbf{G}_3^T(\alpha_4)\mathbf{G}_2^T(\alpha_5)\mathbf{G}_1^T(\alpha_6), \\ \mathbf{U} &\equiv \mathbf{Q}_1(\varphi_1)\mathbf{Q}_2(\varphi_2)\mathbf{Q}_3(\varphi_3), \quad \mathbf{V} \equiv \mathbf{Q}_3^T(\varphi_6)\mathbf{Q}_2^T(\varphi_5)\mathbf{Q}_1^T(\varphi_4).\end{aligned} \tag{249}$$

Matrices $\mathbf{R}$ and $\mathbf{L}$ are Mueller matrices that represent retarders, whereas matrices $\mathbf{U}, \mathbf{V}$ are not Mueller matrices or Stokes matrices. It is easy to see that $\mathbf{U}, \mathbf{V}$ they do not satisfy some of the conditions to be Mueller matrices. For example $\mathbf{Q}_1$ applied to the Stokes vector $(1,1,0,0)^T$ is not a Stokes vector.

The polar decompositions of **M** can be written as

$$\mathbf{M} = \mathbf{KW}; \quad \mathbf{K} \equiv \sqrt{\mathbf{MM}^T} = \mathbf{RUDU}^T\mathbf{R}^T, \quad \mathbf{W} \equiv \mathbf{RUVL},$$
$$\mathbf{M} = \mathbf{W'K'}; \quad \mathbf{W'} \equiv \mathbf{RUVL}, \quad \mathbf{K'} \equiv \sqrt{\mathbf{M}^T\mathbf{M}} = \mathbf{L}^T\mathbf{V}^T\mathbf{DVL},$$
(250)

where, in general, the symmetric components **K**, **K'** and the orthogonal components **W**, **W'** are not Mueller matrices. Nevertheless, it is possible to arrange the products so that the Mueller matrix can be written as (Fig.12)

$$\mathbf{M} = \mathbf{RZL},$$
(251)

where

$$\mathbf{Z} \equiv \mathbf{U}\mathbf{D}_\Delta \mathbf{V}.$$
(252)

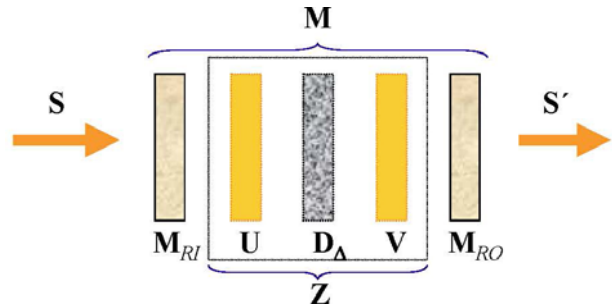

Fig. 12. The kernel form of a Mueller matrix.

We observe that this kernel matrix $\mathbf{Z} \equiv \mathbf{R}^T\mathbf{M}\mathbf{L}^T$ is a product of Mueller matrices and, hence, is a Mueller matrix. Thus, all the three components of the decomposition given by Eq. (251) are Mueller matrices (i.e. satisfy the passivity conditions as well as the covariance conditions). Therefore

$$\mathbf{M} = \mathbf{M}'_{RO}\mathbf{Z}\mathbf{M}'_{RI}, \quad (\mathbf{M}'_{RO} \equiv \mathbf{R}, \; \mathbf{M}'_{RI} \equiv \mathbf{L}).$$
(253)

It has been demonstrated in [J. J. Gil, "Invariant quantities of a Mueller matrix under rotation and retarder transformations," J. Opt. Soc. Am. A **33**, 52-58 (2016)], that there are infinite matrices that, as occurs with **Z**, can be obtained through *dual retarder transformations* of **M**, and that, in particular, the arrow form $\mathbf{M}_A$ obtained through the singular value decomposition of the submatrix **m** of **M**, has the genuine property that $\mathbf{M}_A$ lacks retardance and thus it only contains information on the depolarization, diattenuation and polarizance properties of **M**, besides its mean transmittance $m_{00}$,

$$\mathbf{M} \equiv m_{00}\begin{pmatrix} 1 & \mathbf{D}^T \\ \mathbf{P} & \mathbf{m} \end{pmatrix} = m_{00}\mathbf{M}_{RO}\mathbf{M}_A\mathbf{M}_{RI},$$

$$\mathbf{M}_{Ri} = \begin{pmatrix} 1 & \mathbf{0}^T \\ \mathbf{0} & \mathbf{m}_{Ri} \end{pmatrix}, \; \mathbf{m}_{Ri}^T = \mathbf{m}_{Ri}^{-1}, \; \det \mathbf{m}_{Ri} = +1, \quad i = I, O,$$

$$\mathbf{M}_A \equiv m_{00} \begin{pmatrix} 1 & \mathbf{D}_A^T \\ \mathbf{P}_A & \mathbf{m}_A \end{pmatrix}, \quad \mathbf{m}_A \equiv \begin{pmatrix} 1 & 0 & 0 \\ 0 & 1 & 0 \\ 0 & 0 & \varepsilon \end{pmatrix} \mathbf{m}_{RO} \mathbf{m} \mathbf{m}_{RI} = \begin{pmatrix} a_1 & 0 & 0 \\ 0 & a_2 & 0 \\ 0 & 0 & \varepsilon a_3 \end{pmatrix},$$

$$0 \leq |a_3| \leq a_2 \leq a_1 \leq 1, \quad \varepsilon \equiv \det \mathbf{M} / |\det \mathbf{M}|$$

$$\mathbf{P}_A = \mathbf{m}_{RO} \mathbf{P}, \quad \mathbf{D}_A = \mathbf{m}_{RI} \mathbf{D}.$$

In general, properties relative to diattenuation, polarizance and depolarization are not separable, i.e., there is no serial decomposition where different components exhibit exclusive information on each of the said properties, but they appear necessarily coupled. Nevertheless, what is always possible is to extract, and isolate, the retardation properties by means of the input and output retarders of the arrow decomposition. Therefore, the physical information contained in a Mueller matrix can be described as

1) the mean intensity coefficient $m_{00}$, i.e., the intensity transmittance of $\mathbf{M}$ for input unpolarized light;

2) the *input retarder* $\mathbf{M}_{RI}$, which is determined by three parameters, namely, the *input retardance* $\Delta_I$ and the pair of angles $(\varphi_I, \chi_I)$ determining the azimuth $\varphi_I$ and the ellipticity $\chi_I$ of the eigenstates of $\mathbf{M}_{RI}$;

3) the *output retarder* $\mathbf{M}_{RO}$, which is determined by three parameters, namely, the *output retardance* and the pair of angles $(\varphi_O, \chi_O)$ determining the azimuth $\varphi_O$ and the ellipticity $\chi_O$ of the eigenstates of $\mathbf{M}_{RO}$;

4) the *intrinsic diattenuation vector* $\mathbf{D}_A$, determined by its absolute value $D_A = D$, together with the azimuth and ellipticity $(\varphi_{DA}, \chi_{DA})$ of vector $\mathbf{D}_A$ when it is represented in the Poincaré sphere;

5) the *intrinsic polarizance vector* $\mathbf{P}_A$, determined by its absolute value $P_A = P$, together with the azimuth and ellipticity $(\varphi_{PA}, \chi_{PA})$ of vector $\mathbf{P}_A$ when it is represented in the Poincaré sphere;

6) the three *spherical purity components* $(a_1, a_2, a_3)$ that equal the respective semiaxes of the *intrinsic ellipsoid* of $\mathbf{M}$. Note that $P_S = \sqrt{(a_1^2 + a_2^2 + a_3^2)/3}$.

### 5.14. Physical invariants of a Mueller matrix

*For updated details and additional analyses, see*

J. J. Gil, "Components of purity of a Mueller matrix," J. Opt. Soc. Am. A **28**, 1578-1585 (2011).

J. J. Gil, "Transmittance constraints in serial decompositions of Mueller matrices. The arrow form of a Mueller matrix," J. Opt. Soc. Am. A **30**, 701-707 (2013).

J. J. Gil, "Review on Mueller matrix algebra for the analysis of polarimetric measurements," Journal of Applied Remote Sensing **8**, 081599 (2014).

J. J. Gil, "Invariant quantities of a Mueller matrix under rotation and retarder transformations," J. Opt. Soc. Am. A **33**, 52-58 (2016).

Having determined the serial decomposition of $\mathbf{M}$ in terms of $\mathbf{M}'_{RI}$, $\mathbf{M}'_{RO}$ and $\mathbf{Z}$, we next analyze the properties and physical information contained in matrix $\mathbf{Z}$.

Concerning the matrix $\mathbf{D}_\Delta$, it is easy to show that it is a Stokes matrix. Nevertheless, taking into account Eq. (245), the following inequality is required for $\mathbf{D}_\Delta$ to satisfy the eigenvalue conditions

$$d_0 + d_3 \geq d_1 + d_2, \tag{254}$$

While the additional the condition for $\mathbf{D}_\Delta$ to be passive is

$$d_0 \leq 1. \tag{255}$$

Moreover, **U** and **V** are not Mueller matrices, in such a manner that the serial combination given by Eq. (252) is always a Mueller matrix.

By examining the nature of these different matrices, we see that $\mathbf{D}_\Delta$ always exhibits depolarizing properties, including the case where **M** represents a pure medium. We observe that $P_\Delta(\mathbf{D}_\Delta) < 1$, except for the degenerate case $d_0 = d_1 = d_2 = d_3$ corresponding to a retarder $\mathbf{M} = d_0 \mathbf{M}'_{RO} \mathbf{M}'_{RI}$ with transmittance $d_0$. On the other hand, the non-Mueller matrices **U** and **V** produce *overpolarizing effects*, so that the distances to the origin of the points representing some states on the corresponding *P*-Images (or *DoP* surfaces) exceed the value 1.

As we have seen, all the quantitative depolarization properties are given by the three indices of purity $P_i$, which are defined from the eigenvalues of the coherency matrix **H** associated with **M** [41]. These quantities are invariant under changes of the laboratory reference axis, as well as changes of the generalized basis used for the representation of the states of polarization.

Usually, linear polarizations along the axes *XY* are used as physical reference axes, so that in the Poincaré sphere representation the Stokes parameter $s_1$ corresponds to the *X* axis, $s_2$ corresponds to the *XY* bisector axis and $s_3$ corresponds to right-handed circularly polarized light. Nevertheless, it is possible to construct different physical bases by choosing orthogonal pairs of circularly or elliptically polarized light. All the possible changes of the reference basis for the representation of Stokes vectors correspond to rotations in the Poincaré sphere and, in consequence, are performed by means of similarity transformations whose orthogonal matrix has the form of the **R**, **L** matrices, and not the form of the **U**, **V** matrices.

According to the SVD of **Z**, **U** and **V** are the *output* and *input* orthogonal non-Mueller matrices, whose respective columns are the output and input singular vectors of **Z**. Moreover, if we consider the SVD of the complete Mueller matrix **M**, the orthogonal matrices $\mathbf{M}'_{RI}$ and $\mathbf{M}'_{RO}$ represent respective input and output orthogonal physical transformations, i.e. respective rotations in the Poincaré sphere. Thus, different bases for the representation of input and output Stokes vectors can be specified, so that the form of $\mathbf{M}'_{RI}$ and $\mathbf{M}'_{RO}$ depend on the particular bases chosen. In particular, the input and output equivalent retarders $\mathbf{M}'_{RI}$ and $\mathbf{M}'_{RO}$ depend on three respective free parameters, whereas **Z** depends on ten independent parameters.

Thus, it is easy to show that the indices of purity (and hence the degree of purity) of **Z** are necessarily equal than those of **M**. It is also important to note that **Z** and **M** have equal transmittance for unpolarized light

$$P_i(\mathbf{Z}) = P_i(\mathbf{M}) \quad (i = 1, 2, 3),$$
$$z_{00} = m_{00}.$$
(256)

Concerning the polarizances of **Z**, it is easy to show that, as expected, their values are also physical invariants,

$$P(\mathbf{Z}) = P(\mathbf{M}), \quad D(\mathbf{Z}) = D(\mathbf{M}).$$
(257)

Nevertheless the polarizance vectors **D**, **P** are not physically invariant

$$\mathbf{P}(\mathbf{Z}) \neq \mathbf{P}(\mathbf{M}), \quad \mathbf{D}(\mathbf{Z}) \neq \mathbf{D}(\mathbf{M}),$$
(258)

and, therefore, they should be distinguished from the polarizance vectors $\mathbf{P}_Z, \mathbf{D}_Z$

$$\mathbf{P}_Z \equiv \frac{1}{m_{00}} (z_{10}, z_{20}, z_{30})^T, \quad \mathbf{D}_Z \equiv \frac{1}{m_{00}} (z_{01}, z_{02}, z_{03})^T,$$
(259)

which only depend on the elements of **Z**.

In some works published after the publication of the original version of the present review, in has been pointed out that, in general, $\mathbf{Z}$ preserves certain amount of retardance and that a proper dual retarder transformation is the so-called the *arrow decomposition* $\mathbf{M} = \mathbf{M}_{RO} \mathbf{M}_A \mathbf{M}_{RI}$, where $\mathbf{M}_A$ is the *arrow form* of $\mathbf{M}$,

$$\mathbf{M}_A = m_{00} \begin{pmatrix} 1 & \mathbf{D}_A^T \\ \mathbf{P}_A & \mathbf{m}_A \end{pmatrix},$$

$$\mathbf{m}_A = \mathrm{diag}(a_1, a_2, \varepsilon a_3), \quad a_1 \geq a_2 \geq |a_3| \geq 0,$$

$$\left(\mathbf{P}_A = \mathbf{m}_{RO} \mathbf{P}, \quad \mathbf{D}_A = \mathbf{m}_{RI} \mathbf{D}, \quad \varepsilon \equiv \det \mathbf{M}/|\det \mathbf{M}|\right).$$

[see J. J. Gil, "Transmittance constraints in serial decompositions of Mueller matrices. The arrow form of a Mueller matrix," J. Opt. Soc. Am. A **30**, 701-707 (2013); J. J. Gil, "Invariant quantities of a Mueller matrix under rotation and retarder transformations," J. Opt. Soc. Am. A **33**, 52-58 (2016).]

In consequence, from the arrow decomposition $\mathbf{M} = \mathbf{M}_{RO} \mathbf{M}_A \mathbf{M}_{RI}$ of $\mathbf{M}$, a complete description of the physical quantities embedded in $\mathbf{M}$ is the following:

*a)* 6 parameters of overall birefringence:

- *Input birefringence*, given by $\mathbf{M}_{RI}(\alpha_I, \delta_I, \Delta_I)$.
- *Output birefringence*, given by $\mathbf{M}_{RO}(\alpha_O, \delta_O, \Delta_O)$.

*b)* 10 parameters that are invariant under dual retarder transformations of $\mathbf{M}$:

- Transmittance for unpolarized light (*mean intensity coefficient*) $m_{00}$.
- Indices of purity $P_1, P_2, P_3$, which characterize completely the polarimetric purity of the system.
- Intrinsic polarizance vectors $\mathbf{P}_A$, $\mathbf{D}_A$ [observe that the polarizances $P(\mathbf{M})$, $D(\mathbf{M})$ are the respective magnitudes of these vectors].

The obtainment of this set of physical parameters constitutes an important way to exploit the polarimetric techniques, as well as a tool for analyzing experimental results. In particular, not only the mathematical invariants $(m_{00}, P_1, P_2, P_3)$ derived from the eigenvalues of the coherency matrix $\mathbf{H}$ associated with $\mathbf{M}$, but all the physical invariants $(m_{00}, P_1, P_2, P_3, \mathbf{P}_A, \mathbf{D}_A)$ can be used for the objective characterization of material samples.

Some theorems derived from the above analysis are the following

*a)* $\mathbf{M}_A$ is diagonal if and only if $P = D = 0$

*b)* Any Mueller matrix $\mathbf{M}$ that satisfies $P = D = 0$ can be written as $\mathbf{M} = \mathbf{M}_{RO} \mathbf{M}_{\Delta d} \mathbf{M}_{RI}$. This matrix corresponds to a parallel combination of retarders and depends on up to ten independent parameters.

*c)* $P = D = 1 \Rightarrow P_\Delta = 1$.

*d)* An ideal depolarizer can be synthesized through the parallel combination of retarders given by Eq. (202).

*e)* $P_\Delta^2(\mathbf{M}_A) = P_\Delta^2(\mathbf{M}) = \frac{1}{3}(D^2 + P^2 + 3P_s^2), \quad \left[P_s^2 = (a_1^2 + a_2^2 + a_3^2)/3\right].$

## 6. Some applications of polarimetry

*The rapid advances in both theoretical and experimental polarimetry are leading to a continuous expansion of the scope of its applications, beyond the summary presented below, which just reproduces the one included in the original version of this Review.*

The term *polarimetry* extends over a wide range of scientific, medical and industrial applications. An exhaustive summary of all these applications would require a voluminous treatise. Nevertheless, in order to provide a brief panoramic view, we include here some references to a number of fields where the polarimetric techniques play an important role.

*Light scattering* [8,34,215,216]: remote sensing [217], lidar, atmospheric phenomena, aerosols, hydrosols, surface roughness [26,218-222], particle sizing, particle characterization (contaminants, biological microorganisms…), surface characterization…

*Optical fiber and photonic devices* [143,146,223-228]: communication systems; characterization and control of polarization mode dispersion, fiber optic sensors…

*Synthetic aperture radar* (SAR) polarimetry [59,194,229]: airborne and spaceborne remote sensing, imaging and detection; vegetation, agriculture, crop classification, forestry, sea ice, sandy areas, Geoscience [230], digital terrain models, ground topographic mapping, Meteorology, observation and prediction of hurricanes, detection of devastated areas, oil spills, immersed targets, buried targets, anti-personnel mines [231]…

*Medicine & Biology* [232-236]: study, detection and imaging of immunological reactions, biological tissues, optical coherence tomography [237-240,232,233], human eye [241,242], oral precancer [209], DNA structure…

*Experimentation under microgravity conditions* [243,244].

*Photoelasticity* [127,245,246]

*Astronomy & Astrophysics* [36,247]: X-ray polarimetry, Solar polarimetry, interstellar dust, planetary atmospheres, black holes, satellite missions…

*Optics industry & research*: fabrication [248], characterization of optical components [249], analysis of optical systems [250,251], ray tracing [252], spectral filters design and fabrication [253]...

*Imaging polarimetry* [31,254-257,233,239,241].

*Plasma physics, nuclear fusion, ITER, Tokamak* [35,38]; synchrotron, particle physics...

*Microelectronics Industry & Metrology* [258,259].

*Research in Quantum Physics* [85,96,260-262]: Bell inequalities, optical computing, teleportation, cryptology…

*LCD Technologies,* [263,0,265].

*Thin films, Stratified & Layered Media* [266-268].

*Microwave transmission systems* [269].

### Acknowledgements

The author would like to thank Drs. I. San José, J. M. Correas, C. Ferreira, P. Melero and J. Delso, colleagues of the Group of Optical Polarimetry, for their helpful comments. The author is also obliged to Dr. R. Navarro for his valuable advice and encouragement to write this paper.

**References**

1. R. Loudon, *The quantum theory of light* (Clarendon, Oxford, 1983)
2. C. Brosseau, *Fundamentals of polarized light. A statistical approach* (Wiley, N. Y., 1998)
3. G. G. Stokes, Trans. Cambridge Phil. Soc. **9**, 399 (1852)
4. U. Fano, Phys. Rev. **93**, 121 (1953)
5. R. Barakat, Opt. Acta **32**, 295 (1985)
6. E. Wolf, Nuovo Cimento **13**, 1165 (1959)
7. L. Mandel, Proc. Phys. Soc. **81**, 1104 (1963)
8. C. F. Bohren, D. R. Huffman, *Absorption and Scattering of Light by Small Particles* (Wiley-Interscience, N. Y., 1983)
9. R. C. Jones, J. Opt. Soc Am. **31**, 488 (1941)
10. P. Soleillet, Ann. de Physique **12**, 23 (1929)
11. F. Perrin, J. Chem. Phys. **10**, 415 (1942)
12. H. Mueller, J. Opt. Soc. Am. **38**, 661 (1948)
13. N. G. Parke III, *Matrix optics* Ph. D. thesis, M.I.T., 1948
14. N. G. Parke III, J. Math. Phys. **28**, 131 (1949)
15. W. Swindell, ed., *Benchmark Papers in Optics v1: Polarized Light* (Dowden, Hutchinson and Ross, Stroudsburg, 1975)
16. J. J. Gil, E. Bernabéu, Optik **76**, 67 (1987)
17. S.-Y. Lu, R. A. Chipman, J. Opt. Soc. Am. A **11**, 766 (1994)
18. R. M. A. Azzam, N. M. Bashara, *Ellipsometry and Polarized Light* (North-Holland, N. Y., 1977)
19. M. Schubert, Ann. Phys. **15**, 480 (2006)
20. R. M. A. Azzam, Opt. Lett. **2**, 148 (1978)
21. P. S. Hauge, J. Opt. Soc. Am. **68**, 1519 (1978)
22. R. C. Thompson, J. R. Bottiger, E. S. Fry, Appl. Opt. **19**, 1323 (1980)
23. E. Bernabéu, J. J. Gil, J. Opt. **16**, 139 (1985)
24. D. B. Chenault, J. L. Pezzanity, R. A Chipman, SPIE **1746**, 231 (1992)
25. D. H. Goldstein, R. A. Chipman, eds. *Polarization Analysis and Measurement*, Proceedings of SPIE **1746**, 1992
26. F. Delplancke, Appl. Opt. **36**, 5388 (1997)
27. M. Mujat, A. Dogariu, Appl. Opt. **40**, 34 (2001)
28. K. Ichimoto, K. Shinoda, T. Yamamoto, J. Kiyohara, Publ. Natl. Astron. Obs. Japan **9**, 11 (2006)
29. R. M. A. Azzam, Optik **52**, 253 (1979)
30. F. Gori, Opt. Lett. **24**, 584 (1999)
31. J.-K. Lee, J. T. Shen, A. Heifetz, R. Tripathi, M.S. Shahriar, Opt. Commun. **259**, 484 (2006)
32. S. Tyo, D. L Goldstein, D. Chenault, J. A. Shaw, Appl. Opt. **45**, 5453 (2006)
33. H. C. van de Hulst, *Light scattering by small particles* Ch. 5 (Wiley, N. Y., 1957)
34. F. Moreno, F. González, eds. *Light scattering by Microstructures* (Lecture Notes in Physics, Springer-Verlag, Berlin, 2000)
35. T. Kondoh, A. E. Costley, T. Sugie, Y. Kawano, A. Malaquias, C. I. Walker, Rev. Sci. Instrum. **75**, 3420 (2004)
36. C. Heiles, P. Perillat, M. Nolan, D. Lorimer, R. Bhat, T. Ghosh, M. Lewis, K. O'Neil, C. Salter, S Stanimirovic, Publications of the Astronomical Society of the Pacific, **113**, 1274-1288 (2001). Available from <http://www.journals.uchicago.edu/cgi-bin/resolve?PASP201143PDF>
37. A. L. Fymat, Opt. Eng. **20**, 25 (1981)
38. J. H. Rommers, S. Barry, R. Behn, C. Nieswand, Plasma Phys. Control. Fusion **40**, 2073 (1998)
39. M. Mujat, E. Baleine, A. Dogariu, J. Opt. Soc. Am. A **21**, 2244 (2004)
40. D. T. Chuss, E. J. Wollack, S. H. Moseley, G. Novak, Appl. Opt. **45**, 5107 (2006)
41. J. J. Gil, J. M. Correas, P. A. Melero, C. Ferreira, Monografías del Seminario Matemático García de Galdeano **31**, 161 (2004). Available from <http://www.unizar.es/galdeano/actas_pau/PDFVIII/pp161-167.pdf>
42. T. Tudor, I. Vinkler, Pure Appl. Opt. **7**, 1451 (1998)
43. N. Wiener, Acta. Math. **55**, 182 (1930)
44. G. B. Parrent, P. Roman, Nuovo Cimento **15**, 370 (1960)
45. R. Barakat, J. Opt. Soc. Am. **53**, 317 (1963)
46. U. Fano, Rev. Mod. Phys. **29**, 74 (1957)
47. W. H. McMaster, J. Opt. Soc. Am. **22**, 351 (1954)
48. S. R. Cloude, in *Direct and Inverse Methods in Radar Polarimetry, Part I* edited by W. M. Boerner et al. (Kluwer Academic Publishers, Dordrecht, The Netherlands, 1992)
49. E. L. O'Neill, *Introduction to Statistical Optics* Ch. 9 (Addison-Wesley, Reading, Mass., 1963)
50. A. Al-Qasimi, O. Korotkova, D. James, E. Wolf, Opt. Lett. **32**, 1015 (2007)

51. J. Ellis, A. Dogariu, J. Opt. Soc. Am. A **21**, 988 (2004)
52. A. Luis, Phys. Rev. A **75**, 053806 (2007)
53. R. Barakat, Opt. Acta **30**, 1171 (1983)
54. R. Barakat, C. Brosseau, J. Opt. Soc. Am. A **10**, 529 (1992)
55. R. Barakat, Opt. Commun. **123**, 443 (1996)
56. C. Brosseau, D. Bicout, Phys. Rev. E **50**, 4997 (1994)
57. S. R. Cloude, E. Pottier, Opt. Eng. **34**, 1599 (1995)
58. F. Le Roy-Brehonnet, B Le Jeune, P Y Gerligand, J Cariou, J Lotrian, Pure Appl. Opt. **6**, 385 (1997)
59. W.-M. Boerner, in *Radar Polarimetry and Interferometry 2007* . (Educational Notes RTO-EN-SET-081bis, Paper 12. Neuilly-sur-Seine, France: RTO p. 12-1). Available form <http://www.rto.nato.int/abstracts.asp>
60. A. F. Fercher, P. F. Steeger, Opt. Acta **28**, 443 (1981)
61. R. Barakat, J. Opt. Soc. Am. A **4**, 1256 (1987).
62. C. Brosseau, J. Opt. Soc. Am. A **6**, 649 (1989)
63. C. Brosseau, R. Barakat, E. Rockower, Opt. Commun. **82**, 204 (1991)
64. C. Brosseau, J. Mod. Opt. **39**, 1167 (1992)
65. C. Brosseau, Appl. Opt. **34**, 4788 (1995)
66. S. Pancharatnam, Proc. Ind. Acad. Sci. A **44**, 247 (1956)
67. S. Pancharatnam, Proc. Ind. Acad. Sci. A **44,** 398 (1956)
68. F. Gori, Opt. Lett. **23**, 241 (1998)
69. J. M. Movilla, G. Piquero, R. Martínez-Herrero, P. M: Mejías, Opt. Commun. **149**, 230 (1998)
70. J. Tervo, T. Setälä, A. T. Friberg, Opt. Express **11**, 1137 (2003)
71. J. Ellis, A. Dogariu, Opt. Lett. **29**, 536 (2004)
72. J. Tervo, T. Setälä, A. T. Friberg, J. Opt. Soc. Am. A **21**, 2205 (2004)
73. P. Réfrégier, F. Goudail, Opt. Express **13**, 6051 (2005)
74. R. Castañeda, Opt. Commun. **267**, 4 (2006)
75. O. Korotkova, E. Wolf, Opt. Lett. **30**, 1 (2005)
76. E. Wolf, Phys Lett. A **312**, 263 (2003).
77. A. Luis, "Properties of spatial-angular Stokes parameters" Opt. Commun. **251**, 243 (2005)
78. A. Luis, Opt. Commun. **263**, 141 (2006)
79. A. Luis, Opt. Commun. **246**, 437 (2005)
80. O. Korotkova, E. Wolf, J. Mod. Opt. **52**, 2659 (2005)
81. O. Korotkova, E. Wolf, J. Mod. Opt. **52**, 2673 (2005)
82. E. Collett, Am J. Phys. **38**, 563 (1969)
83. A. Luis, Phys. Rev. A **66**, 013806 (2002)
84. A. Luis, "Polarization distribution and degree of polarization for three-dimensional quantum light fields" Phys. Rev. A **71**, 063815 (2005)
85. A. Luis, N. Korolkova, Phys. Rev. A **74**, 043817 (2006)
86. A. Luis, Opt. Commun. **273**, 173 (2007)
87. P. Roman, Nuovo Cimento **13**, 2546 (1959)
88. A. Papoulis, *Probability, random variables and stochastic processes* (Mac Graw-Hill, Singapur, 1984)
89. M. R. Dennis, J. Opt. A: Pure Appl. Opt. **6**, S26 (2004)
90. J. H. Hannay, J. Mod Opt. **45**, 1001 (1998)
91. T. Saastamoinnen, J. Tervo, J. Mod. Opt. **51**, 2039 (2004)
92. J. Ellis, A. Dogariu, Opt. Commun. **253**, 257 (2005)
93. T. Carozzi, R. Karlsson, J. Bergman, Phys. Rev. E **61**, 2024 (2000)
94. T. Setälä, A. Shevchenko, M. Kaivola, A. T. Friberg, Phys. Rev. E **66**, 016615 (2002)
95. A. Luis, Opt. Commun. **253**, 10 (2005)
96. A. P. Alodzhants, S. M. Arakelian, Opt. Spectrosc. **97**, 453 (2004)
97. A. P. Alodzhants, S. M. Arakelian, J. Mod. Opt. **46**, 475 (1999)
98. T. Setälä, M. Kaivola, A. T. Friberg, Phys. Rev. Lett. **88**, 1239021 (2002)
99. R. Dändliker, P. Tortora, L. Vaccaro, A. Nesci, J. Opt. A Pure Appl. Opt. **6**, S18 (2004)
100. J. Ellis, A. Dogariu, Phys. Rev. Lett. **95**, 203905 (2005)
101. J. Ellis, A. Dogariu, S. Ponomarenko, E. Wolf, Opt. Lett. **29** 1536 (2004)
102. J. M. Correas, P. A. Melero, J. J. Gil, Monografías del Seminario Matemático García de Galdeano **27**, 23 (2003). Available from <http://www.unizar.es/galdeano/actas_pau/PDF/233.pdf>
103. R. A. Horn, C. R. Johnson, *Matrix Analysis* (Cambridge Univ. Press, 1985)

104. J.C. Samson, Geophys. J. R. Astron. Soc. **34**, 403 (1973)
105. J. Ellis, A. Dogariu, S. Ponomarenko, E. Wolf, Opt. Commun. **248**, 333 (2005)
106. F. T. Hioe, J. Mod. Opt. **53**, 1715 (2006)
107. P. Réfrégier, M. Roche, F. Goudail, J. Opt. Soc. Am. A **23**, 124 (2006)
108. P. Réfrégier, F. Goudail, J. Opt. Soc. Am. A **23**, 671 (2006)
109. M. R. Dennis, J. Opt. Soc. Am. A **24**, 2065 (2007)
110. A. Luis, Phys. Rev. A. **71**, 1023810 (2005)
111. P. Vahimaa, J. Tervo, J. Opt. A: Pure Appl. Opt. **6**, S41 (2004)
112. W. A. Holm, R. M. Barnes in *Proceedings of the 1988 IEEE USA radar conference*, p. 249
113. S. R. Cloude, E. Pottier, IEEE Trans. GRS **34**, 498 (1996)
114. P. Réfrégier, F. Goudail, P. Chavel, A. Friberg, J. Opt. Soc. Am. A **21**, 21124 (2004)
115. P. Réfrégier, J. Morio, J. Opt. Soc. Am. A **23**, 3036 (2006)
116. R. Simon, Opt. Commun. **77**, 349 (1990)
117. A. P. Loeber, J. Opt. Soc. Am. **72**, 650 (1982)
118. L. N. Dlugnikov, Opt. Acta, **31**, 803 (1984)
119. B. Chakraborty, J. Opt. Soc. Am. A **3**, 1422 (1986)
120. G. Indebetouw, Optik **85**, 78 (1990)
121. R. Simon, Opt.Commun. **42**, 293 (1982)
122. J. W. Hovenier, Appl. Opt. **33**, 8318 (1994)
123. R. C. Jones, J. Opt Soc Am. **32**, 486 (1942)
124. R. Barakat, J. Mod. Opt. **34**, 1535 (1987)
125. C. Brosseau, R. Barakat, Opt. Commun. **84**, 127 (1991)
126. A. B. Kostinski, R. C. Givens, J. Mod. Opt. **39**, 1947 (1992)
127. P. S. Theocaris, E. E. Gdoutos, *Matrix theory of photoelasticity* Ch 4 (Springer-Verlag, Berlin, 1979)
128. N. Vansteenkiste, P. Nignolo, A. Aspect, J. Opt. Soc. Am.A **10**, 2240 (1993)
129. R. J. Potton, Rep. Prog. Phys. **67**, 717 (2004).
130. J. J. Gil, Ph. D. thesis, Facultad de Ciencias, Univ. Zaragoza, Spain, 1983. Available from <http://www.pepegil.es/PhD-Thesis-JJ-Gil.pdf>
131. J. J. Gil, E. Bernabéu, Opt. Acta **33**, 185 (1986)
132. H. Hurwitz, R. C. Jones, J. Opt. Soc. Am. **31**, 493 (1941)
133. J. R. Priebe, J. Opt. Soc. Am. **59**, 176 (1969)
134. P. H. Richter, J. Opt. Soc. Am **69**, 460 (1979)
135. C. Whitney, J. Opt. Soc. Am. **61**, 1207 (1971)
136. R. Barakat, Eur. J. Phys. **19**, 209 (1998)
137. S. T. Tang, H. S. Kwok, J. Opt. Soc. Am. A **18**, 2138 (2001)
138. J. J. Gil. E. Bernabéu, Opt. Pur. Apl. **15**, 39 (1982)
139. R. C. Jones, J. Opt. Soc. Am. **38**, 671 (1948)
140. N. Go, J. Phys. Soc. Jpn. **23**, 88 (1967)
141. R. Barakat, J. Opt. Soc. Am. A **13**, 158 (1996)
142. R. M. A. Azzam, J. Opt. Soc. Am. **68**, 1756 (1978)
143. C. Brosseau, Opt. Lett. **20**, 1221 (1995)
144. J. F. Mosiño, O. Barbosa-García, M. A. Meneses-Nava, L. A. Díaz-Torres, E. de la Rosa-Cruz, J. T. Vega-Durán, J. Opt. A: Pure Appl. Opt. **4**, 419 (2002)
145. C. D. Poole, R. E. Wagner, Electron. Lett. **22**, 1029 (1986)
146. W. Shieh, IEEE Phot. Tech. Lett. **11**, 677 (1999)
147. E. Collett, Am. J. Phys. **39**, 517 (1971)
148. A. S. Marathay, J. Opt. Soc. Am. **55**, 969 (1965)
149. S.-Y. Lu, R. A. Chipman, J. Opt. Soc. Am. A **13**, 1106 (1996)
150. C. R. Fernández-Pousa, I. Moreno, N. Bennis, C. Gómez-Reino, C. Ferreira, Opt. Commun. **183**, 347 (2000)
151. R. M. A. Azzam, J. Opt. Soc. Am. **62**, 1252 (1972)
152. Z-F Xing, J. Mod Opt. **39**, 461 (1992)
153. S. N. Savenkov, O. I Sydoruk, R. S. Muttiah, J. Opt. Soc. Am. A **22**, 1447 (2005)
154. G. S. Agarwal, Opt. Commun. **82**, 213 (1991)
155. S. N. Savenkov, V. V. Marienko, E. A. Oberemok, Phys. Rev. E **74**, 056607 (2006)
156. M. V. Berry, M. R. Dennis, J. Opt. A: Pure Appl. Opt. **6**, S24 (2004)
157. H. Takenaka, Nouv. Rev. d'Optique **4**, 37 (1973)
158. Sudha, A. V. Gopala, J. Opt. Soc. Am. A **18**, 3130 (2001)
159. C. S. Brown, A. E. Bak, Opt. Eng. **34**, 1625 (1995)
160. R. Sridhar, R. Simon, J. Mod Opt. **41**, 1903 (1994)
161. S. R. Cloude, Optik **75**, 26 (1986).
162. S. R. Cloude, J. Electromag. Waves Appl. **6**, 947 (1992)
163. S. R. Cloude, SPIE **2256**, 292 (1994)
164. E. S. Fry, G. W. Kattawar, Appl. Opt. **20**, 3428 (1981)

165. K Kim. L Mandel, E. Wolf, J. Opt. Soc. Am. A **4**, 433 (1987)
166. S. R. Cloude, SPIE **1166**, 177 (1989)
167. C. V. M. van der Mee, J. Math Phys. **34**, 5072 (1993)
168. J. J. Gil, J. Opt. Soc. Am. A **17**, 328 (2000)
169. C. V. M. van der Mee, J. W. Hovenier, J. Math. Phys. **33**, 3574 (1992)
170. D. I. Nagirner, Astron. Astrophys. **275**, 318 (1993)
171. M. S. Kumar, R Simon, Opt Commun. **11**, 2305 (1994)
172. C. R. Givens, A. B. Kostinski, J. Mod. Opt. **40**, 471 (1993)
173. A. B. Kostinski, C. R. Givens, J. M. Kwiatkowski, Appl. Opt. **32**, 1646 (1993)
174. A. V. Gopala , K. S. Mallesh, Sudha, J. Mod. Optics **45**, 955 (1998)
175. A. V. Gopala, K. S. Mallesh, Sudha, J. Mod. Optics **45**, 989 (1998)
176. J. J. Gil, E. Bernabeu, Opt. Acta **32**, 259 (1985)
177. J. W. Hovenier, H. C. van de Hulst, C. V. M. van der Mee, Astron. Astrophys. **157**, 301 (1986)
178. R. Simon, J. Mod. Opt. **34**, 569 (1987)
179. G. Arfken, *Mathematical methods for physicists* Ch. 4 (Academic Press, 1970)
180. A. Aiello, G. Puentes, J. P. Woerdman, arXiv:quant-ph/0611179v1 (2006)
181. Sudha, A.V. Gopala Rao, A. R. Usha Devi, A.K. Rajagopal, arXiv:0704.0147v2 [physics.optics] (2007)
182. D. G. M. Anderson, R. Barakat, J. Opt. Soc. Am. A **11**, 2305 (1994)
183. I. Kuscer, M. Ribaric, Opt. Acta **6**, 42 (1959)
184. K. D. Abhyankar, A. L. Fymat, J. Math Phys **10**, 1935 (1969)
185. R. Barakat, Opt. Commun. **38**, 159 (1981)
186. C. Brosseau, C. R. Givens, A. B. Kostinski, J. Opt. Soc Am. A **10**, 2248 (1993)
187. J. J. Gil, in *Light scattering from microestructures* (Springer: Lecture Notes in Physics, 2000)
188. R. Espinosa-Luna, Appl. Opt. **46**, 6047-6054 (2007)
189. C. Brosseau, Optik **85**, 83 (1990)
190. A. B. Kostinski, Appl. Opt. **31**, 3506 (1992)
191. F. Le Roy-Brehonnet, B. Le Jeune, Prog. Quant. Electr. **21**, 109 (1997)
192. F. Le Roy-Brehonnet, B Le Jeune, P Eliès, J Cariou, J Lotrian, J. Phys. D: Appl. Phys. **29**, 34-38 (1996)
193. A. Aiello, J. P. Woerdman, Phys. Rev Lett. **94**, 090406 (2005)
194. R. A. Chipman, Appl. Opt. **44**, 2490 (2005)
195. R. Touzi, W.M. Boerner, J.S. Lee, E. Lueneburg, Can. J. Remote Sensing **30**, 380 (2004). Available from <www.radarsat2.info/outreach/resources/Review-of-Polarimetry.pdf>
196. C. Brosseau, Opt. Commun. **131**, 229 (1996)
197. F. Le Roy-Bréhonnet, B. Le Jeune, P. Y. Gerligand, J. Cariou, J. Lotrian, Pure Appl. Opt. **6**, 385 (1996)
198. B. DeBoo, J. Sasian, R. Chipman, Opt. Express **12**, 4941 (2004)
199. C. Ferreira, I. San José, J. J. Gil, J. M. Correas, Monografías del Seminario Matemático García de Galdeano **33**, 115 (2006). Available from <http://www.unizar.es/galdeano/actas_pau/PDF IX/FerSanGilCor05.pdf>
200. R. A. Chipman, in *The Handbook of Optics*, Chap. 22 (McGraw-Hill, N. Y., 1994)
201. G. Puentes, D. Voigt, A. Aiello, J. P. Woerdman, Opt. Lett. **30**, 3216 (2005).
202. J. J. Gil, SPIE **5622**, 725 (2004)
203. E. C. G. Sudarshan, Phys. Rev. **121**, 920 (1961)
204. G. Kimura, Phys. Lett. A **314**, 339 (2003)
205. G. L. Long, Y-F Zhou, J-Q. Jin, Y. Sun,1, H-W. Lee, Found. Phys. **36**, 1217 (2006)
206. S. Lu, A. P. Loeber, J. Opt. Soc. Am. **65**, 248 (1975)
207. W. A. Shurcliff, *Polarized light* (Harvard U. Press, Cambridge, Mass. 1962)
208. S-Y. Lu, R. A. Chipman, Opt. Commun. **146**, 11 (1998)
209. J. Chung, W. Jung, M. J. Hammer-Wilson, P. Wilder-Smith, Z. Chen, Appl. Opt. **46**, 3038 (2007)
210. M. K. Swami, S. Manhas, P. Buddhiwant, N. Ghosh, A. Uppal, P. K. Gupta, Opt. Express **14**, 9324 (2006)
211. C. Collet, J. Zallat, Y. Takakura, Opt. Express **12**, 1271 (2004)
212. J Morio, F Goudail, Opt. Lett **29**, 2234 (2004)
213. R. Ossikovski, A. De Martino, S. Guyot, Opt. Lett. **32**, 689 (2007)
214. M. Anastasiadou, S. Ben Hatit, R. Ossikovski, S. Guyot, A. De Martino, J. Eur Opt. Soc. Rapid Pub. **2**, 07018 (2007)

215. M. I. Mishchenko, L. D. Travis, A. A. Lacis, *Scattering, Absorption and Emission of Light by Small Particles* (Cambridge University Press, Cambridge, UK, 2002).

216. M.I. Mishchenko, J.W. Hovenier, L.D. Travis, *Light scattering by nonspherical particles* (Academic Press, San Diego, 1999)

217. G. Videen, Y, Yatskiv, M. I. Mishchenko eds. *Photopolarimetry in Remote Sensing* Proceedings of the NATO Advanced Study Institute, Yalta, 2003

218. B. J. DeBoo, J. M. Sasian, R. A. Chipman, Appl. Opt. **44**, 5434 (2005)

219. R. Espinosa-Luna , G. Atondo-Rubio, A. Mendoza-Suárez, Opt. Commun. **257**, 62 (2006)

220. G. Videen, J.-Y. Hsu, W. S. Bickel, W. L. Wolfe, J. Opt. Soc. Am. A **9**, 1111 (1992)

221. R. Espinosa-Luna , A. Mendoza-Suárez, G. Atondo-Rubio, S. Hinojosa , J. Rivera-Vázquez, J. T. Guillén-Bonilla, Opt. Commun. **259**, 60 (2006)

222. S. Hinojosa, R. Espinosa-Luna, Opt. Commun. **250**, 213 (2005)

223. E. Collett, *Polarized light in Fiber Optics* (PolaWave Group, Lincroft, N.J., 2003)

224. J. P. Gordon, H. Kogelnik, Proc. Natl. Acad. Sci. USA **97**, 4541-4550 (2000)

225. H. Dong, P. Shum, M. Yan, J. Q. Zhou, G. X. Ning, Opt. Express **14**, 5067 (2006)

226. P. Olivard, P. Y. Gerligand, B. Le Jeune, J. Cariou, J. Lotrian, J. Phys. D: Appl. Phys. **32**, 1618 (1999)

227. S. X. Wang, A. M. Weiner, Opt. Lett. **29**, 923 (2004)

228. F. Tang, X.-Z. Wang, Y. Zhang, W. Jing, (*to be published in Applied Optics, accepted march 2007*)

229. ESA Proceedings of POLinSAR2003: Workshop on Applications of SAR Polarimetry and Polarimetric Interferometry. Available from <http://earth.esa.int/workshops/polinsar2003/pr.html>

230. F. T. Ulaby, ed. *Radar Polarimetry for Geoscience Applications* (Artech House Remote Sensing Library, 1990)

231. L. Sagués, J. M. López-Sánchez, J. Fortuny, X. Fàbregas, A. Broquetas, A. J. Sieber, in *Ultra-Wideband, Short-Pulse Electromagnetics 5* P. D. Smith and S. R. Cloude eds. (Springer US 2002)

232. S. L. Jacques, R. J. Roman, K. Lee, J. Biomed. Opt. **7**, 329 (2002).

233. J. S. Baba, J.-R. Chung, A. H. DeLaughter, B. D. Cameron, G. L. Coté, J. Biomed. Opt. **7**, 341 (2002)

234. A. Lompado, M. H. Smith, V. Krishnaswamy, SPIE **4263**, 161 (2001)

235. S. Guyot, M. Anastasiadou, E. Deléchelle, A. De Martino, Opt. Express **15**, 7393 (2007)

236. F. Boulvert, Y. Piederrière, G. Le Brun, B. Le Jeune, J. Cariou, Opt. Commun. **272**, 534 (2007)

237. V. V. Tuchin; J. A. Izatt; J. G. Fujimoto, eds. *Coherence Domain Optical Methods and Optical Coherence Tomography in Biomedicine VIII* (Proceedings of SPIE **5316**, 2004)

238. M. Todorovic, S. Jiao, L. V. Wang, G. Stoica, Opt. Lett. **29**, 2402 (2004)

239. S. Jiao, L. V. Wang, J. Biomed. Opt. **7**, 350 (2002)

240. S. Jiao, L. V. Wang, Opt. Lett. **27**, 101 (2002)

241. J. M. Bueno, F. Vargas-Martín, Appl. Opt. **41**, 116 (2002)

242. J. M. Bueno, J. Opt. A: Pure Appl. Opt. **6**, S91 (2004)

243. S. N. Savenkov, L. T. Mishchenko, R.S. Muttia, Y. A. Oberemok, I.A. Mishchenko, J. Quant. Spectrosc. Radiat. Transfer **88**, 327 (2004)

244. A. C. Levasseur-Regourd, in *Photopolarimetry in Remote Sensing* (NATO Science Series II: Mathematics, Physics and Chemistry Vol. 161, 2005)

245. H. Yang ; S. Gibson; R. Tomlinson, Exp. Mech. **46**, 619 (2006)

246. G. C. Gaunaurd, Opt. Eng. **42**, 467 (2003)

247. Stefan Larsson, Mark Pearce, Nucl. Instr. and Meth. A **525**, 148 (2004)

248. J. Wolfe, R. A. Chipman, Opt. Express **12**, 3443 (2004)

249. P.-Y. Gerligand, M. Smith, R. A. Chipman, Opt. Express **4**, 420 (1999)

250. R.A. Chipman, Opt. Eng. **28**, 90 (1989)

251. R.A. Chipman, SPIE **1166**, 79 (1989)

252. R.A. Chipman, Opt. Eng. **34**, 1636 (1995)

253. I. Abdulhalim, Opt. Commun. **267**, 36 (2006)

254. J. L Pezzaniti, R. C. Chipman, Opt. Eng. **34**, 1558 (1995)

255. P. Réfrégier, F. Goudail, N. Roux, J. Opt. Soc. Am. A **21**, 2292 (2004)

256. D. Lara, C. Dainty, Appl. opt. **45**, 1917 (2006)

257. C. Kieleck, M. Floc'h, G. Le Brun, J. Cariou, J. Lotrian, Pure Appl. Opt. **6**, 749 (1997)

258. T. Novikova, A. De Martino, S. B. Hatit, B. Drévillon, Appl. Opt. **45**, 3688 (2006)
259. T. Novikova, A. De Martino, P. Bulkin, Q. Nguyen, B. Drévillon, Opt. Express **15**, 2033 (2007)
260. A. Ekert, A. Zeilinger, *The Physics of Quantum Information*, edited by D. Bouwmeester (Springer, Berlin, 2000)
261. G. E. Stedman, J. Mod. Opt. **36**, 461 (1989)
262. M. Takeuchi, T. Takano, S. Ichihara, A. Yamaguchi, M. Kumakura , T. Yabuzaki, Y.Takahashi, Appl. Phys. B **83**, 33 (2006)
263. C. Bohley, T, Scharf, J. Opt. A: Pure Appl. Opt. **6**, S77 (2004)
264. I. Dahl, Meas. Sci. Technol. **12**, 1938 (2001)
265. J. L. Pezzaniti, S. C. McClain, R. A. Chipman, S.-Y. Lu, Opt. Lett. **18**, 2071 (1993)
266. K. L. Woon, M. O'Neill, G. J. Richards, M. P. Aldred, S. M. Kelly, J. Opt. Soc. Am. A **22**, 760 (2005)
267. P. Yeh, *Optical waves in layered media* (Wiley Series in Pure and Applied Optics, 1988)
268. C. Chou, H.-K. Teng, C.-J. Yu, H.-S. Huang, Opt. Commun. **273**, 74 (2007)
269. H. Liu, X. Zhang, K. Chen, Opt. Commun. **236**, 109 (2004)